\documentclass{article}
\def\twocol{"n"}

\usepackage[utf8]{inputenc}
\usepackage{graphicx}
\usepackage{natbib}
\usepackage{hyperref}
\usepackage{xr}

\usepackage[english]{babel}
\usepackage{float}
\usepackage{amsmath}
\usepackage{fullpage}
\usepackage{caption}
\usepackage{ifthen}
\usepackage[format=hang]{subfig}
\usepackage{threeparttablex}
\usepackage{tabularx}
\usepackage{longtable}
\usepackage{booktabs}
\usepackage{geometry}
\usepackage{lscape}
\usepackage{afterpage}
\usepackage{xcolor}
\usepackage{placeins}

\setcitestyle{authoryear,open={(},close={)}} 
\setcitestyle{citesep={;}}

\definecolor{magenta_custom}{rgb}{1, 0, 1}
\definecolor{green_custom}{rgb}{0, 0.7, 0}

\graphicspath{ {images/} }

\begin{document}

\title{Jupiter's ultraviolet auroral bridge: the influence of the solar wind on polar auroral morphology}
\author{
    L. A. Head \thanks{Corresponding author: LA.Head@uliege.be (email address)},
    D. Grodent,
    B. Bonfond,
    A. Sulaiman,
    A. Moirano,\\
    G. Sicorello,
    S. Elliott,
    M. F. Vogt,
    C. K. Louis,
    N. Kruegler,\\
    J. Vinesse,
    T. K. Greathouse
}

\abstract{
    Jupiter's ultraviolet aurora frequently shows a number of arcs between the dusk-side polar region and the main emission, which are denoted as ``bridges''.
    This work presents a largely automated detection and statistical analysis of bridges over 248 Hubble-Space-Telescope observations, alongside a multi-instrument study of crossings of magnetic field lines connected to bridges by the Juno spacecraft during its first 30 perijoves.
    Bridges are observed to arise on timescales of $\sim$2 hours, can persist over a full Jupiter rotation, and are conjugate between hemispheres.
    The appearance of bridges is associated with compression of the magnetosphere, likely by the solar wind.
    Low-altitude bridge crossings are associated with upward-dominated, broadband electron distributions, consistent with Zone-II aurorae, as well as with plasma-wave emission observed by Juno-Waves, in agreement with existing theoretical models for the generation of polar-region aurorae.
    Main-emission crossings where no bridges are visible also show characteristics associated with bridges (more upward electron flux, plasma-wave emission), which is not the case for main-emission crossings with visible bridges, as though bridges remain present but spatially indistinguishable from the main emission in the former case.
    In all, compression of the magnetosphere may work to spatially separate the Zone-I and Zone-II regions of the main emission, in the form of Zone-II bridges.
}

\maketitle

\FloatBarrier
\section{Introduction}

Jupiter's ultraviolet (UV) aurora is home to a number of distinct and discrete features.
Of these, Jupiter's UV auroral ``bridges'' \citep{pardocantos+:2019} are one of the largest and most consistently present, though, thus far, they have typically been discussed in the context of wider studies of the aurora \citep[e.g.][]{palmaerts+:2023,greathouse+:2021,nichols+:2009b} and their underlying processes not yet the subject of a dedicated study. 
This work will exclusively address the auroral bridge and thus provide details on its observational characteristics, origins, and relation to other features in Jupiter's aurora.

\begin{figure}[htb]
    \centering
    \ifthenelse{\equal{\twocol}{"y"}}{
        \includegraphics[width=\linewidth]{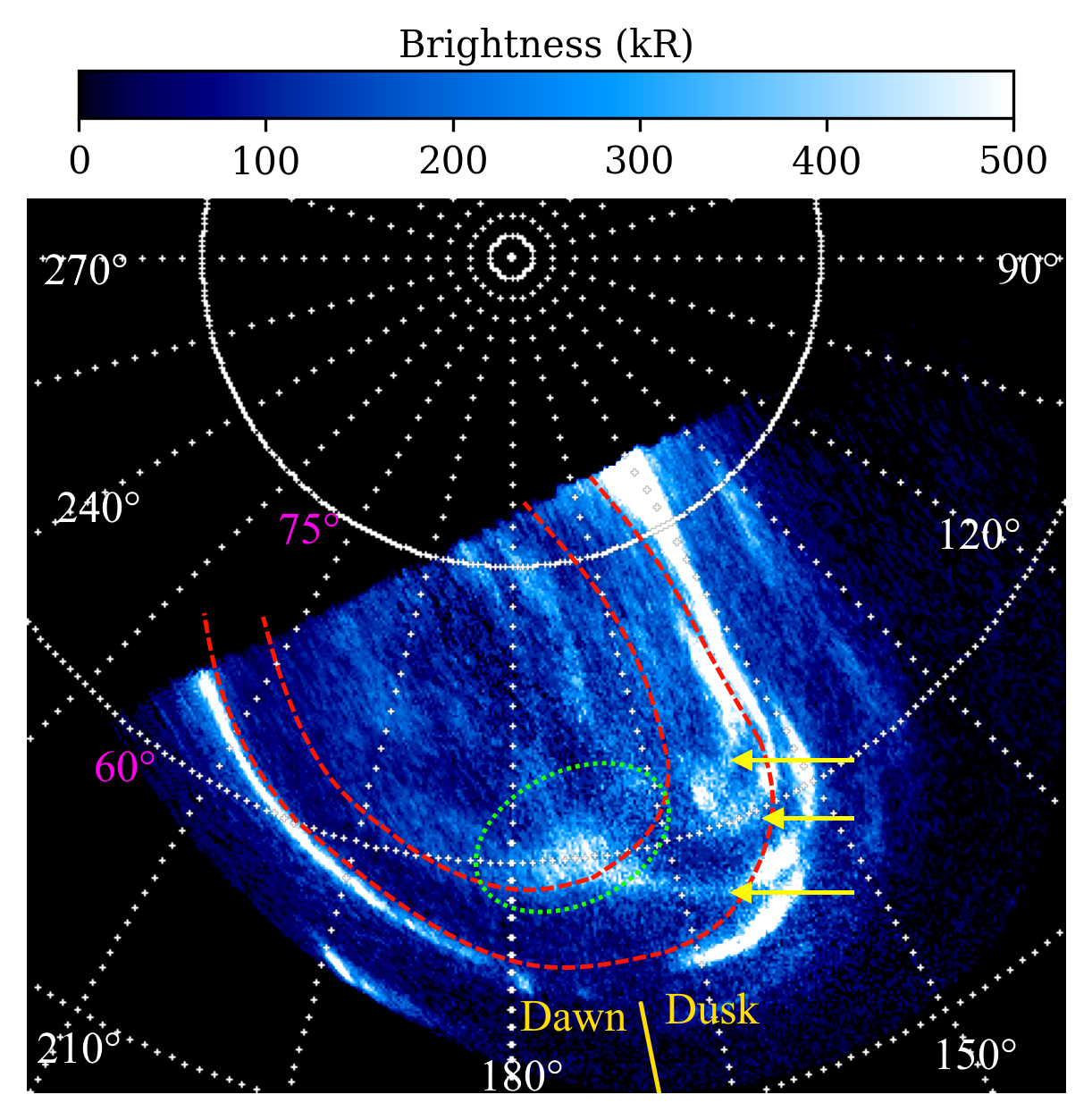}
    }
    {
        \includegraphics[width=\linewidth]{Plot_ConvolvedPolarview_odxc01okq_brightness_annotated.png}
    }
    \caption{
        An 300-second image of the northern jovian UV aurora captured by HST during the GO-15638 campaign (exposure ID: odxc01okq). A 15°-by-15° grid in System-III longitude and planetocentric latitude is included; the System-III longitude of certain meridians are given in white, and certain planetocentric latitudes in magenta. The average subsolar longitude during this exposure (170\textdegree) is denoted by a solid yellow line, and the positions of the dawn and dusk hemispheres are included to guide the reader. The approximate location of the polar collar is enclosed by red dashed lines, and that of the noon active region by a green dashed ellipse. Bridges are highlighted with yellow arrows.
    }
    \label{fig:bridge_example}
\end{figure}

In this work, ``bridge'' refers to any arc-like feature in Jupiter's auroral polar region that spans (though not necessarily fully) the polar collar \citep{greathouse+:2021} between Jupiter's main auroral emission (ME) \citep[e.g.][]{grodent+:2003} and the day-side active region \citep{nichols+:2017}. 
An example image of the aurora showing multiple bridges is given in Figure \ref{fig:bridge_example}.
It can be seen that these bridges are present within the dusk-side polar collar and that they connect the ME to the noon active region.
This feature has previously been referred to as the ``arc-like feature of the polar active region'' \citep{grodent:2015} or ``arcs parallel to the main oval'' \citep{nichols+:2017}; this work will use the term ``bridge'' to highlight its bridge-like nature between the ME and polar region as well as to distinguish this feature from other arcs in the polar region, such as the Polar Auroral Filaments (PAFs) \citep{nichols+:2009}. 

Thus far, the only dedicated study of the auroral bridge was a Master's thesis \citep{pardocantos+:2019} which analysed three images of the aurora taken by the Space Telescope Imaging Spectrograph (STIS) instrument aboard the Hubble Space Telescope (HST) that contained bridges.
In these cases, the bridges were found to map to the dusk-side magnetosphere between 10 and 22 magnetic local time (MLT) and be largely confined to distances greater than 60~R$_{J}$.
Based on its approximate mapped position in the magnetosphere, the bridge was suggested to arise from vorticity in the dusk-side plasma flow \citep{fukazawa+:2006} caused by Kelvin-Helmholtz instabilities; however, neither Kelvin-Helmholtz instabilities nor dusk-side reconnection (as suggested by e.g. \citealt{grodent+:2003,cowley+:2003}) are individually sufficient to explain the appearance of bridges \citep{nichols+:2017} and the question of their origins remains open.

\citet{greathouse+:2021} also observed bridges in Juno UltraViolet Spectrograph (Juno-UVS) images of both hemispheres of the aurorae of Jupiter. 
Due to the fact that the aurora is not symmetric around Jupiter's axis of rotation, especially in the northern hemisphere, HST observations are typically made when the largest proportion of Jupiter's aurora is visible - that is, around System-III subsolar longitudes of 150\textdegree\ in the north and 30\textdegree\ in the south - which introduces a local-time bias.
However, Juno, as a spacecraft in polar orbit around Jupiter, can image the aurora at many subsolar longitudes, thus removing the local-time bias.
``Bright, dusk-side [...] concentric arcs'' \citep{greathouse+:2021} in the polar collar (which we identify as bridges in this work) were observed independently of System-III longitude, indicating that bridges have their origins in the dusk-side magnetosphere.
The emergence of bridges has also been tentatively associated with solar-wind compression of Jupiter's magnetosphere \citep{nichols+:2007,nichols+:2009b,nichols+:2017}.
\citet{nichols+:2009b} also indicates that the bridge is stable over at least 3 hours, which is supported by later Juno observations of two bridges stable over at least 5 hours \citep{palmaerts+:2023}.
This suggests, based on the typical timescale for solar-wind compression events, that bridges may have lifetimes on the order of several days.
Dusk-side polar-collar arcs (identified as bridges) have also been observed to disrupt the smooth morphology of the dusk-side ME on occasion \citep{groulard+:2024}, whereas, in other instances, it appears alongside an unperturbed ME \citep{nichols+:2009b}.
This may indicate that the process that gives rise to bridges can (but is not required to) interact with the source process of the ME.
A correlation between solar-wind compression of the magnetosphere and the appearance of dusk-side polar arcs in the aurora was identified in a limited number of cases during Juno's approach to Jupiter \citep{nichols+:2017}.


From particle measurements made by Juno, the auroral region can be split into several zones \citep{mauk+:2020}; of these, Zone I (ZI) and Zone II (ZII), typically associated with the ME, are most relevant to this work.
ZI is associated with predominantly planetward- or downward-moving electrons (or upward field-aligned currents; FACs).
ZII is immediately poleward of the ZI subregion and is dominated by upward-moving electrons (downward FACs).
ZI and ZII are typically located alongside one another, such that the ME has been associated with an inversion of the FACs, upward then downward as Juno crosses the ME into the polar region \citep{al_saati+:2022}.
Auroral emission in the polar region, particularly the ZII aurora, has been suggested to be the consequence of broadband, bidirectional electron acceleration by upward-travelling electrostatic waves (ESWs) \citep{sulaiman+:2022}, generated by the upward electron beams \citep{elliott+:2018,elliott+:2020} observed by Juno \citep{mauk+:2017}.
This is noteworthy since analogous processes on Earth and Saturn do not give rise to appreciable auroral emission \citep{sulaiman+:2022}.

This work presents an investigation into Jupiter's auroral bridge using two methods: firstly, a largely automatic analysis of a large number of HST-STIS images of the aurora to determine statistical properties of bridges; and secondly, a multi-instrument analysis of Juno data from the first 30 perijoves to put the bridge into a wider magnetospheric context.
The results from these two investigations are then combined to discuss the bridge in the context of existing auroral frameworks.

\section{Observations}

In the first part of this work, HST-STIS UV observations between 2012 and 2023 are considered, notably those from the campaigns GO-11649, 12883, 13035, 14105, 14634, 15638, 16193, and 16675.
For the automated bridge detection discussed in section \ref{sec:automatic_detection}, only northern-hemisphere visits are considered, due to the favourable viewing geometry in the northern hemisphere for observation of the dusk-side polar region (where bridges are located) from Earth orbit, equivalent to 248 unique HST-STIS visits or 143 hours of observation.
These time-tagged images were accumulated into 10-second frames using the CALSTIS calibration tools from the Space Telescope Science Institute \citep{katsanis+mcgrath:1998}, then converted to brightness in kilo-Rayleigh (kR) assuming a colour ratio of 2.5 \citep{gustin+:2012} and fitted to the ellipsoid of Jupiter as per \citep{bonfond+:2009}.

Composite UV images from Juno-UVS (68-210~nm; \citealt{gladstone+:2017_uvs}) are also used in this work.
Juno has a highly elliptical polar orbit, allowing it to view Jupiter's aurora in both hemispheres.
As a spin-stabilised spacecraft, Juno cannot point its instruments freely; instead, Juno-UVS observes ``strips'' of Jupiter's aurora as the spacecraft rotates, which can be collated to create wider maps of the aurora. 
For each pass of the Juno spacecraft over Jupiter's poles (a perijove; designated as e.g. PJ1-N for perijove 1, northern hemisphere), an exemplar map was produced from Juno-UVS data.
This exemplar map uses a 100-spin ($\sim$50-minute) stack of UVS data that is centred as close as possible to the perijove time whilst covering at least 75\% of the auroral region \citep{bonfond+:2018}.
Radiation noise from the impinging of high-energy electrons on the detector is also removed \citep{bonfond+:2021}.
A detailed description of the production of the exemplar map is given in \citet{head+:2024}. 
This results in a representative view of the aurora in each hemisphere during each perijove. 
In total, this corresponds to 58 images of the aurora for the first 30 perijoves, two per perijove excluding PJ2, when Juno was placed into safe mode.

In addition to image data from Juno-UVS, data from other Juno instruments are used in this work, notably from the FluxGate Magnetometer (MAG-FGM; \citealt{connerney+:2017}), the Juno Energetic-particle Detector Instrument (JEDI; \citealt{mauk+:2017_jedi}), and the Waves instrument \citep{kurth+:2017}. 
Technical details of each instrument are described within their associated reference.

\section{Methods}

\subsection{Automatic detection of bridge-like arcs}
\label{sec:automatic_detection}

\begin{figure}[thbp]
    \centering
    \ifthenelse{\equal{\twocol}{"y"}}{
        \includegraphics[width=0.95\linewidth]{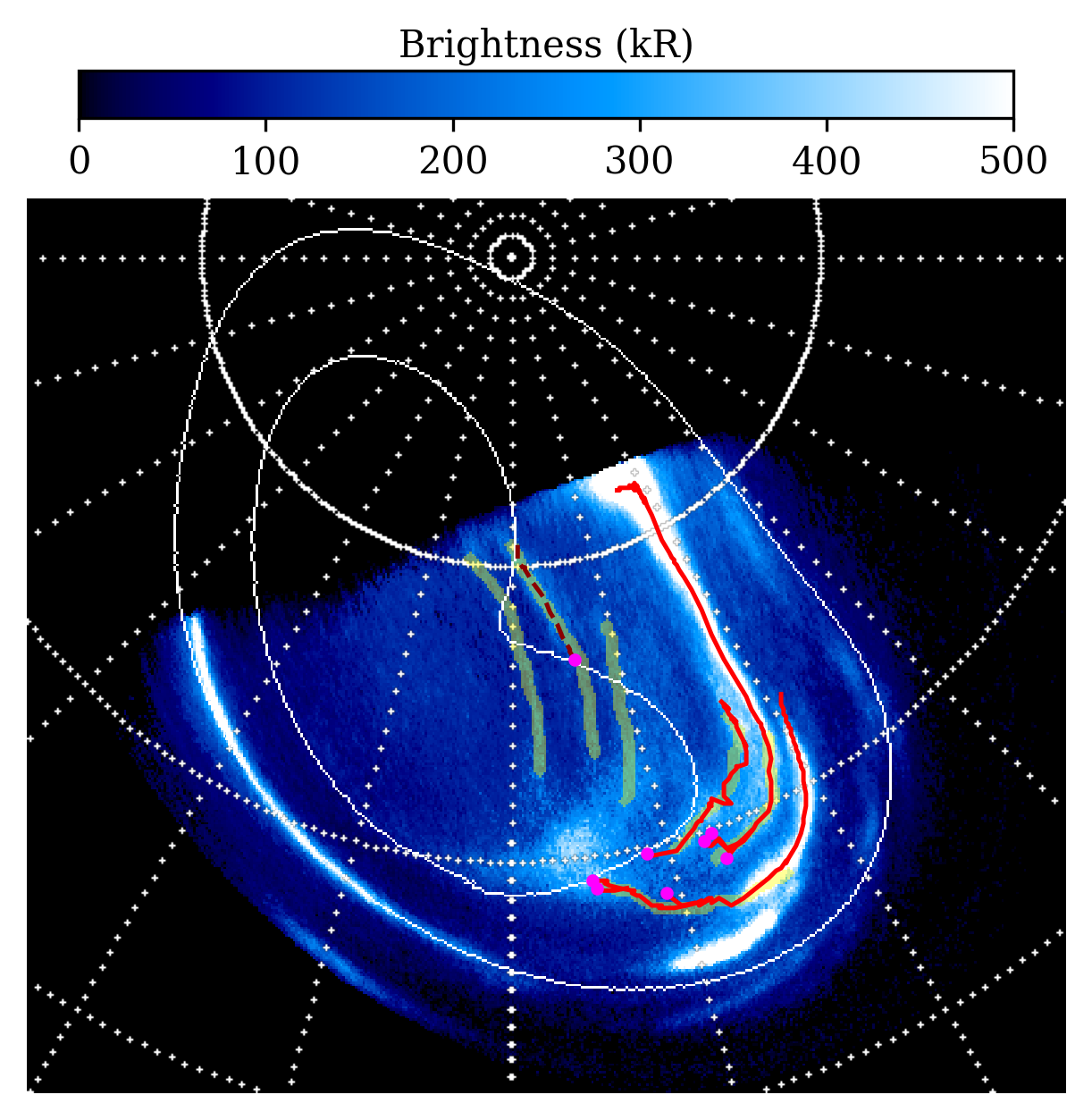}
    }
    {
        \includegraphics[width=0.48\linewidth]{Plot_DetectedVsManualArcs_odxc01okq_postfilter.png}
    }
    \caption{
        Results of the bridge-detection algorithm after filtering. Red solid lines denote detected arcs identified as bridges. Dark-red dashed lines denote arcs identified as PAFs. The seed point of each arc is in magenta. Manually designated arcs are in yellow. White contours give the region of validity of the \citet{vogt+:2011} JRM33 flux-equivalence mapping along closed field lines.
    }
    \label{fig:bridge_detection}
\end{figure}

In the first part of this work, an automated method is employed to detect bridge-like arcs in 248 HST-STIS images of the northern aurora.
Each STIS time-tag image series is typically split into 30-second frames, which presents a compromise between image noise reduction and preservation of auroral dynamics.
These images are first mapped into a polar projection - as though observed from directly above the axis of rotation of Jupiter, with a System-III longitude of 0\textdegree\ to the top of the image and 90\textdegree\ to the right - to ensure ease of comparison between images whilst preserving the physical size of features in the aurora. 
For this work, each STIS image series was collated into a single frame by taking the pixelwise median of the image stack; since prior investigations have indicated that bridge morphology does not greatly change over the 40 minutes of a HST exposure \citep{nichols+:2009b}, this reduces the required computational effort whilst ensuring that each image represents a unique instance of auroral morphology.
Though bridges have been observed to remain fixed in local time \citep{pardocantos+:2019} rather than in System-III longitude (and thus are observed to slowly advance in longitude over the course of a STIS exposure), they remain clearly visible in the collated images, as shown in Figure \ref{fig:bridge_example}.
To more clearly highlight the arc-like form of the bridges and hence to make automatic detection of these structures more feasible, each collated image was convolved with a Gaussian-arc kernel to produce an ``arcness'' map of the aurora, as in Figure \ref{fig:bridge_detection_suppmat}a; see section 3.2 of \citet{head+:2024} for a thorough description of this method.

Bridges are defined as arcs that connect the day-side active region to the ME via the polar collar (though they may not span this gap fully).
This implies that they must traverse a significant ($>$50 R$_{J}$) radial distance in the magnetosphere, since the ME is surmised to originate from a region between 20 and 40~R$_{J}$ from Jupiter \citep{cowley+bunce:2001}, whereas the active region is firmly within the polar aurora and hence maps to more distant regions of the magnetosphere.
This behaviour can be leveraged to automatically detect auroral arcs that are likely to be bridges.
The details of this method are described in Appendix A of the supplementary material.
Briefly, a number of fixed-radius contours are magnetically mapped from the magnetosphere to the ionosphere, and the locations where auroral arcs cross these contours are used as starting points to follow the shape of each arc in the ionosphere.
Random-forest filtering is used alongside manual bridge-arc designations to remove artefacts of the method.
An example of the detected arcs after this filtering is given in Figure \ref{fig:bridge_detection}.

\subsection{Juno multi-instrument analysis}
\label{sec:juno_multiinstrument_analysis}
In the second part of this work, data from multiple Juno instruments are combined to build a full picture of the bridge. These quantities are determined as follows:
\begin{itemize}
    \item Field-aligned currents: calculated from FGM data compared against the latest JRM33 magnetic-field model \citep{connerney+:2022} (implemented using the \texttt{JupiterMag} Python wrapper; \citealt{jupitermag}, \citealt{wilson+:2023}) and extrapolated via magnetic-flux conservation to the assumed auroral altitude of 400 km, after \citet{al_saati+:2022}; see section \ref{sec:fac_method} of the supplementary material.
    \item Alfv\'{e}nic Poynting flux: from FGM data, extrapolated via magnetic-flux conservation to the assumed auroral altitude of 400 km, after the method presented by \citet{gershman+:2019}.
    \item Electron energy and pitch-angle distribution: from JEDI measurements \citep{mauk+:2017_jedi}, where data from all detectors have been stacked.
    \item Juno-Waves spectrum: from Juno-Waves data; in this work, only data from the LFR-Lo channel (50~Hz to 20~kHz) is presented.
\end{itemize}
Non-UVS data are sourced from the Automated Multi-Dataset Analysis (AMDA) database maintained by the Centre des Donn\'{e}es de la Physique des Plasmas \citep{genot+:2021} and accessed using the \texttt{speasy} Python library \citep{jeandet+:2024}, with the exception of JEDI data which were downloaded using the JMIDL tool provided by John Hopkins University.

Due to the relatively low number of cases (58 over 30 PJs), the determination of the positions of bridges in each exemplar image, and hence the Juno bridge traversal timestamps, was done manually (using the arc-convolved exemplar images, similar to Figure \ref{fig:bridge_detection}a, to highlight arc locations) to avoid the introduction of artefacts or the omission of bridges by the automatic detection algorithm.
The Juno footprint is considered to cross a manually identified arc when within 5 px ($\sim$600 km); this is a relatively broad threshold to ensure that the full arc crossing is included.
However, since the auroral morphology in the base image is determined from stacked UVS spectral scan centred at a particular time, it may not be representative of the auroral morphology at the time of a supposed Juno bridge traversal. 
This effect can be counteracted by looking at the instantaneous UVS map of the aurora at the time of the suspected bridge crossing to ensure that the bridge is approximately in the same position as in the exemplar image, and making adjustments to the crossing timestamps if necessary.
The mean (and standard deviation) difference between the exemplar-map time and bridge crossings is 44 minutes ($\pm$23 minutes), so most bridge crossings are reliably determined from the initial manual estimates. 
Additionally, since the auroral arcs of bridges and the ME are typically crossed perpendicularly, the instantaneous UVS Juno-footprint brightness should show a peak during the supposed crossings, which can be used as an additional timestamp-validity check.
In this work, the first 30 perijoves are considered, since after the 30th orbit, Juno's orbit is such that incomplete maps of the northern aurora become more and more common, making it difficult to identify mesoscale features like bridges, and passes over the southern aurora occur at increasingly greater altitudes, lowering the resolution of southern auroral maps and again making identification of the bridge challenging.
Details of the automatically detected bridge and ME crossings are given in Table \ref{tab:uvs_cases} of the supplementary material.

\FloatBarrier
\section{Results}

\subsection{HST-STIS large-scale analysis}
\label{sec:HST_analysis}

\begin{figure}[tbhp]
    \centering
    \ifthenelse{\equal{\twocol}{"y"}}{
        \includegraphics[width=\linewidth]{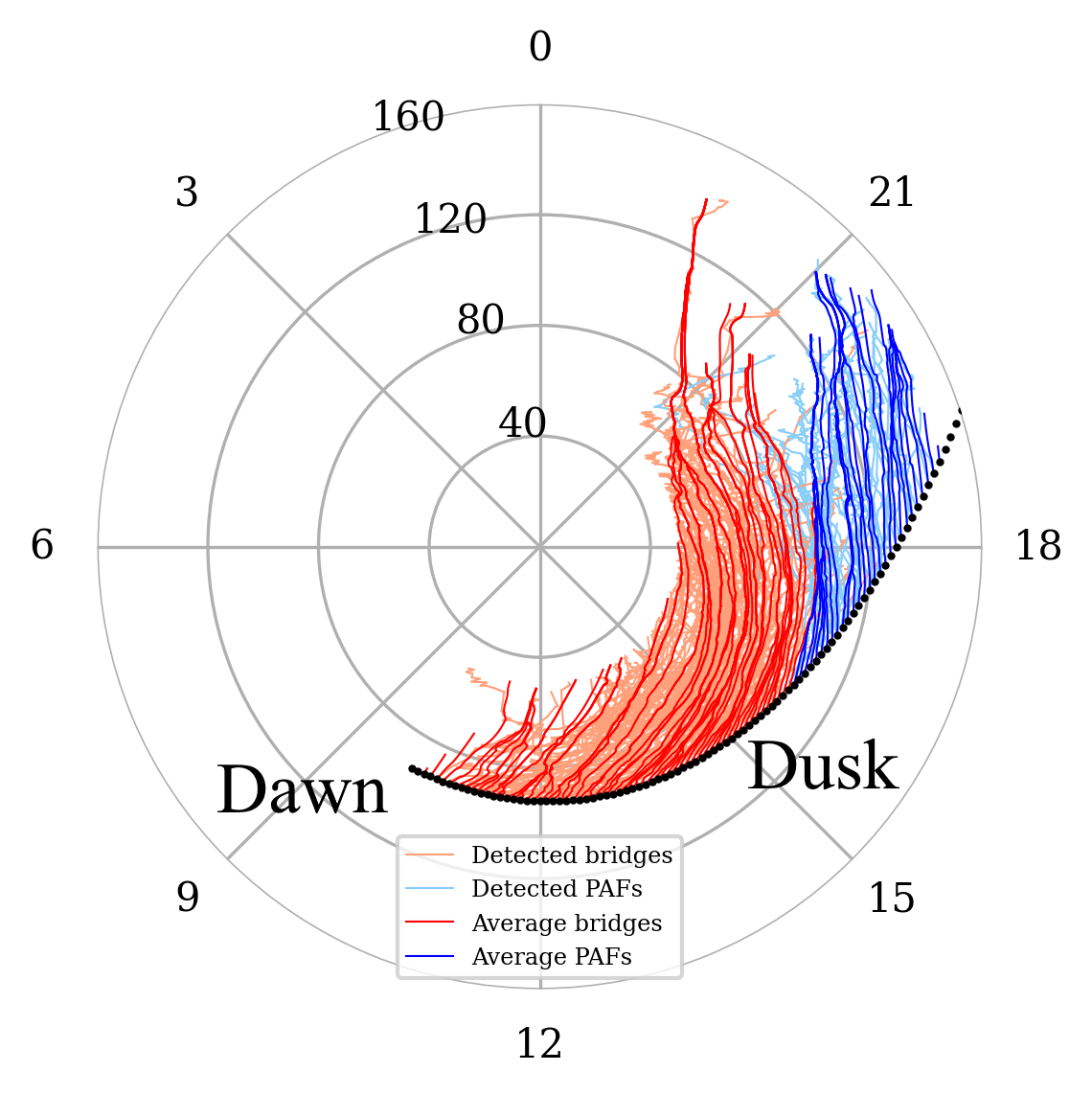}
    }
    {
        \includegraphics[width=0.48\linewidth]{STIS_N_Plot_BridgesLocation_vogt_brightness_annotated.png}
    }
    \caption{
         Automatically detected bridge/PAF-like arcs (in pale red/blue) in all northern-hemisphere HST-STIS observations considered in this work, mapped via flux-equivalence mapping to the magnetospheric equator (radius, local time). Radii are given in R$_{J}$. A set of average bridge/PAF contours are given in dark red/blue; their starting locations on the \citet{joy+:2002} expanded magnetopause (solar-wind ram pressure = 0.039~nPa; stand-off distance 92 R$_{J}$) are given in black.
    }
    \label{fig:HST_mapping}
\end{figure}

\begin{figure}[tbhp]
    \centering
    \ifthenelse{\equal{\twocol}{"y"}}{
        \captionsetup[subfigure]{width=0.95\linewidth}
        \subfloat[2017-05-19 02:37:37]{
            \includegraphics[width=\linewidth]{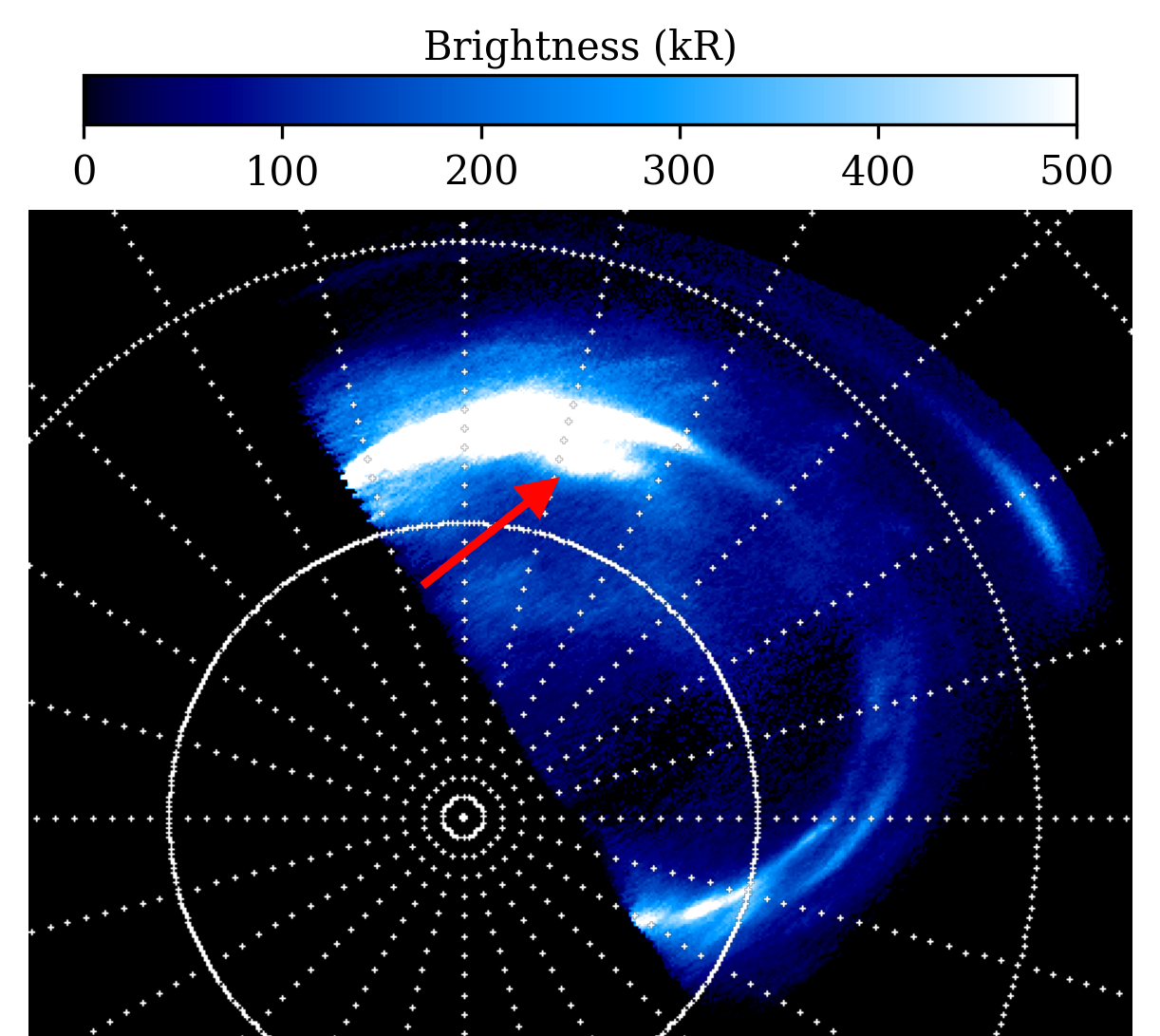}
        } \\
        \subfloat[2017-05-19 12:09:43]{
            \includegraphics[width=\linewidth]{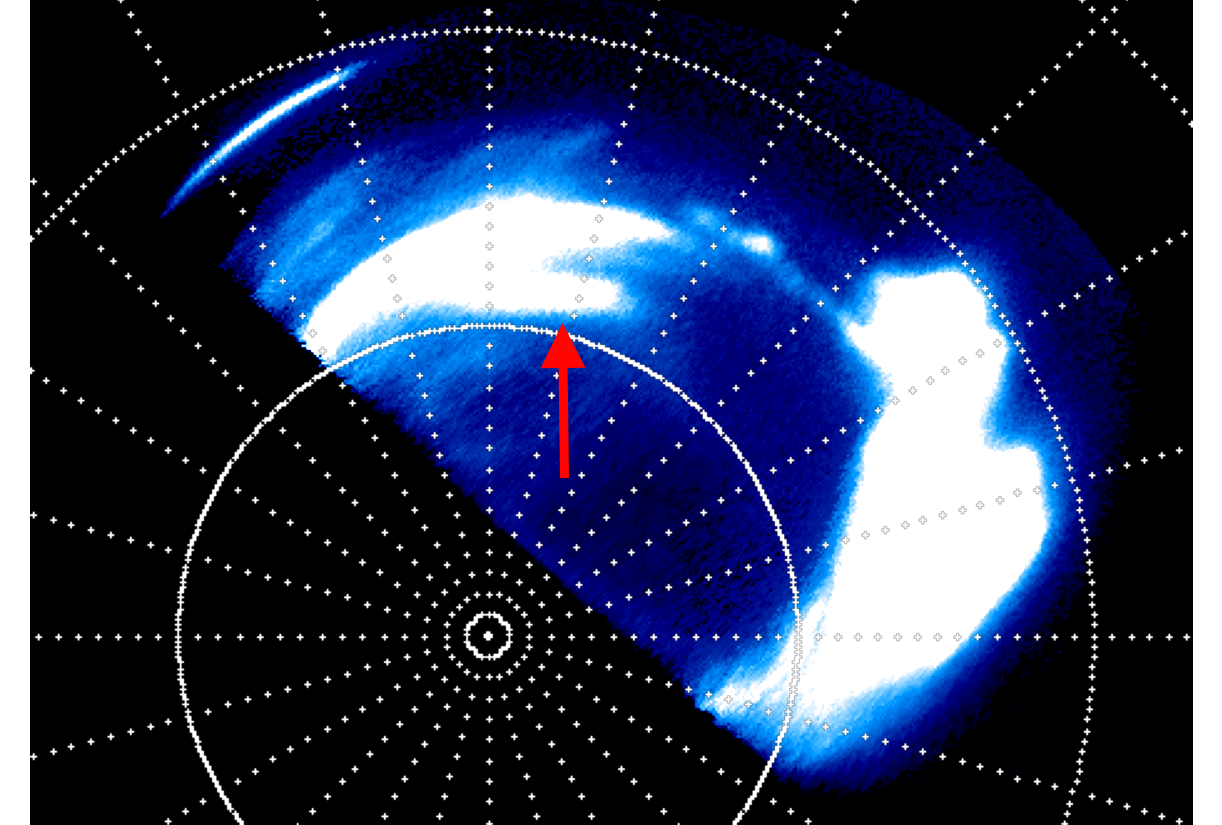}
        }
    }
    {
        \captionsetup[subfigure]{width=0.42\linewidth}
        \subfloat[2017-05-19 02:37:37]{
            \includegraphics[width=0.48\linewidth]{Plot_ConvolvedPolarview_od8k92weq_brightness_annotate.png}
        }
        \subfloat[2017-05-19 12:09:43]{
            \includegraphics[width=0.48\linewidth]{Plot_ConvolvedPolarview_od8k0ay5q_brightness_annotate.png}
        }
    }
    \caption{
         Observation of a persistent bridge-like arc over a full Jupiter rotation in the southern aurora during the GO-14634 HST campaign. Bridge location is highlighted by the red arrow. This bridge is also present in the intervening HST and UVS (PJ6) observations.
    }
    \label{fig:bridge_lifetime}
\end{figure}

\begin{figure}[tbhp]
    \centering
    \ifthenelse{\equal{\twocol}{"y"}}{
        \captionsetup[subfigure]{width=0.95\linewidth}
        \subfloat[2019-02-13 21:16:57]{
            \includegraphics[width=\linewidth]{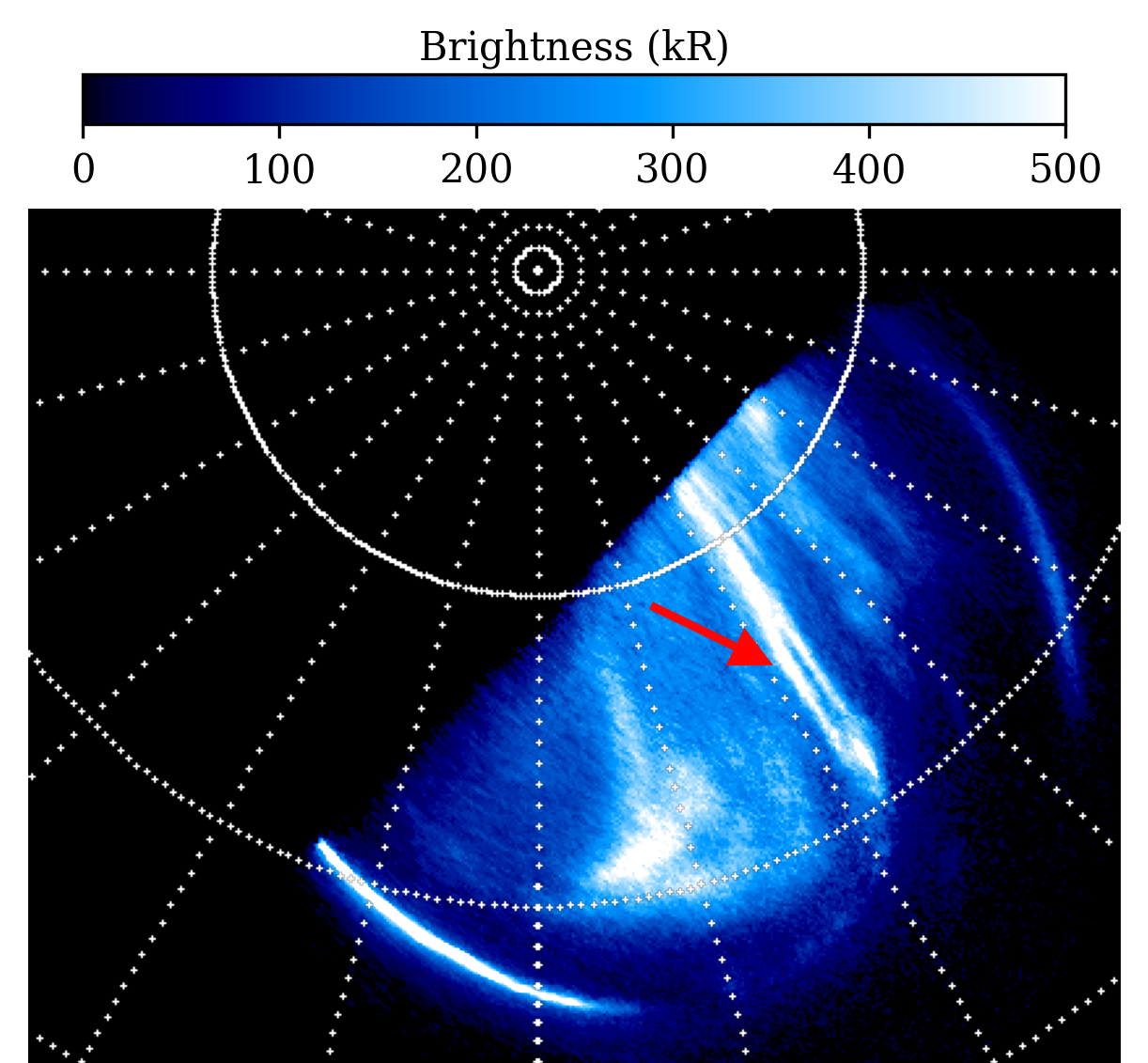}
        } \\
        \subfloat[2019-02-13]{
            \includegraphics[width=\linewidth]{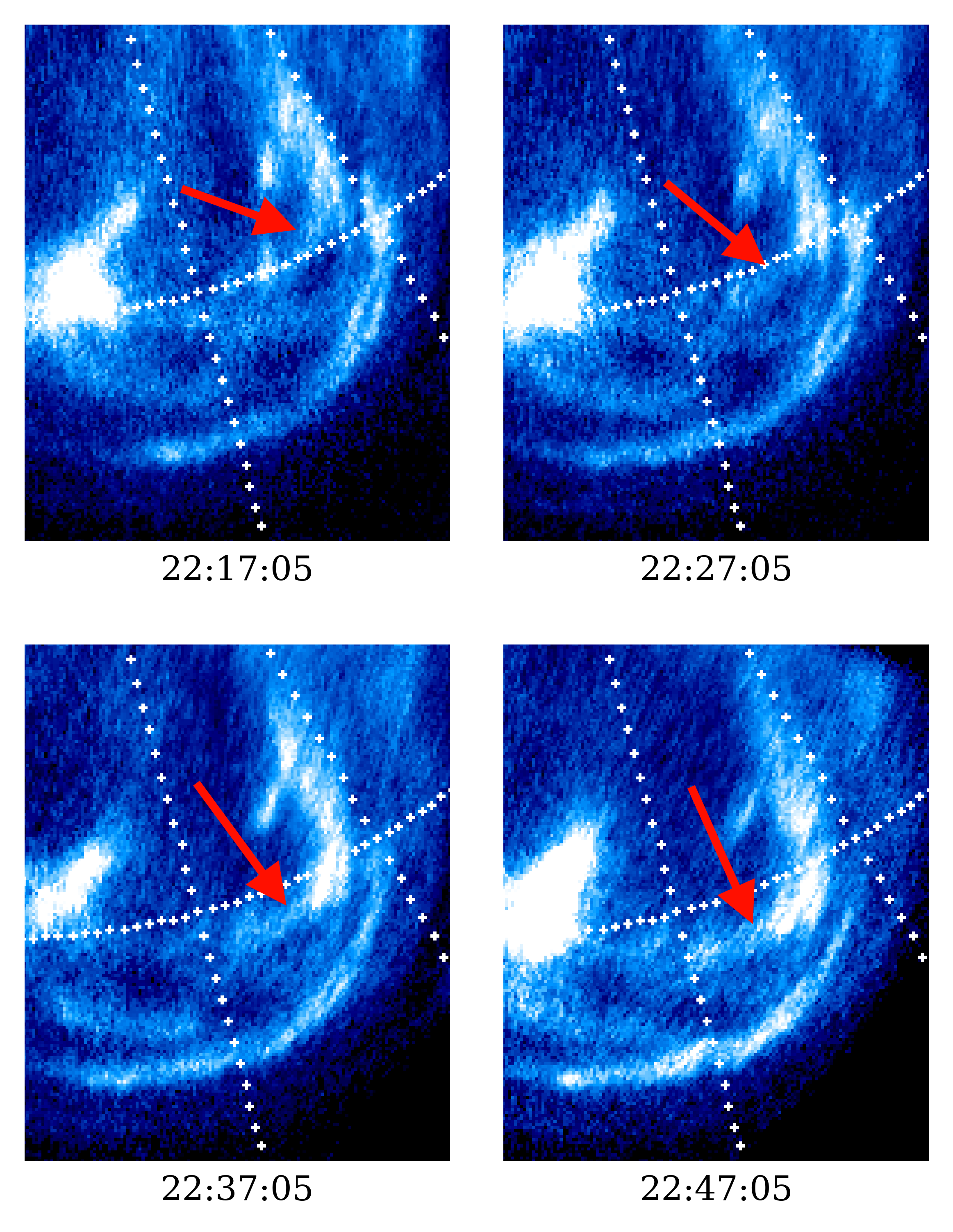}
        }
    }
    {
        \captionsetup[subfigure]{width=0.42\linewidth}
        \subfloat[2019-02-13 21:16:57]{
            \includegraphics[width=0.48\linewidth]{Plot_ConvolvedPolarview_odxc07q5q_brightness_annotate.png}
        }
        \subfloat[2019-02-13 22:51:15]{
            \includegraphics[width=0.48\linewidth]{MakeFig5_odxc08q9q.png}

        }
    }
    \caption{
         Observation of the growth of a bridge-like arc in the northern aurora during the GO-15638 HST campaign. Bridge location is highlighted by the red arrow. 
    }
    \label{fig:bridge_onset}
\end{figure}

Detected bridges in the ionosphere were magnetically mapped to the equatorial plane in the magnetosphere using the \citet{vogt+:2011} flux-equivalence mapping.
As can be seen in Figure \ref{fig:HST_mapping}, the majority of detected bridges (light red) map to the dusk-side magnetosphere between 10 and 20~MLT, as expected from previous work \citep{pardocantos+:2019}.
Detected arcs that map to the magnetopause beyond 16~MLT are considered to be PAFs and excluded from this analysis, to be discussed separately later in this work, though a continuum of arcs exist between bridges and PAFs and hence the choice of this cutoff is approximate.

To aid understanding of the average properties of these bridges, ``average-bridge'' contours (dark red, and blue for PAFs, in Figure \ref{fig:HST_mapping}) were calculated by taking a number of starting points on the magnetopause (black in Figure \ref{fig:HST_mapping}), following the average orientation of detected bridges (weighted by the inverse square of the distance to the point to produce a ``local-average'' orientation), and continuing this process to propagate each contour into the magnetosphere.
These contours indicate that bridges typically curve duskward and radially inwards toward Jupiter.
The bridge mapping shown in Figure \ref{fig:HST_mapping} depends on the magnetic field model used to map poleward of the ME. 
Although the quantitative details are affected by the choice of such model (compare Figure \ref{fig:HST_mapping} with Figure \ref{fig:HST_mapping_flt} using field-line tracing instead of flux-equivalence mapping), our results are broadly consistent with the expected location of the bridges.
Bridges are thus named because they (at least partially) bridge the polar collar between the active region (which can be assumed to relate to the magnetopause, or at least to large radial distance in the day-side magnetosphere; \citealt{nichols+:2007}) and the ME (which is expected to arise much closer to Jupiter, at $\sim$30~R$_{J}$; \citealt{cowley+bunce:2001}).


Of all the HST cases analysed in this work, there exists only one quasi-continuous set of observations (notably, surrounding PJ6) that tracks the evolution of a bridge over a full Jupiter rotation; the start and end series of this set are shown in Figure \ref{fig:bridge_lifetime}.
The bridge has persisted over the $\sim$10-hr span of this set of observations with very little change in morphology, having moved only slightly equatorward.
This result is consistent with the findings of \citet{palmaerts+:2023} and \citet{nichols+:2009b} in which bridges are observed to be stable over at least 3 hours.
This latter work also associated the appearance of bridges with magnetospheric compression by the solar wind, noted to be maintained over several days, which is consistent with bridges that can survive a full Jupiter rotation.
A case in which a bridge was observed to develop between two HST observations is given in Figure \ref{fig:bridge_onset}; see the supplementary material for GIF images of these cases.
In this case, the ME was observed to first exist as two parallel arcs, which had evolved into a clear bridge-like arc separate from the ME one hour later, though this interpretation assumes that the two double-arc phenomena are related, impossible to confirm in the absence of intervening observations. 
In any case, this hour-scale onset is consistent with the suggested solar-wind influence on the presence of bridges \citep{nichols+:2017}, since it is expected to vary on timescales of hours \citep{chane+:2017}; 
however, some internal processes (such as dawn storms) are also known to vary over hour timescales \citep{bonfond+:2021} and so this timescale cannot be used as a confirmation of the influence of the solar wind without further supporting evidence.


\begin{figure}[htb]
    \centering
    \ifthenelse{\equal{\twocol}{"y"}}{
        \includegraphics[width=\linewidth]{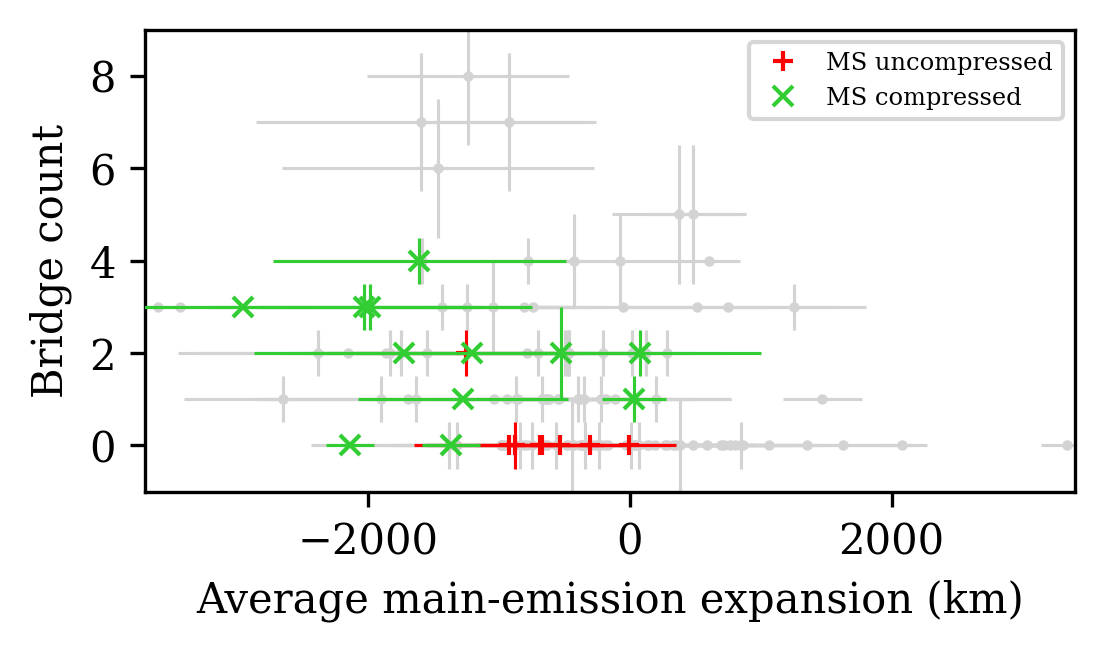}
    }
    {
        \includegraphics[width=0.48\linewidth]{STIS_N_Plot_BridgesVsMainEmissionExpansion_vogt_convolve_bridges_count.png}
    }
    \caption{
         Detected average expansion of the ME from \citet{head+:2024} vs. the total detected bridge count for each northern HST-STIS series considered in this work. Negative expansions imply a contracted ME. Bridge-count error bars are determined as described in Appendix A. Green crosses denote those cases where the magnetosphere was compressed, and red pluses those cases where the magnetosphere was uncompressed; grey points denote cases where the compression state of the magnetosphere is unknown. 
    }
    \label{fig:bridge_vs_ms_size}
\end{figure}

The appearance of bridges can also be directly related to the compression of the magnetosphere.
Figure \ref{fig:bridge_vs_ms_size} shows the expansion of the ME against the total detected bridge count for each northern HST series considered in this work, where ``expansion'' refers to the average equatorward distance (in km) of the ME from its average position, as per \citet{head+:2024}.
The use of bridge count over e.g. total projected bridge length in the magnetosphere may potentially misrepresent the aurora in the case where many small bridges are detected; as shown in Figure \ref{fig:bridge_length_vs_count}, there exists a broadly linear relation between detected bridge count and total projected bridge length, and so we conclude that both parameters are reasonable measures of the quantity of bridges in the aurora.
Figure \ref{fig:bridge_vs_ms_size} shows no clear relation between bridge count and ME expansion.
However, by isolating those cases where the state of compression of the magnetosphere is known from Juno magnetopause-crossing altitudes \citep{yao+:2022,louis+:2023}, a clear distinction between cases with compressed and uncompressed magnetospheres is apparent.
A Pearson-$\chi^2$ test on these points with categories [bridge count > 0, bridge count = 0] and [compressed magnetosphere, uncompressed magnetosphere] indicates that compression of the magnetosphere is associated with the presence of bridges beyond the 99th-percentile level ($\chi^2$~=~9.73, p-value 0.002). 
This association between magnetospheric compression and bridge count is evident even from a cursory inspection of the aurora in these cases, given in Figure \ref{fig:compressed_aurora_images} of the supplementary material. 
While the size of the magnetosphere may vary due to a number of factors, such as the solar-wind dynamic pressure \citep{chane+:2017} and the mass-outflow rate from Io \citep{bagenal+delamere:2011}, this result, combined with the previously determined hour-scale variability of bridge morphology, is compatible with the idea that the solar wind exerts some measure of influence on the appearance of bridges in the aurora, which is in line with the results of \citet{nichols+:2017}.
However, this scenario cannot be robustly confirmed without simultaneous measurement of the solar wind near Jupiter, which is left to a future work.
The arcs used in the above analysis were those with magnetopause local times less than 16 MLT; these arcs can be confidently said to be bridges, whereas there is some ambiguity with PAFs in those arcs that map to the distant magnetosphere beyond 16 MLT.
A similar analysis was performed separately for the assumed PAFs (blue in Figure \ref{fig:HST_mapping}) and no difference was found in the total detected arc length for compressed and uncompressed magnetospheres (see Figure \ref{fig:paf_vs_ms_size} in the supplementary material).
This suggests that PAFs are not exactly the same feature as the auroral bridge and that they do not show a dependence on the solar wind, in agreement with \citet{nichols+:2009}.
However, the methods presented in this work are not necessarily suited to the analysis of these polar-cap features, which are often found within the region of open magnetic flux after \citet{vogt+:2011}, and so more specialised work should be carried out to confirm this conclusion.

\subsection{Juno multi-instrument analysis}
\FloatBarrier
\subsubsection{Case study - PJ9-S}

\begin{figure}[tbhp]
    \centering
    \ifthenelse{\equal{\twocol}{"y"}}{
        \includegraphics[width=\linewidth]{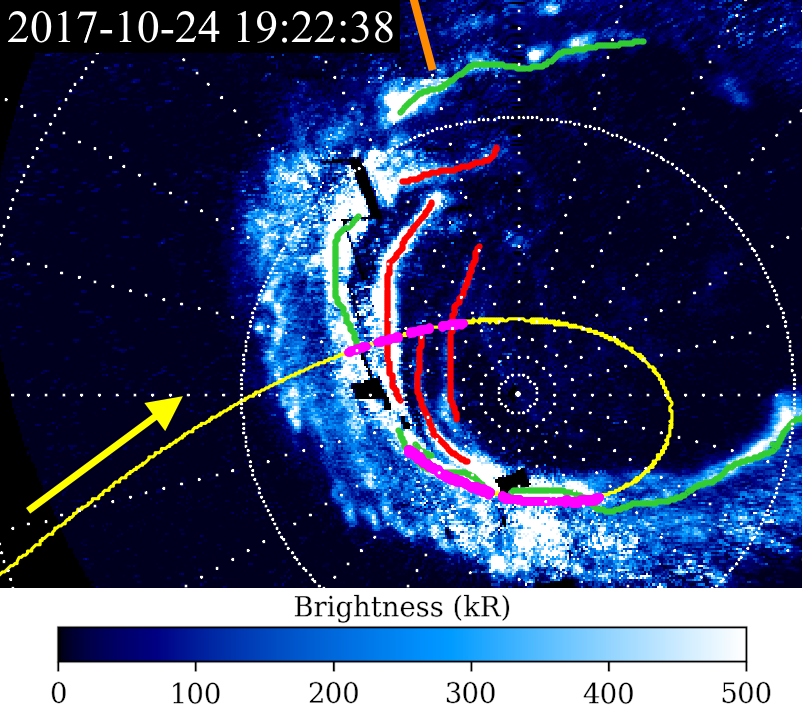}
    }
    {
        \includegraphics[width=0.48\linewidth]{JunoTraversal_PJ9_S_crossingMap_brightness_annotate.png}
    }
    \caption{
         Mapped Juno-footprint trajectory for PJ9-S overlain on the UVS exemplar image (central spin  timestamp in top-left). The yellow arrow indicates the direction of travel of Juno. The sun position is marked by a solid orange line. Bridges are given in red, and the ME in green; Juno crossings of these arcs are given in magenta. 
    }
    \label{fig:PJ9_S_trajectory}
\end{figure}

\begin{figure}[tbhp]
    \ifthenelse{\equal{\twocol}{"y"}}{
        \includegraphics[width=\linewidth]{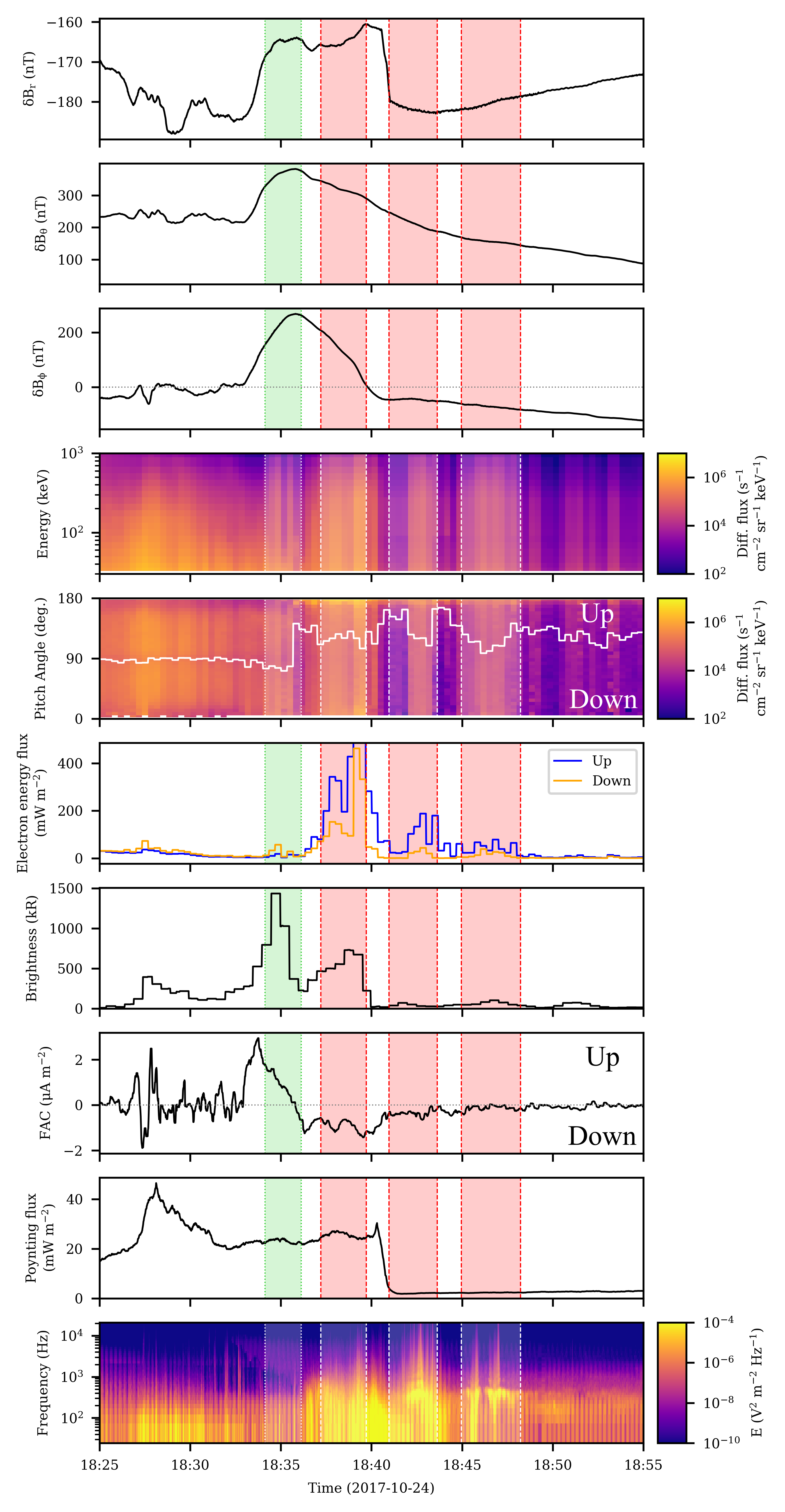}
    }
    {
        \includegraphics[width=0.48\linewidth]{PJ9_S_Plot_JunoInstrumentData_crossing_bridges_annotated.png}
    }
    \caption{
         Juno instrument data for PJ9-S. The ME crossing is first (green, dotted) followed by three bridge crossings (red, dashed). From top to bottom: residual magnetic field components (FGM data - JRM33); JEDI electron energy; JEDI electron pitch-angle distribution (average given by solid white line); JEDI field-aligned (0\textdegree-20\textdegree, 160\textdegree-180\textdegree) electron energy flux; UVS footprint brightness; calculated ionospheric FACs; calculated ionospheric Alfv\'{e}nic Poynting flux (the ``step'' in the flux is related to a change in operating range of the FGM instrument); Waves-E LFR-Lo spectral density. 
    }
    \label{fig:PJ9_S_multiinstrument}
\end{figure}


By comparing the projected Juno ionospheric footprint against manually determined bridge positions in the base image of each hemisphere during each perijove, approximate bridge-crossing timestamps can be determined and compared against data and derived parameters from Juno.
Note that mapping of the Juno footprint in the ionosphere was performed using field-line tracing of the JRM33 field model rather than flux-equivalence mapping, since Juno is at low altitude, minimising the effect of model inaccuracies in the polar region, and this method can map the footprint within the entire polar region.
An example of this analysis is given in Figures \ref{fig:PJ9_S_trajectory} and \ref{fig:PJ9_S_multiinstrument} for PJ9-S, in which Juno passed first over the ME (green) and then over three clear bridge-like structures in the dusk-side polar collar (red) at low altitude ($\sim$2 R$_{J}$), where the positions of these features are determined as per section \ref{sec:juno_multiinstrument_analysis}. 
The crossing showed the expected FAC inversion \citep{al_saati+:2022,mauk+:2020}; Juno observed upward FACs (majority downward-travelling electrons) as it first passed through the ME, then downward FACs (majority upward-travelling electrons) as it continued into the polar collar.
It is notable that the ME crossing was associated with uniquely upward FACs, rather than an inversion, indicating that the ME was a uniquely ZI feature during this crossing \citep{mauk+:2020}.
Unlike the ME crossing of PJ7-N described in \citep{kurth+:2018}, there was no significant Waves-E emission; instead, we see that Juno started to observe significant plasma-wave emission once it entered into the polar collar, with high-frequency peaks occurring during bridge crossings, which then abated after the third bridge. 
The calculated FACs remained downward for all three of these bridge crossings, though no particular peaks or signatures were observed that can be associated with the bridge.
In particular, the lack of FAC inversion signatures during the bridge crossings suggests two things: firstly, that bridges are mechanically distinct from the ME, although they are morphologically related insofar as they can disrupt the morphology of the ME \citep{nichols+:2009}; and secondly, that vorticity in the dusk-side magnetosphere is likely not the source of the bridges, as had been previously suggested \citep{fukazawa+:2006,pardocantos+:2019}, since this would be expected to give rise to noticeable FAC signatures \citep{delamere+:2013, johnson+:2021}.
Additionally, there was no bridge-crossing signature in the derived Alfv\'{e}nic Poynting flux which indicates that the mechanisms responsible for auroral moon footprints \citep{gershman+:2019} are not responsible for the bridges.
During the first bridge crossing, Juno observed a clearly broadband, bidirectional, field-aligned electron distribution, though the upward electron energy flux was greater than the downward flux.
This trend is stronger for the latter two crossings; while the electron distributions remain field-aligned and broadband, there is a clear preference for upward-travelling electrons.
However, only a small downward electron flux would be required to produce the low auroral brightnesses associated with these bridge crossings, regardless of the upward electron flux; peaks in the downward electron energy flux are present and coincide with the peaks in auroral brightness. 
It is notable that, while plasma-wave emission can clearly be associated with these bridge crossings (and not with the ME crossing), the intensity of this emission does not appear to correlate with the auroral brightness seen during the bridge crossings.
The plasma-wave emission is also not constant during the crossings, though these peaks are not associated with any flaring behaviour in the bridges.
This may indicate a more complex relationship between these plasma waves and generation of the aurora, if they are indeed related to the processes that give rise to bridges, rather than merely coincident.


\FloatBarrier
\subsubsection{Analysis of the first 30 perijoves}

A similar analysis was performed for the first 30 perijoves, equivalent to 58 auroral crossings.
Bridges are present in the dusk-side polar collar during 39 of these traversals, and Juno passes over at least one arc in 26 of these cases. 
11 of these traversals show aurorae with no clear bridges.
The remaining 8 cases are those with large gaps in UVS coverage and are ignored.
Bridges are present in a large fraction (at least 67\%) of the first 30 perijoves.
Of these perijoves, the presence or absence of bridges is mostly maintained between the northern and southern crossing; if Juno observes bridges in the northern aurora, it also typically sees bridges in the southern aurora $\sim$2 hours later.
This is in line with previous work, which suggested that bridges are stable over timescales of several hours \citep{nichols+:2009b}, and indicates that the processes that give rise to bridges are conjugate between hemispheres and thus likely occur on closed field lines.
These bridges are noted to occur at similar local times and have comparable geometries, though more accurate magnetic-field models are required to determine whether the bridges lie along the same field lines in the north and south.
Only two perijoves (PJ4, PJ9) showed bridges that appeared or disappeared over the course of a perijove.
During PJ4, a long, faint bridge was visible in the northern aurora around 21 MLT, which had disappeared completely by Juno's pass over the southern aurora.
During PJ9, the dusk-side polar collar was free of bridges (though the dusk-side ME was slightly non-continuous) during the northern pass; by the time of Juno's southern pass, three distinct bridges had developed.
This, combined with the result that bridges usually appear conjugate between the two hemispheres, indicates that the process that causes bridges can occur over hour-long timescales, consistent with the result shown in Figure \ref{fig:bridge_onset}.

Bridge crossings at low altitudes were also generally associated with downward FACs (average for cases with altitudes below 3~R$_{J}$ = -0.1$\pm$0.3~$\mu$A m$^{-2}$) but without a particular signature (a peak during the crossing, for example), as well as broadband upward-dominated bidirectional electron distributions, as suggested by the results from PJ9-S.
ME crossings, however, were more usually associated with upward FACs (average for cases with altitudes below 3~R$_{J}$ = 0.2$\pm$0.7~$\mu$A m$^{-2}$) or FAC inversions, though the large uncertainties make it challenging to differentiate the two features from their average currents alone.

\begin{figure}[tbhp]
    \centering
    \ifthenelse{\equal{\twocol}{"y"}}{
        \includegraphics[width=\linewidth]{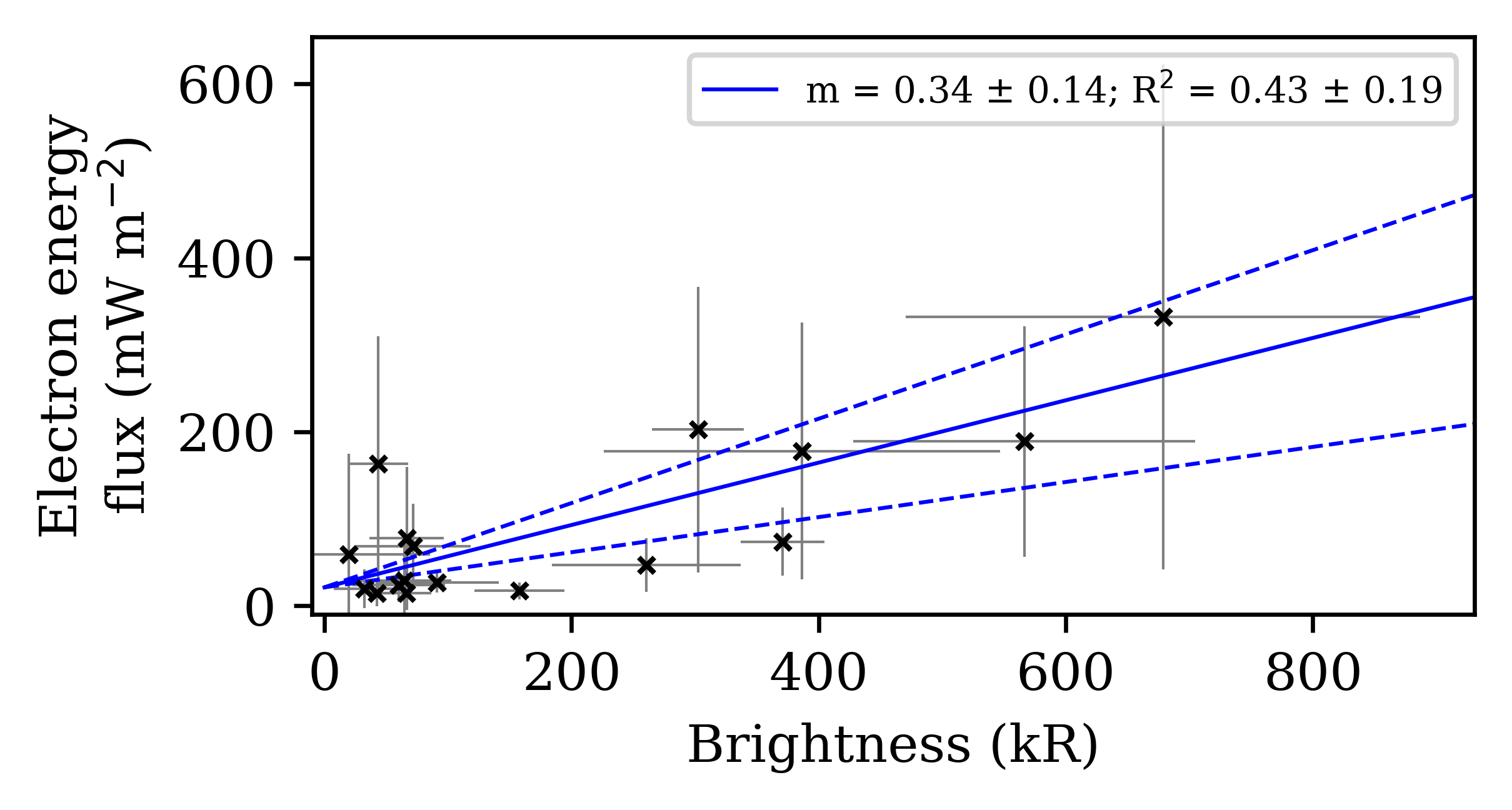}
    }
    {
        \includegraphics[width=0.48\linewidth]{UVS_both_med_noFilt30_Plot_BrightnessvsElectronFlux_bridges_down.png}
    }
    \caption{
         Mean Juno-footprint UV auroral brightness vs. mean JEDI downward electron energy flux observed during low-altitude ($<$3 R$_{J}$) bridge crossings between PJ1 and PJ30. Uncertainties (grey) are estimated using the 50th and 100th-percentile values observed during crossings. The gradient and R$^2$ value of the least-squares linear relation (solid blue line) is given in the legend. 
    }
    \label{fig:brightness_vs_electron_energy}
\end{figure}


Figure \ref{fig:brightness_vs_electron_energy} shows an approximately linear relationship (that passes through the origin) between the mean UV auroral brightnesses and downward electron energy fluxes seen by Juno during bridge crossings.
This is an indication that the downward electrons seen by Juno are indeed the source of the auroral emission associated with bridges.
An electron energy flux of 1 mW m$^{-2}$ is equivalent to an auroral brightness of 2.9 kR.
This is a slightly lower brightness than previous estimates for the aurora \citep{mauk+:2017, nichols+cowley:2022}.
This may be due to subtle implementation differences with previous work, or a genuine difference from the canonical ``1 mW m$^{-2}$ = 10 kR'' in the case of bridges. 
This also assumes that the entire electron energy flux seen by Juno at the typical crossing altitude of $\sim$2 R$_{J}$ contributes toward the auroral brightness of the bridge.
In the presence of a vertically extended acceleration region at or below the altitude of Juno \citep[e.g.][]{sulaiman+:2022}, some of the downward electron flux may be bidirectionally re-accelerated, reducing the downward flux that reaches the auroral layer.

\begin{figure}[tbhp]
    \centering
    \ifthenelse{\equal{\twocol}{"y"}}{
        \includegraphics[width=\linewidth]{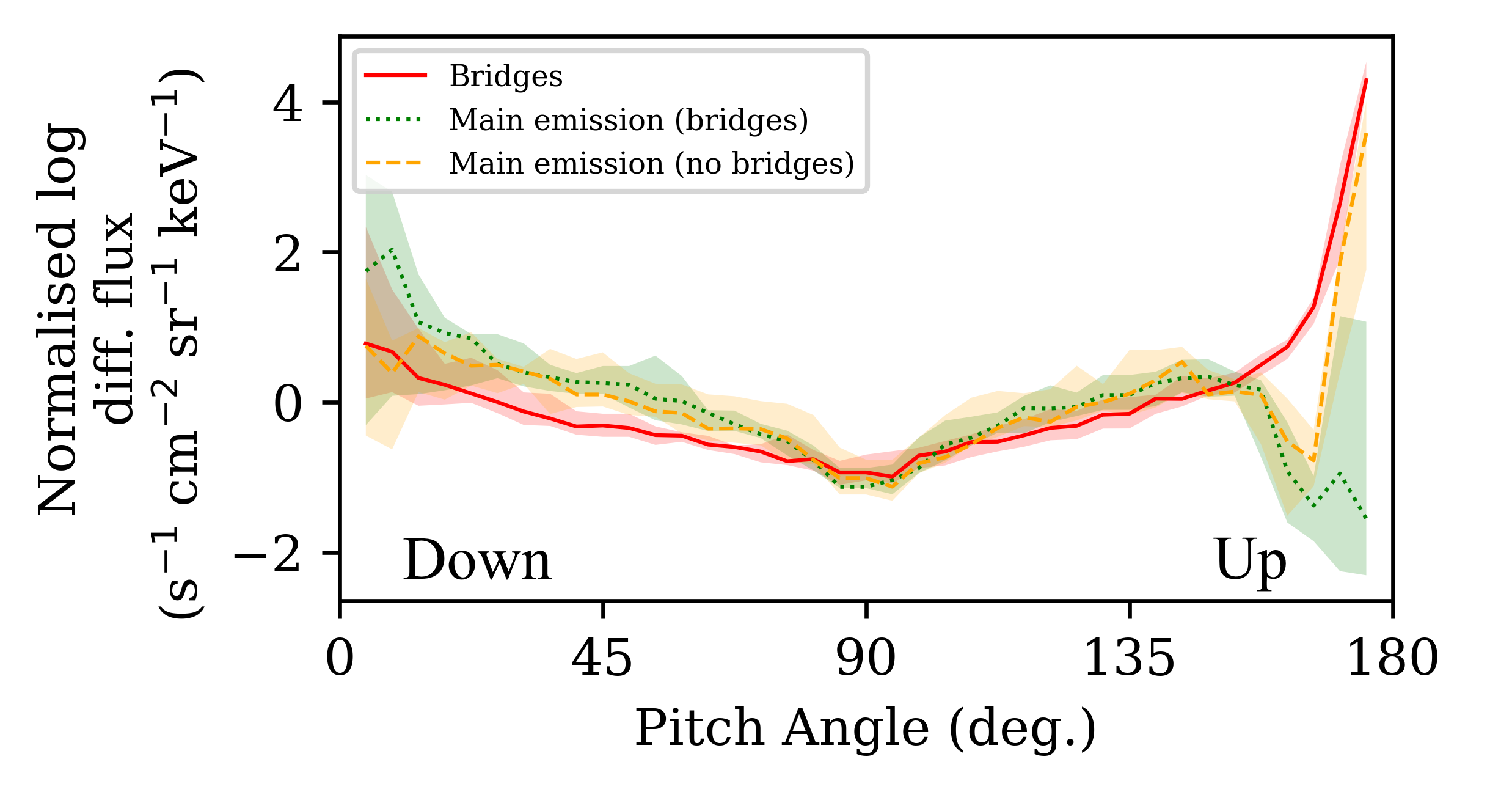}
    }
    {
        \includegraphics[width=0.48\linewidth]{UVS_both_med_noFilt30_Plot_JunoInstrumentData_instrAverage_annotate.png}
    }
    \caption{
         Median-average JEDI electron flux vs. pitch angle profiles during low-altitude ($<$3 R$_{J}$) bridge crossings (red, solid) and ME crossings, both with (green, dotted) and without (orange, dashed) local bridges, observed by Juno. The shaded regions denote the 25-to-75th percentile range.
    }
    \label{fig:bridge_averages}
\end{figure}

In addition to the above relation between bridge auroral brightness and downward electron flux, the typical electron pitch-angle profiles seen during bridge and ME crossings (red in Figure \ref{fig:bridge_averages}) are also noteworthy; the crossings used in this analysis are given in Figures \ref{fig:pitch_angle_average_bridges} to \ref{fig:pitch_angle_average_ME_nobridge} and Table \ref{tab:uvs_cases} of the supplementary material.
These pitch-angle distributions are the average normalised profiles of bridge/ME crossings rather than the average distribution, to account for large differences in total electron flux, since the profile is the aspect of interest.
Bridges are dominated by upward-travelling, field-aligned electrons, though with a considerable downward component.
ME crossings have been split into cases where the ME crossing is immediately preceded/succeeded by a bridge (whether this is crossed by Juno or not) and cases with no bridges in the vicinity of the ME crossing.
ME crossings with bridges (green) show electron distributions that are dominated by downward-going electrons. 
Cases without bridges (orange) have broadly symmetrical field-aligned electron populations, though there is a preference toward upward-travelling electrons.
It is first worth noting that the presence or absence of bridges affects the properties of the ME auroral-electron population, not simply its morphology.
Secondly, in the absence of a bridge, the ME electron population takes on a more ``bridge-like'' character, with a greater proportion of upward-travelling electrons.
This hints that the ``bridge'' is still present within the aurora but indistinguishable from the ME.

\begin{figure}[tbhp]
    \centering
    \ifthenelse{\equal{\twocol}{"y"}}{
        \includegraphics[width=\linewidth]{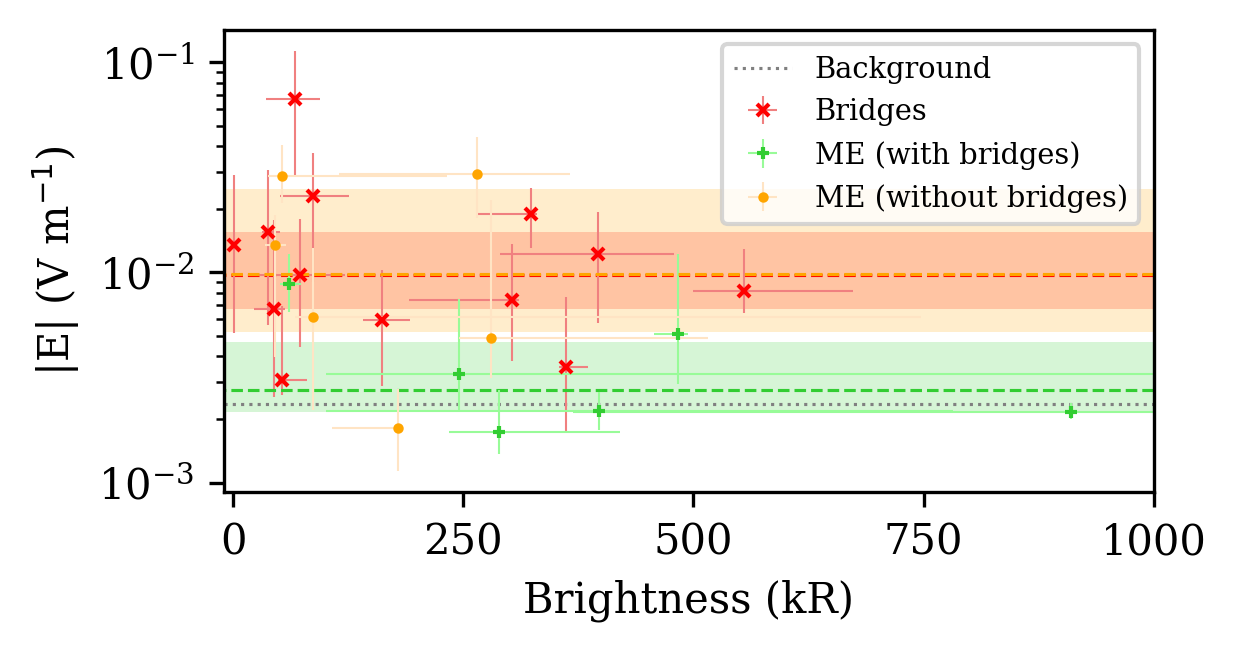}
    }
    {
        \includegraphics[width=0.48\linewidth]{UVS_both_med_noFilt30_Plot_JunoWavesBVsBrightness.png}
    }
    \caption{
         Median high-frequency ($>$1000 Hz) WAVES-E LFR-Lo intensity vs median UV Juno-footprint brightness during dusk-side (12$<$MLT$<$18), low-altitude ($<$3 R$_{J}$) bridge crossings (red, cross) and ME crossings, both with (green, plus) and without (orange, dot) local bridges, observed by Juno. Error bars give the 25th- and 75th-percentile values during each crossing. The median WAVES-E intensity of each distribution is given by a dashed line, and the 25-to-75th percentile range by the shaded areas. The background intensity is given by a dotted grey line.
    }
    \label{fig:waves_e_brightness}
\end{figure}

%

The three types of crossing also differ in the observed Waves-E LFR-Lo intensity, especially at higher frequencies; see Figure \ref{fig:waves_e_brightness}, where only the higher frequency portion ($>$1000 Hz) of the Waves-E LFR-Lo spectra has been used to capture the ``spiking'' behaviour of the Waves-E signal seen in Figure \ref{fig:PJ9_S_multiinstrument}.
The high-frequency Waves-E intensity is noticeably elevated compared to the polar-region background level (dotted grey line) for bridge crossings (red), in line with previous results \citep{sulaiman+:2022}, and crossings of the ME when no bridges are present (orange).
ME crossings with bridges present (green) have noticeably lower high-frequency Waves-E LHR-Lo intensities comparable with the background, as seen in PJ9-S, and there is a clear distinction (in the 25-to-75th-percentile range) from the other two crossing types.
If increased Waves-E intensity can be associated with bridge crossings, as seen during PJ9-S and as indicated by Figure \ref{fig:waves_e_brightness}, then the ME appears to take on bridge-like traits in the absence of a discernable bridge, again indicating that the ``bridge'' emission is still present within the ME.
As in PJ9-S, although this Waves-E intensity is increased during bridge crossings, it is not correlated with auroral brightness, which may be due to a strong dependence on altitude or a genuine complex non-linearity in the associated acceleration process(es).

\section{Discussion}
\label{sec:discussion}


The results of this work indicate that bridges show similarities with ZII aurora, as defined by \citet{mauk+:2020}.
They preferentially coincide with the downward-FAC region poleward of the ME and they show bidirectional electron distributions dominated by upward-travelling electrons, as in Figure \ref{fig:bridge_averages}.
This tentative association between ZII aurora and bridges is notable because it would imply that the spacing between ZI (upward FAC) and ZII (downward FAC) aurora is variable, to such an extent that there is sometimes a considerable gap between the two (in this case, between the bridge and the rest of the ME).
The presence or absence of bridges is also suggested to affect the properties of the ME.
When bridges are present, the ME is dominated by downward-travelling electrons (see Figure \ref{fig:bridge_averages}), as expected of ZI aurora. 
When bridges are instead absent, the ME electron distribution is typically more symmetric.
The proposed interpretation is that bridges are ZII aurora that have become spatially separated from the ZI aurora, such that, in the absence of bridges, the ZII aurora is spatially indistinguishable from the ZI ME, leading to a more symmetric electron distribution during these ME crossings.
This hypothesis is further supported by the stronger broadband plasma-wave signatures (considered to be a signature of ZII aurora; \citealt{sulaiman+:2022}) seen during ME crossings in the absence of bridges (compared to those where bridges are present), as though the ZI and ZII aurorae are spatially adjacent (the ``typical'' configuration; \citealt{mauk+:2020}).
Even if future work weakens the association between bridges and ZII aurora, the fact that the presence of bridges affects not only the morphology but also the electron populations of the ME is itself notable.

Pending sampling of the low-altitude dawn aurora by Juno, the distinction between ZI and ZII aurora is expected to be a phenomenon that is present in the entire ME, though perhaps with some considerable local-time dependence \citep{sulaiman+:2022}, which may seem at odds with the idea that the bridge, a uniquely pre-dusk feature, can be identified as ZII aurora.
It is suggested that the process that gives rise to the separation between ZI and ZII in the case of bridges may occur more easily or to a greater extent in the pre-dusk magnetosphere.
This is compatible with analysis by \citet{jenkins+:2024}, in which the Alfv\'{e}n radius is shown to be greater than 60 R$_{J}$ only between 10 and 20 MLT, limiting the information that can be conveyed to the aurora from the distant magnetosphere beyond 60 R$_{J}$ via Alfv\'{e}n waves outside of these sectors; the bridge source process may similarly be hampered in the dawn and night sectors, explaining the lack of observed bridges in these regions of the aurora.
This interpretation is tentative and would require both modelling work and further data from Juno, as well as a better understanding of the processes that give rise to bridges.

The results of this work also support the ZII auroral-generation scenario described by \citet{elliott+:2018} and \citet{sulaiman+:2022}.
In this scenario, electric-potential structures above the ionosphere create upward electron beams that generate large-amplitude ESWs, as has been demonstrated \citep{elliott+:2020}.
These ESWs provoke bidirectional broadband electron acceleration that leads to auroral emission.
This scenario explains the presence of high-frequency Waves-E emission observed by Juno-Waves during bridge crossings (frequency-domain representations of large-amplitude ESWs; \citealt{sulaiman+:2022}) and upward-dominated bidirectional broadband electrons observed by JEDI (the combination of upward electron beams and stochastic acceleration from ESWs).
The association of enhanced Waves-E emission with bridge crossings indicated in this work, rather than with signatures in the FACs or Alfv\'{e}nic Poynting flux, supports this scenario. 
ESWs, observed as enhanced Waves-E emission, are also associated with ME crossings in the absence of bridges, but not when bridges are present, again supporting the hypothesis that the ``bridge'' aurora remains present even when spatially indistinguishable from the ME, consistent with the association between bridges and ZII.
While ESWs appear to be associated with bridge crossings, the expected acceleration processes are poorly understood and highly non-linear and, though Waves-E intensity peaks are systematically seen during bridge crossings, there is a lack of correlation between auroral brightness and Waves-E intensity, and so the above scenario remains speculative.


The results of this work also suggest that the solar wind exerts some influence on the morphology of the ZI and ZII aurorae.
If magnetospheric compression can be predominantly attributed to the solar wind (rather than a variable plasma-outflow rate from Io), the agreement between bridge count and the compression state of the magnetosphere, shown in Figure \ref{fig:bridge_vs_ms_size}, is consistent with the result of \citet{nichols+:2017}, in which a bridge-like morphology was observed in the dusk-side polar collar during a solar-wind shock measured by Juno during its approach to Jupiter.
The results regarding the onset times and lifetimes for bridges (that they can arise within a few hours and can last for longer than a Jupiter rotation) also fit this hypothesis, since hour-scale variations are compatible with simulations of the effect of solar-wind shocks on the magnetosphere \citep{chane+:2017}.
Therefore, if ZI aurora can be spatially associated with upward Birkeland currents and ZII aurorae with the downward currents necessary to close the loop \citep{sulaiman+:2022}, then the solar wind is suggested to affect the local ionospheric or magnetospheric conditions required for current-loop closure, thus affecting the morphology of the ZI and ZII aurorae.
It may, to some extent, do this by modifying the magnetic topology within Jupiter's middle magnetosphere.
Compression of the magnetosphere by the solar wind has already been shown to play a significant role in contracting the ME via modification of the current sheet and associated magnetic-field contribution, rather than by modifying the ME magnetospheric source radius \citep{nichols+:2009,head+:2024}.
This present work may indicate that magnetospheric compression by the solar wind pushes the magnetic field lines inward, which would move ME features poleward as well as increase the ionospheric separation between adjacent features in the magnetosphere.
This scenario, though tentative, is consistent with bridges that ``grow'' from the ME rather than from the active region (Figure \ref{fig:bridge_onset}).
Additionally, three ME crossings (PJ3-N, PJ14-N, PJ22-N) were identified by \citet{al_saati+:2022} in which the ZII (downward FAC) region was located equatorward of the ZI region. 
These three cases all also show expanded MEs (\citealt{head+:2024}; supporting material), associated with an uncompressed magnetosphere, which may work in the opposite direction to pull the ZII aurorae equatorward of the ZI aurora, in contrast with previous predictions \citep{cowley+:2007}.
More work is required to investigate the exact mechanisms for the variation in morphology of the ZI and ZII aurorae, but, in all, this work suggests that the solar wind may exert an indirect yet significant influence on Jupiter's aurorae.


\FloatBarrier
\section{Conclusions}

The findings of this work can be summarised as follows:
\begin{itemize}
    \item Bridges, observed as dusk-side polar-collar arcs in Jupiter's UV aurora, are frequently seen in both hemispheres, in both HST-STIS and Juno-UVS images.
    \item They are observed to appear and disappear on timescales of hours and have been seen to persist over a full Jupiter rotation, in line with previous conclusions made using limited data.
    \item The appearance of bridges in the aurora is associated with compression of the magnetosphere.
    \item Crossings of bridges by Juno are preferentially associated with downward FACs and broadband field-aligned electron distributions dominated by upward-travelling electrons, indicating that bridges are possibly ZII aurorae, as defined by \citet{mauk+:2020}, that have become noticeably separated from the (ZI) ME. Where Juno passes at low ($<$3~R$_{J}$) altitude over the aurora, bridge crossings are also often associated with ESWs observed by Juno-Waves. This aligns with the scenario presented by \citet{elliott+:2018} and \citet{sulaiman+:2022}, in which upward-travelling ESWs induce bidirectional broadband auroral electron acceleration.
    \item The ME is typically associated with ESWs and largely symmetric, bidirectional electron distributions when bridges are absent from the aurora, and with predominantly downward electron populations without considerable ESWs when bridges are present. This suggests that the ME may exist as either an adjacent ZI/ZII aurora, or as a uniquely ZI aurora when the ZII aurora has become spatially separated in the form of bridges.
    \item Finally, the identification of the bridge as ZII aurora spatially separated from the ME, alongside the observed dependence of the appearance of bridges on the state of compression of the magnetosphere, implies that the solar wind can exert influence on the morphology of Jupiter's UV aurora.  
\end{itemize}

Future numerical experiments are required to investigate the effect of various types of solar-wind compression on auroral morphology and provide a theoretical explanation for the results described in this work.


\section*{Data availability statement}
Juno data can be obtained from the NASA Planetary Data System (\url{https://pds-atmospheres.nmsu.edu/data_and_services/atmospheres_data/JUNO/juno.html}).

\section*{Acknowledgements}
We are grateful to NASA and contributing institutions which have made the Juno mission possible. This work was funded by NASA's New Frontiers Program for Juno via contract with the Southwest Research Institute. This publication benefits from the support of the French Community of Belgium in the context of the FRIA Doctoral Grant awarded to L. A. Head. B. Bonfond is a Research Associate of the Fonds de la Recherche Scientifique - FNRS. M. F. Vogt was supported by NASA grants 80NSSC17K0777 and 80NSSC20K0559. Data analysis was performed with the AMDA science analysis system provided by the Centre de Donn\'{e}es de la Physique des Plasmas (CDPP) supported by CNRS, CNES, Observatoire de Paris and Universit\'{e} Paul Sabatier, Toulouse. A. Moirano is supported by the Fonds de la Recherche Scientifique - FNRS under Grant(s) No T003524F.

\bibliography{references}

\begin{appendix}






\onecolumn

\renewcommand{\thefigure}{A.\arabic{figure}}
\renewcommand{\thetable}{C.\arabic{table}}

\section{Description of methods}

\subsection{Bridge detection}
\label{sec:method_explanation}

\begin{figure}[ht]
    \centering

    \captionsetup[subfigure]{width=0.42\linewidth}
    \subfloat[]{
        \includegraphics[width=0.48\linewidth]{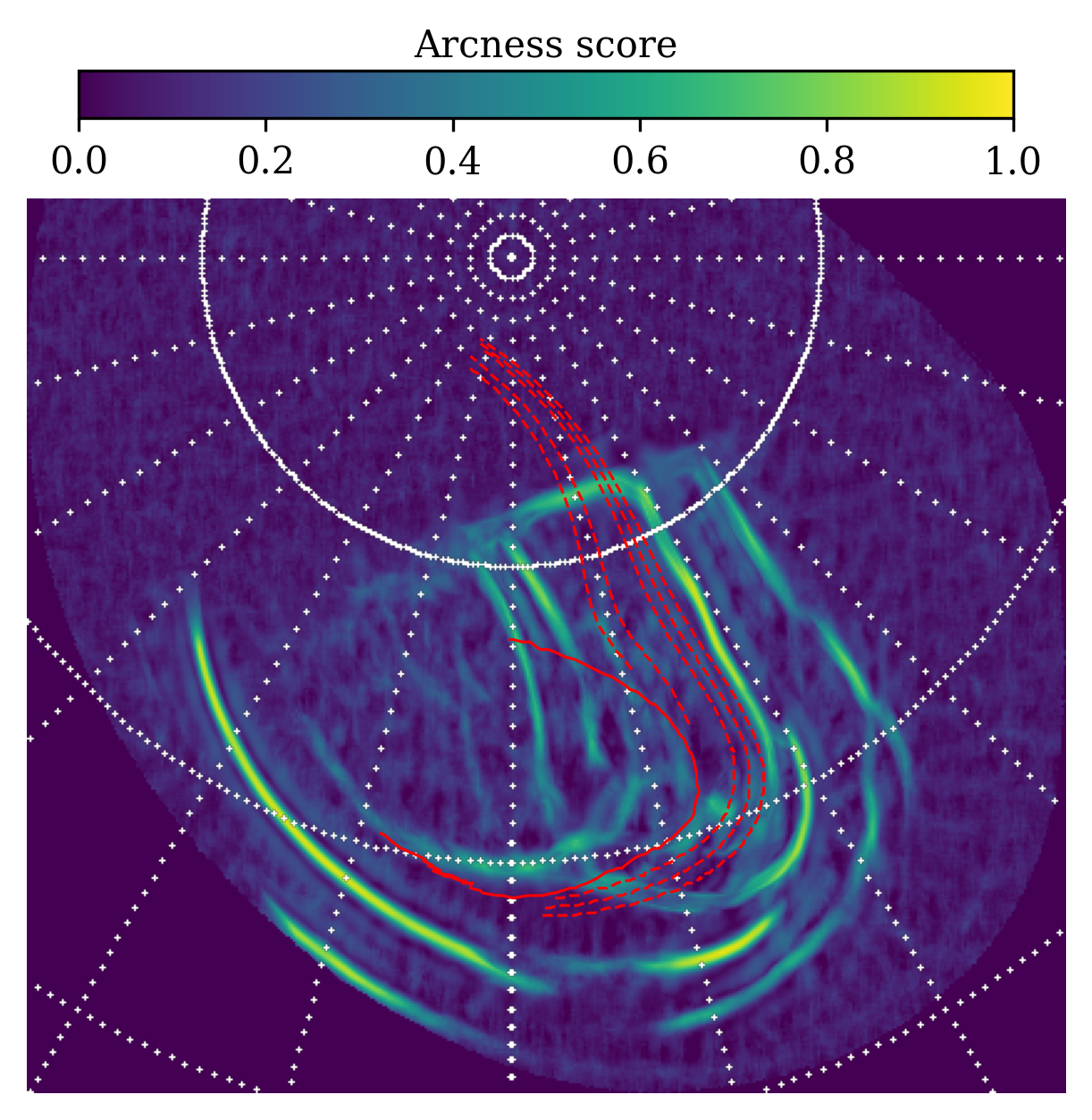}
    }
    \subfloat[]{
        \includegraphics[width=0.48\linewidth]{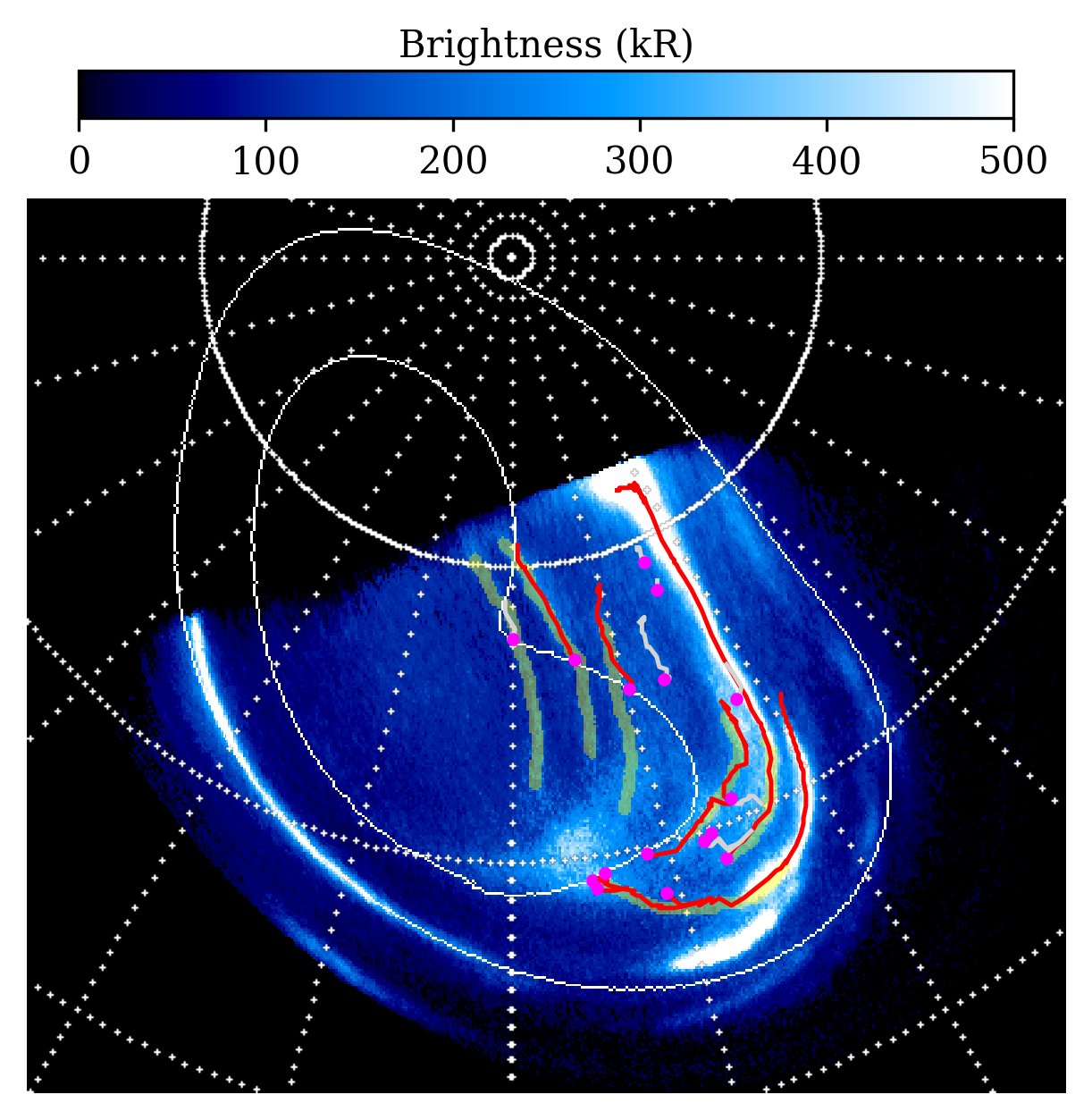}
    }
    \caption{
        (a) Arc convolution of the auroral image shown in Figure \ref{fig:bridge_example}. Red lines denote the mapped magnetopause (solid) and fixed-radius contours (dashed) described in the text. (b) Results of the bridge-detection algorithm after filtering. Red lines denote arcs accepted by the filter, and grey the arcs that have been discarded. The seed point of each arc is given in magenta. Manually designated arcs are given in yellow. White contours give the region of validity of the \citet{vogt+:2011} JRM33 flux-equivalence mapping along closed field lines. 
    }
    \label{fig:bridge_detection_suppmat}
\end{figure}

Bridges are defined as arcs that connect the day-side active region to the main emission via the polar collar (though they may not span this gap fully).
This implies that they traverse a significant radial distance in the magnetosphere, since the main emission is surmised to originate from a region between 20 and 40 R$_{J}$\citep{cowley+bunce:2001}, whereas the active region is firmly within the polar aurora and hence maps to more distant regions of the magnetosphere.
Hence, a number of fixed-radius contours were projected to the ionosphere (75, 80, 85, 100, and 110 R$_{J}$), as well as the \citet{joy+:2002} expanded magnetopause (stand-off distance 92 R$_{J}$; used as the outer boundary of validity in the \citealt{vogt+:2011} flux-equivalence model) which was observed to approximately coincide with the boundary between the polar collar and the day-side active region; see red contours in Figure \ref{fig:bridge_detection_suppmat}a. 
Here, the flux-equivalence mapping of \citet{vogt+:2011} was used with the JRM33 magnetic-field model \citep{connerney+:2022} since it is expected to be more reliable than field-line tracing in the polar aurora, though the choice of mapping method was a posteriori determined to not affect the conclusions of this work.
The brightness along these mapped contours was determined and a broad uniform filter (0.3 MLT) applied.
Peaks in the smoothed brightness were presumed to coincide with arcs that cross these contours.
For each of these ``seed'' points, the arc was propagated in a stepwise fashion, by taking all pixels within a 11$\times$11-pixel area around the seed pixel, calculating an effective path brightness for each pixel by averaging the brightness of all pixels between the seed and the target pixel (to ensure that the arc propagation algorithm does not jump over regions of dark pixels to arrive at a bright pixel), and adding the pixel with the greatest effective path brightness to the arc.
This process is repeated for all seed pixels until one of the following conditions is met: the new best target pixel is already part of a different arc, at which point the two arcs can be merged; the target pixel maps to a magnetospheric radius that is more than 10 R$_{J}$ from the radius of the previous pixel, which was determined to often coincide with a failure of the algorithm; the target pixel maps to a radius less than 50 R$_{J}$ in the magnetosphere, at which point the algorithm is at risk of simply detecting the main emission; the target pixel has an arcness score below 0.2.

To ensure that the algorithm is indeed detecting bridges, manual designations were created (thick yellow arcs in Figure \ref{fig:bridge_detection_suppmat}b) for the northern-hemisphere image series in the STIS campaign GO-15638, which contains a total of 66 manually identified bridges over 42 image series.
At this stage, the bridge-detection algorithm performs poorly; it does detect 90\% of the manually identified bridges, but only 16\% of the automatically detected bridges have a manual equivalent.
Using the \texttt{scikit-learn} Python library, a random-forest classifier with a typical train-test split of 8:2 was trained on a number of detected-arc parameters (seed magnetosphere radial distance, seed magnetosphere local time, average arc brightness, average arc ``arcness'' value, total projected length of the arc in the magnetosphere).
The optimal hyperparameters were determined using a randomised search with tree counts between 10 and 1000 and maximum tree depths between 3 and 50; a tree count of 912 and a maximum tree depth of 42 were found to be optimal, and resulted in a model test accuracy of 0.82 when applied to the test data.
This classifier was then applied to the full set of automatically detected bridges as a filter, which increased the proportion of detected arcs with manual equivalents (of those cases where arcs had been identified manually) from 16\% to 96\%.
An example of the filtered bridge arcs is given in Figure \ref{fig:bridge_detection_suppmat}b. 

To estimate the uncertainty in the detected bridge count introduced by this method, a sensitivity analysis was performed by varying the cutoff arcness score between 0.1 and 0.3. 
The uncertainties in the detected bridge count shown in Figure 6 indicate the range of detected bridge counts introduced by varying this cutoff parameter.

\subsection{Description of the field-aligned-current calculation}
\label{sec:fac_method}

The method used to estimate the field-aligned-current (FAC) densities close to Jupiter uses data from the Juno FluxGate Magnetometer (FGM) instrument and is essentially an application of the method described in \citet{al_saati+:2022}.
The interested reader is invited to consult the thorough description of this method provided in the supporting material of \citet{al_saati+:2022}.
The method makes several key assumptions.
Firstly, the ionosphere/auroral layer is modelled as an infinitely thin shell; this is a reasonable approximation given the thickness of this layer compared to the scale of the jovian magnetosphere.
Jupiter's magnetic field is assumed to be dipolar, and hence the magnetic field lines radial close to the planet.
The main emission is taken to run (locally) along a contour of constant magnetic latitude, and that magnetospheric-coupling parameters are (locally) constant along the field line and along the main emission, and only vary in the direction perpendicular to the contour of the main emission.
Juno is assumed to be moving roughly perpendicularly to the background magnetic field, and perturbations in this magnetic field are assumed to be spatial, not temporal, since Juno cannot distinguish between the two \citep{sulaiman+:2023}.
Under these assumptions, Amp\`{e}re's law (in a vacuum and in the absence of a time-varying electric field)
\begin{equation}
    \vec{J}=\frac{1}{\mu_0}\vec{\nabla}\times \vec{B},
\end{equation}
where $\vec{J}$ is the current density at Juno, $\mu_0$ is the permeability of free space, and $\vec{B}$ is the magnetic field at Juno, can be reduced to
\begin{equation}
    J_{\parallel}=\frac{1}{\mu_0} \frac{\partial(\sin{\theta}  \delta B_{\phi})}{r \sin{\theta} \partial \theta},
\end{equation}
where $\delta \vec{B}$ denotes the magnetic-field residuals obtained by subtracting a magnetic-field model from the FGM magnetic-field measurements (JRM09 in \citet{al_saati+:2022}, JRM33 in this work).
By conservation of magnetic flux,
\begin{equation}
    J_{\parallel,iono}=\frac{|B_{iono}|}{|B|}J_{\parallel},
\end{equation}
where $X_{iono}$ refers to values determined for the ionosphere.
\citet{al_saati+:2022} also perform a Fourier filtering on their magnetic-field residuals to remove superfluous frequencies that lay far from the frequency of their signal(s) of interest.

This work makes the same assumptions as \citet{al_saati+:2022}, barring the assumption of a dipolar magnetic field.
Instead, the full magnetic field is taken into account when calculating $\vec{\nabla}\times \delta\vec{B}$, and so the field-aligned currents are given by
\begin{equation}
    J_{\parallel}=(\frac{1}{\mu_0} \vec{\nabla}\times \delta\vec{B})\cdot \hat{B}_{JRM33}.
\end{equation}
This does not greatly increase the computational effort required to estimate the field-aligned currents.
Additionally, no Fourier filtering is performed on the magnetic-field residuals since so particular signal frequency is targeted by this work.
The implementation of this method was checked against the results presented in \citet{al_saati+:2022} and found to agree exactly in all cases, and thus the method used in this work is considered to be verified against \citet{al_saati+:2022}, though the assumptions used in this work risk misrepresenting the true state of the auroral currents and should be investigated more thoroughly in a dedicated work.

\newpage
\renewcommand{\thefigure}{B.\arabic{figure}}
\FloatBarrier
\section{Supplementary figures}

\begin{figure}[h]
    \centering

    \captionsetup[subfigure]{width=0.18\linewidth}
    \subfloat[]{
        \includegraphics[width=0.18\linewidth]{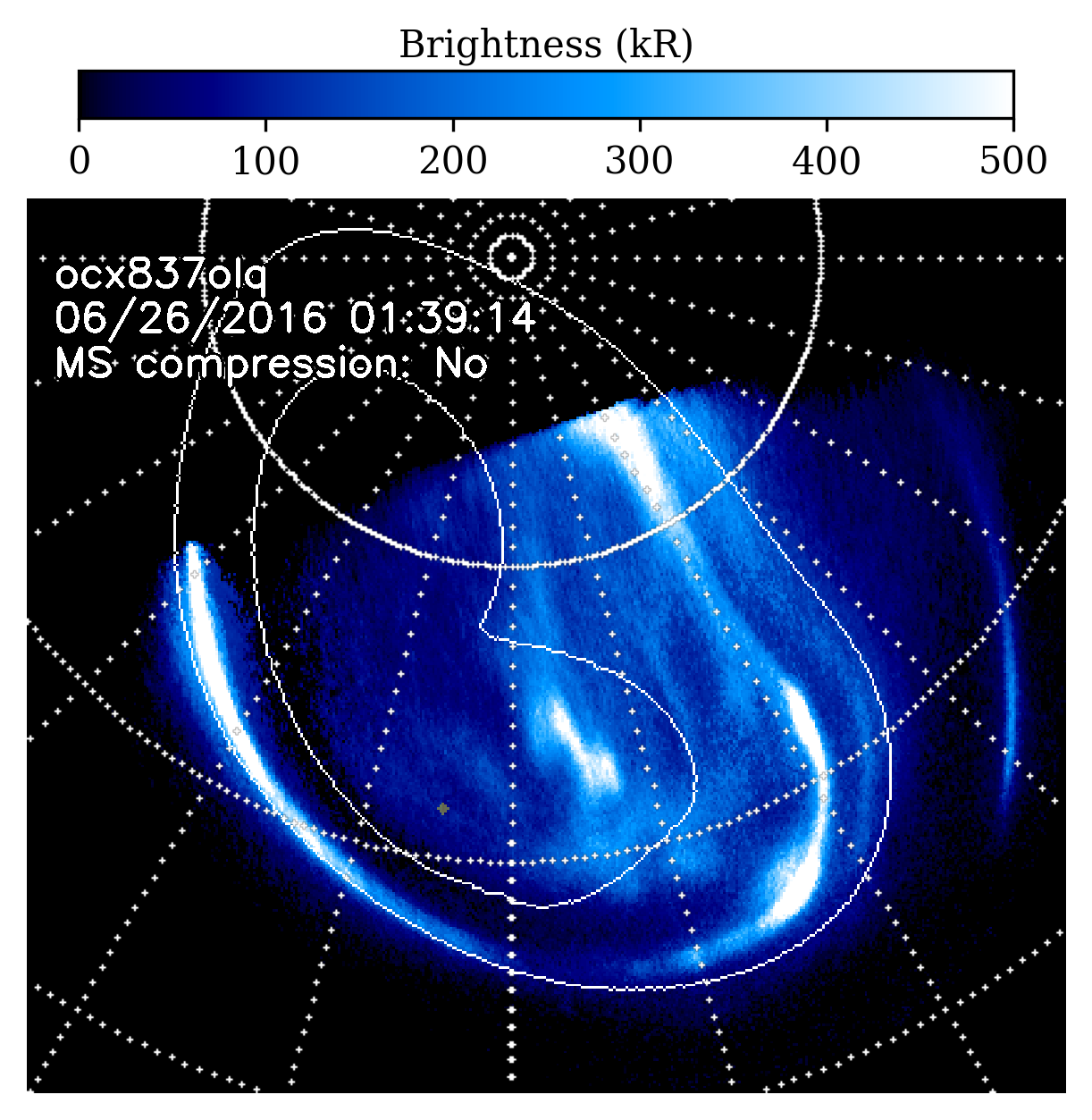}
    }
    \subfloat[]{
        \includegraphics[width=0.18\linewidth]{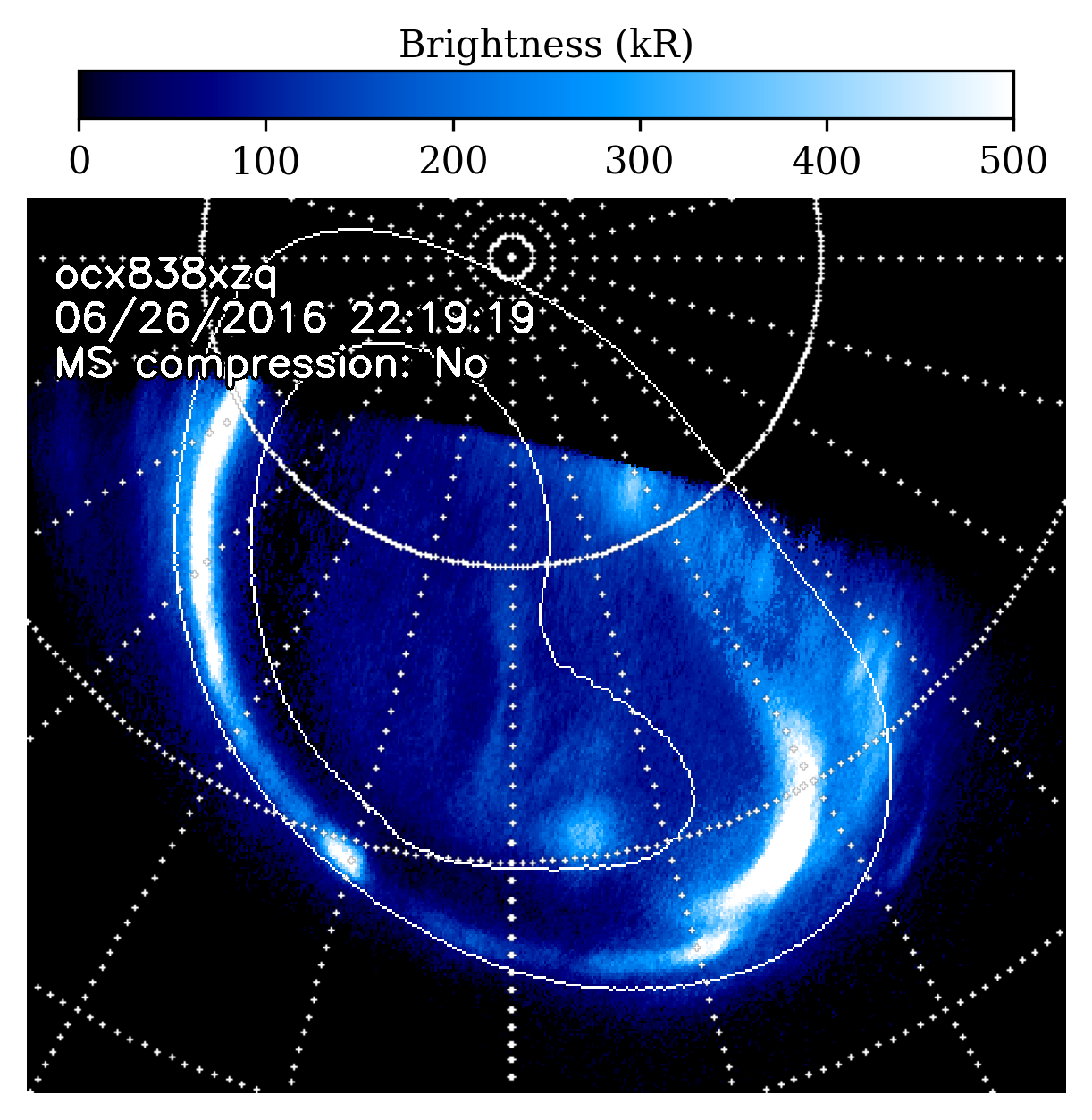}
    } 
    \subfloat[]{
        \includegraphics[width=0.18\linewidth]{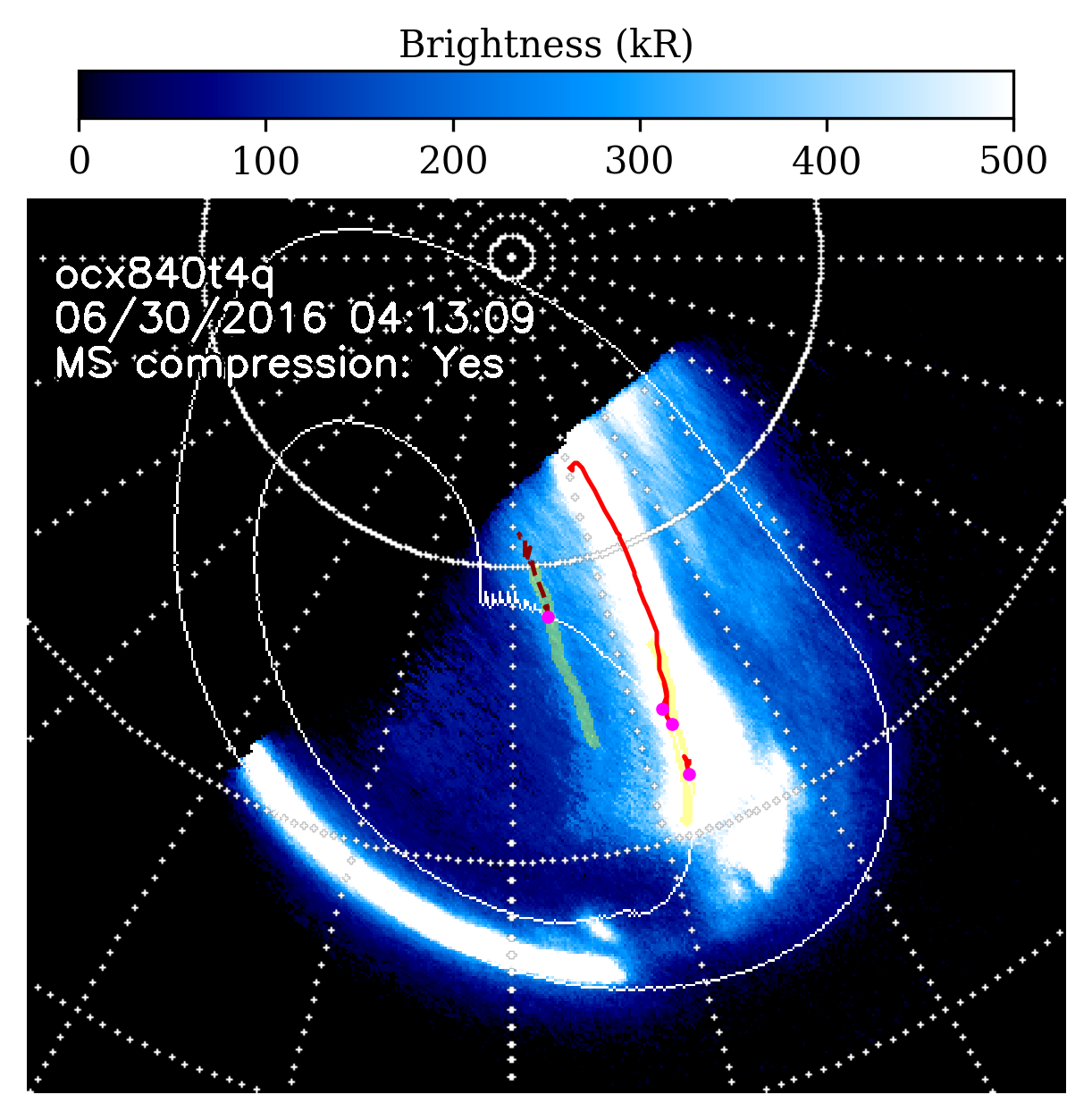}
    }
    \subfloat[]{
        \includegraphics[width=0.18\linewidth]{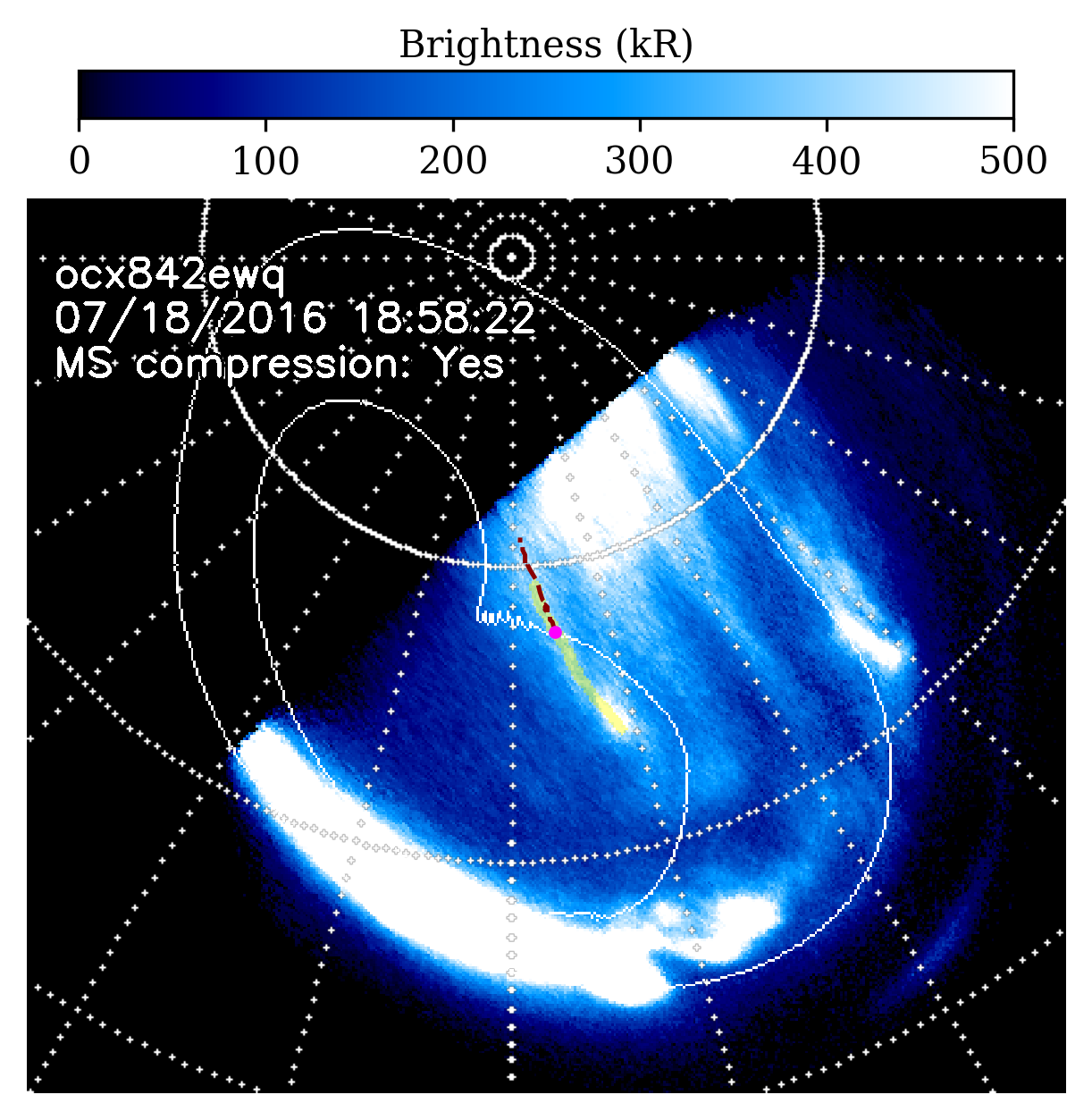}
    }
    \subfloat[]{
        \includegraphics[width=0.18\linewidth]{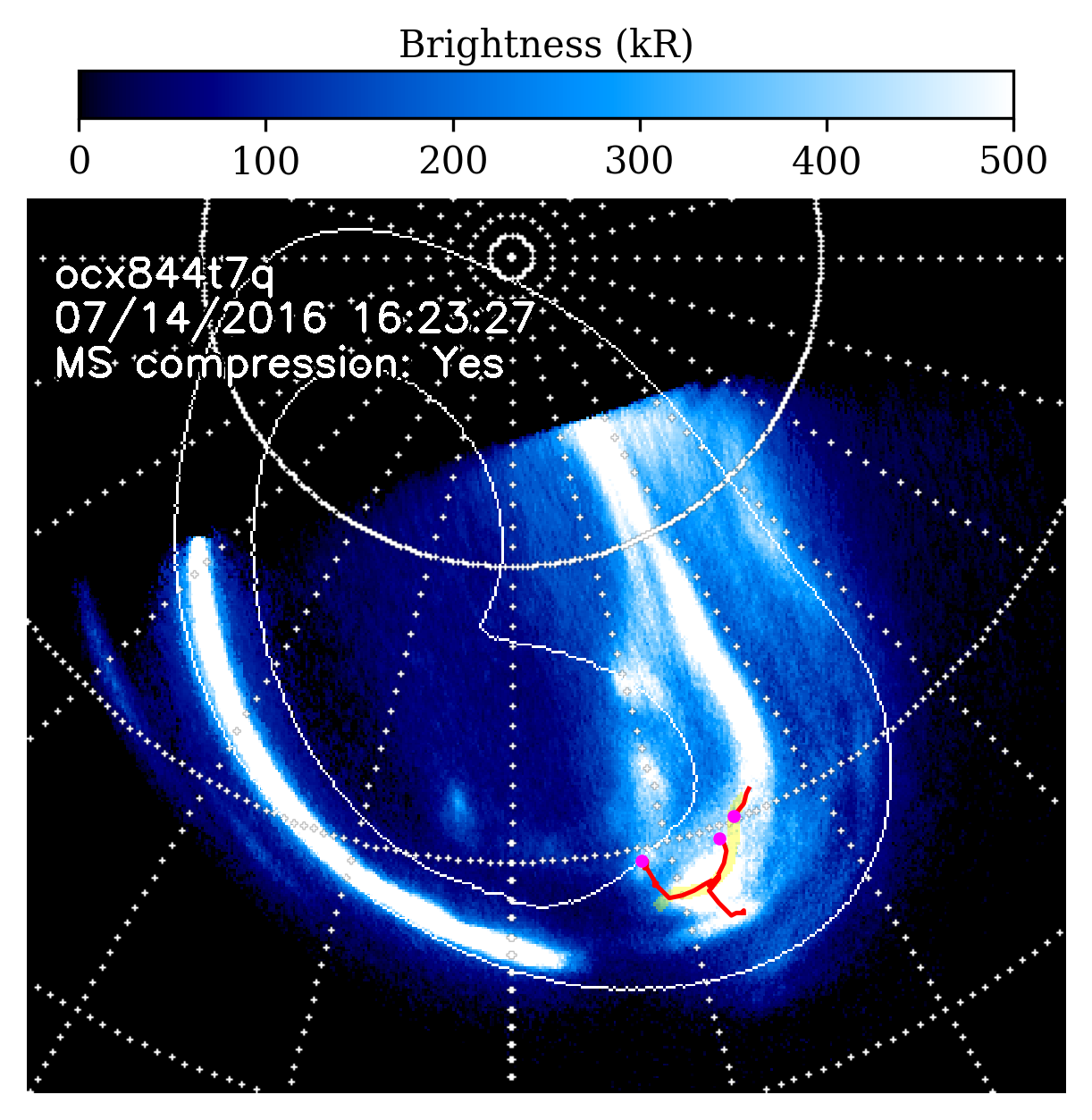}
    }\\
    \subfloat[]{
        \includegraphics[width=0.18\linewidth]{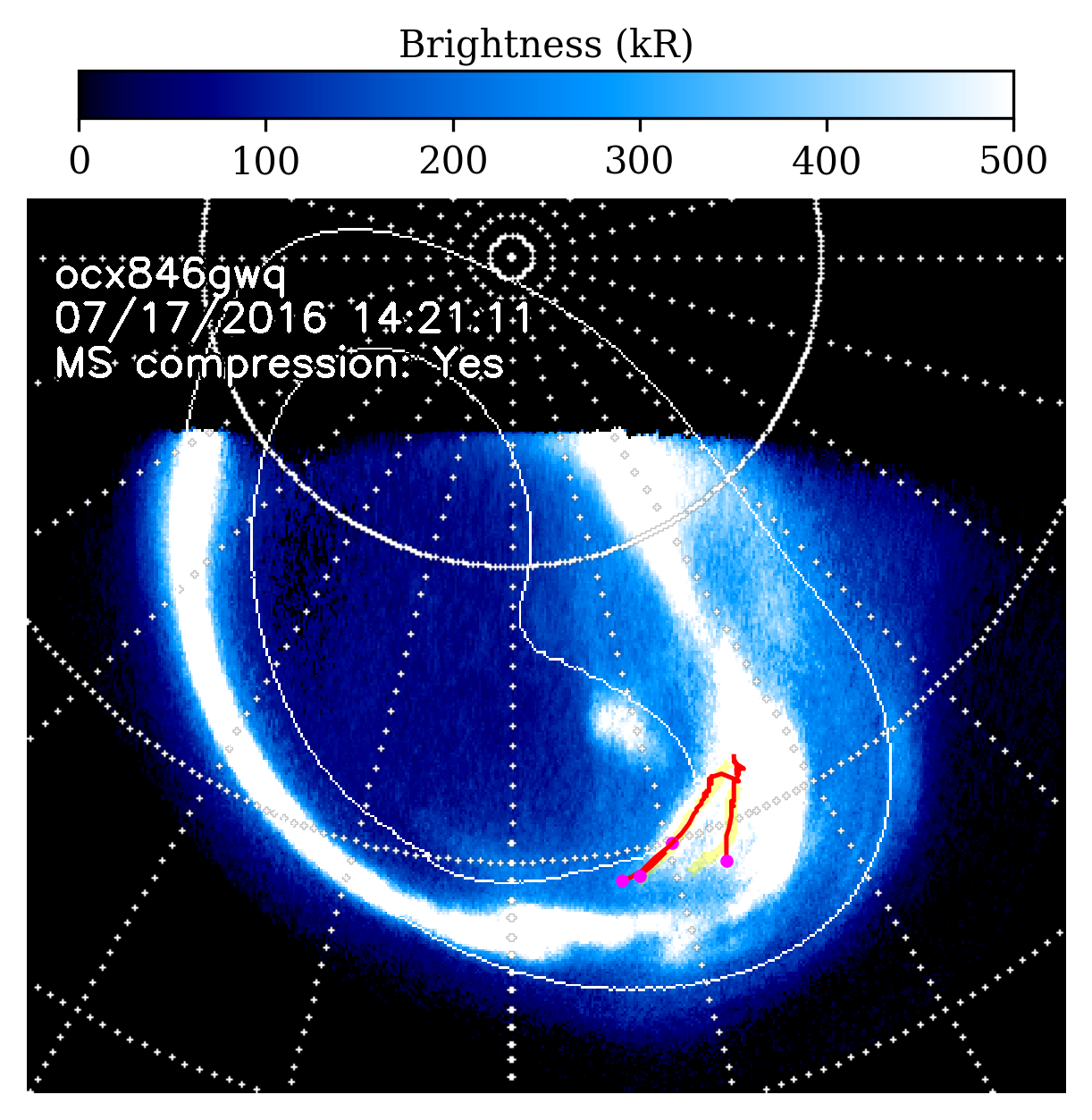}
    } 
    \subfloat[]{
        \includegraphics[width=0.18\linewidth]{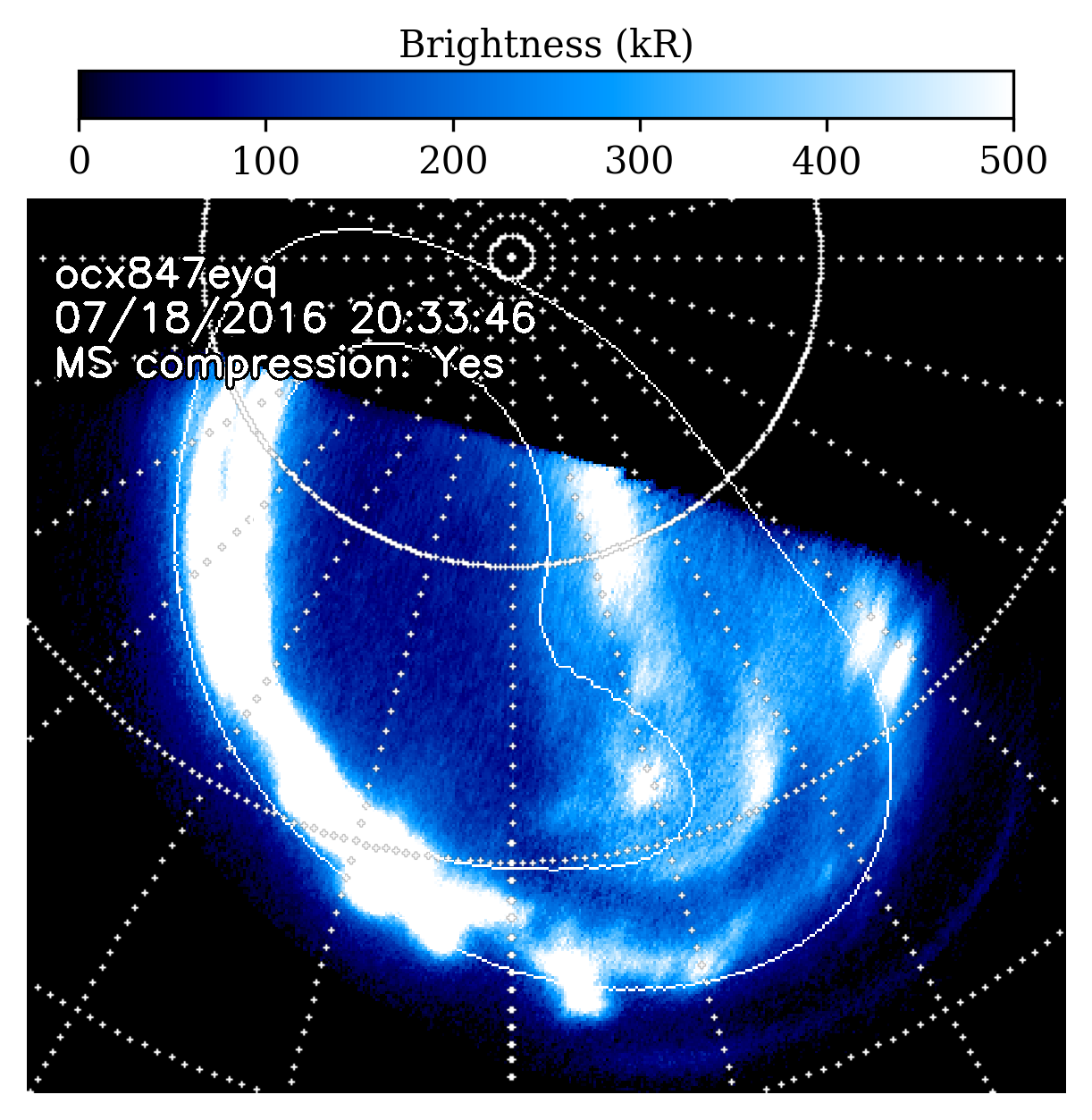}
    }
    \subfloat[]{
        \includegraphics[width=0.18\linewidth]{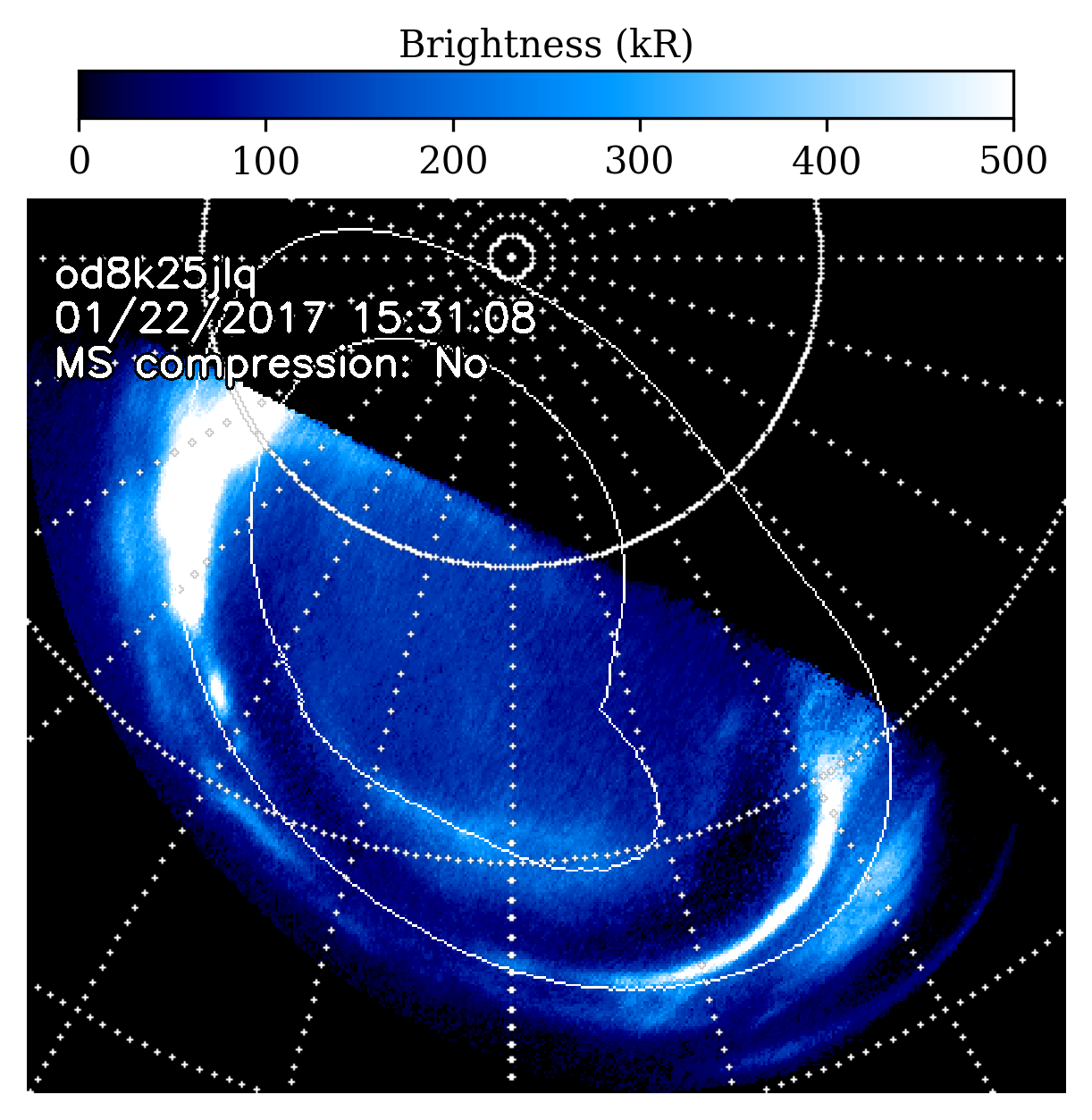}
    } 
    \subfloat[]{
        \includegraphics[width=0.18\linewidth]{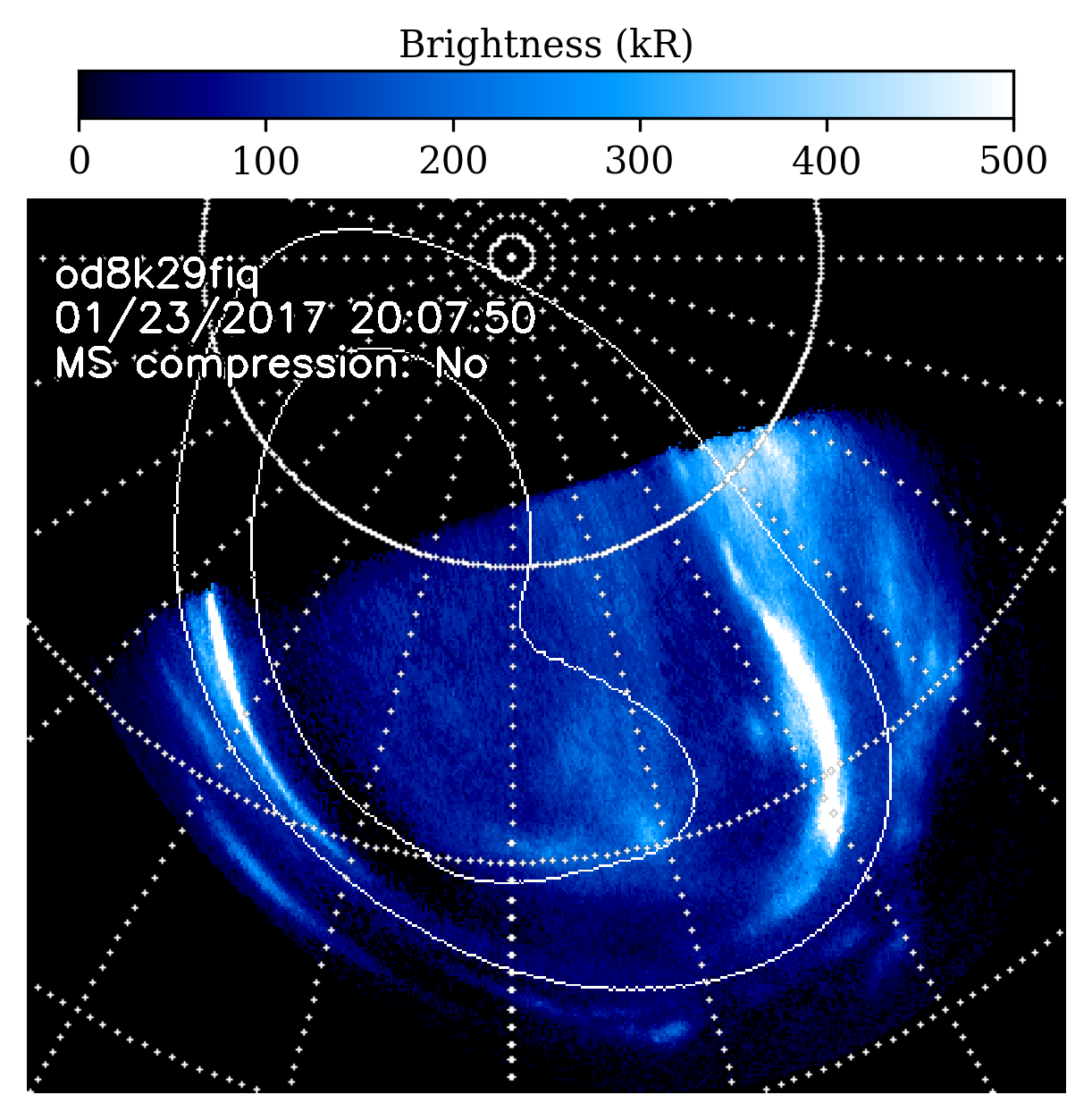}
    }
    \subfloat[]{
        \includegraphics[width=0.18\linewidth]{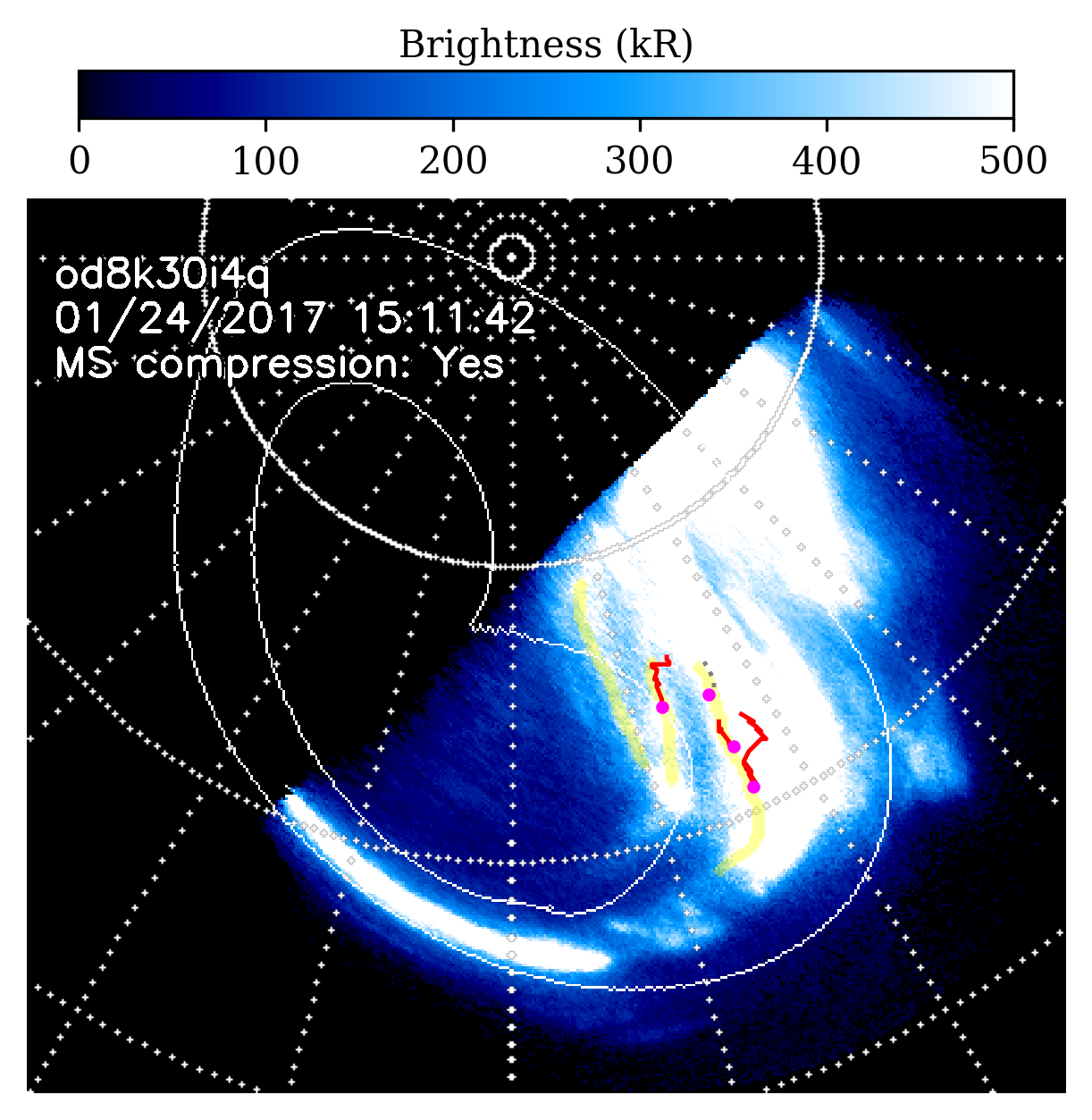}
    } \\
    \subfloat[]{
        \includegraphics[width=0.18\linewidth]{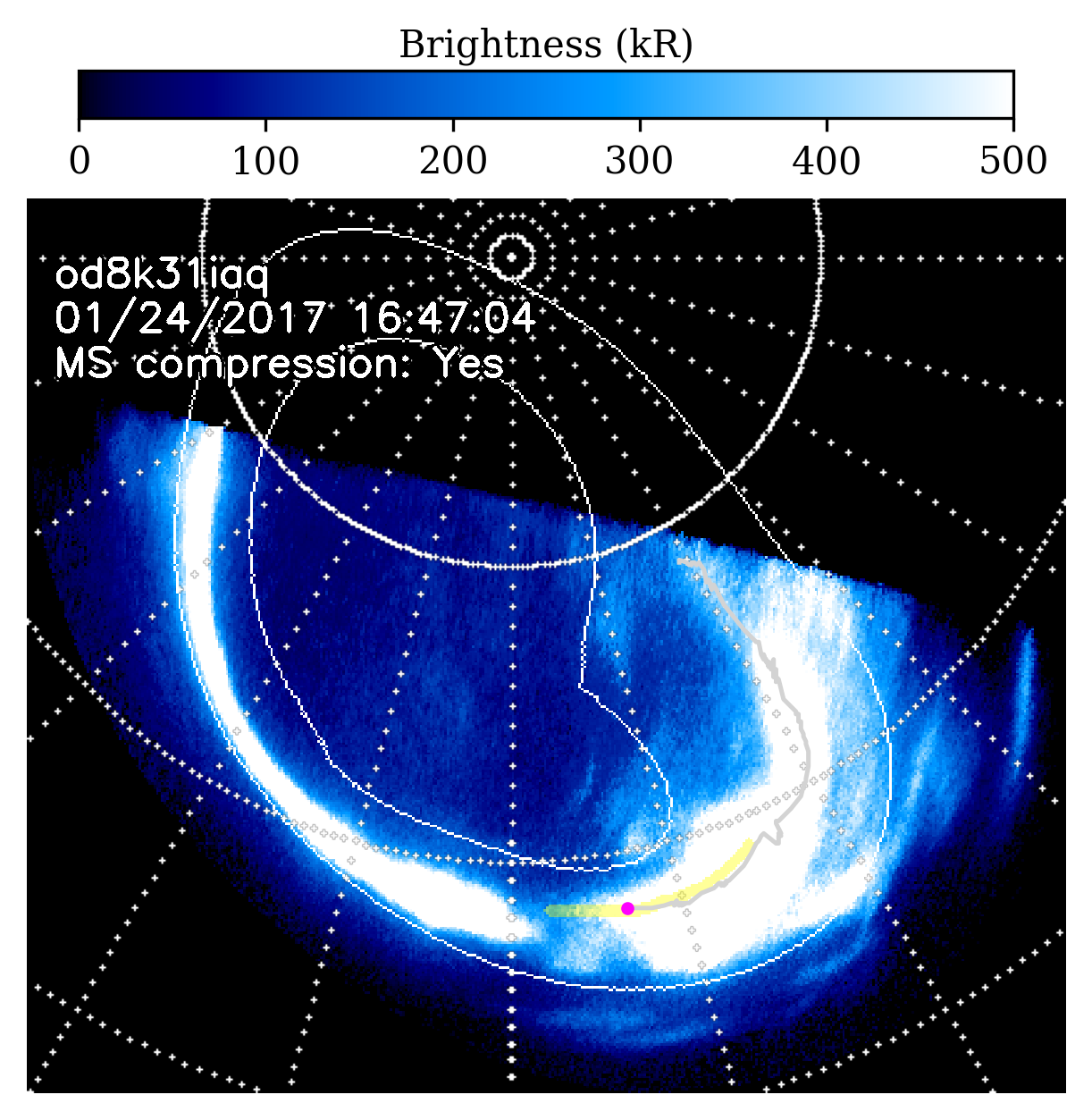}
    }
    \subfloat[]{
        \includegraphics[width=0.18\linewidth]{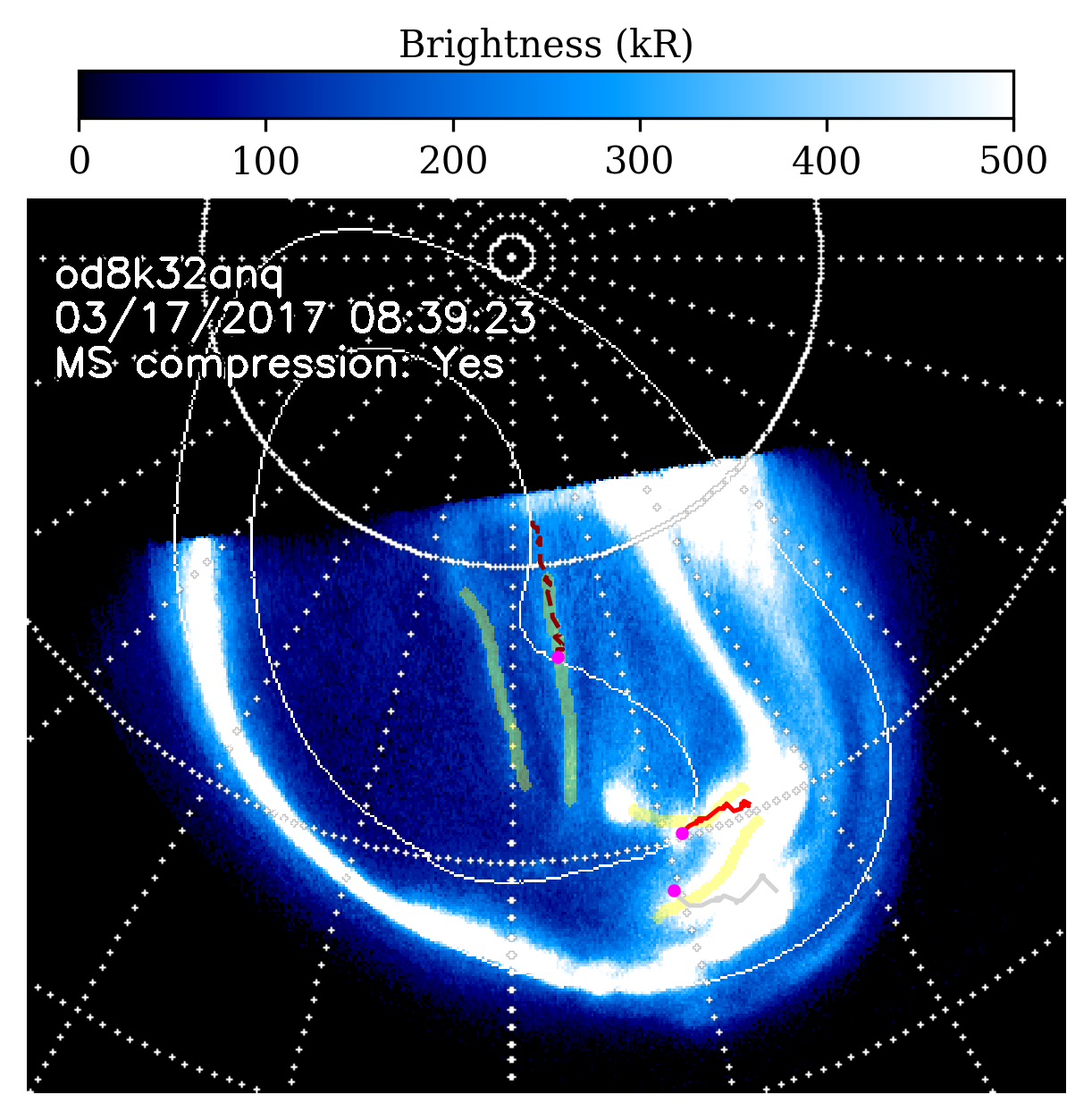}
    } 
    \subfloat[]{
        \includegraphics[width=0.18\linewidth]{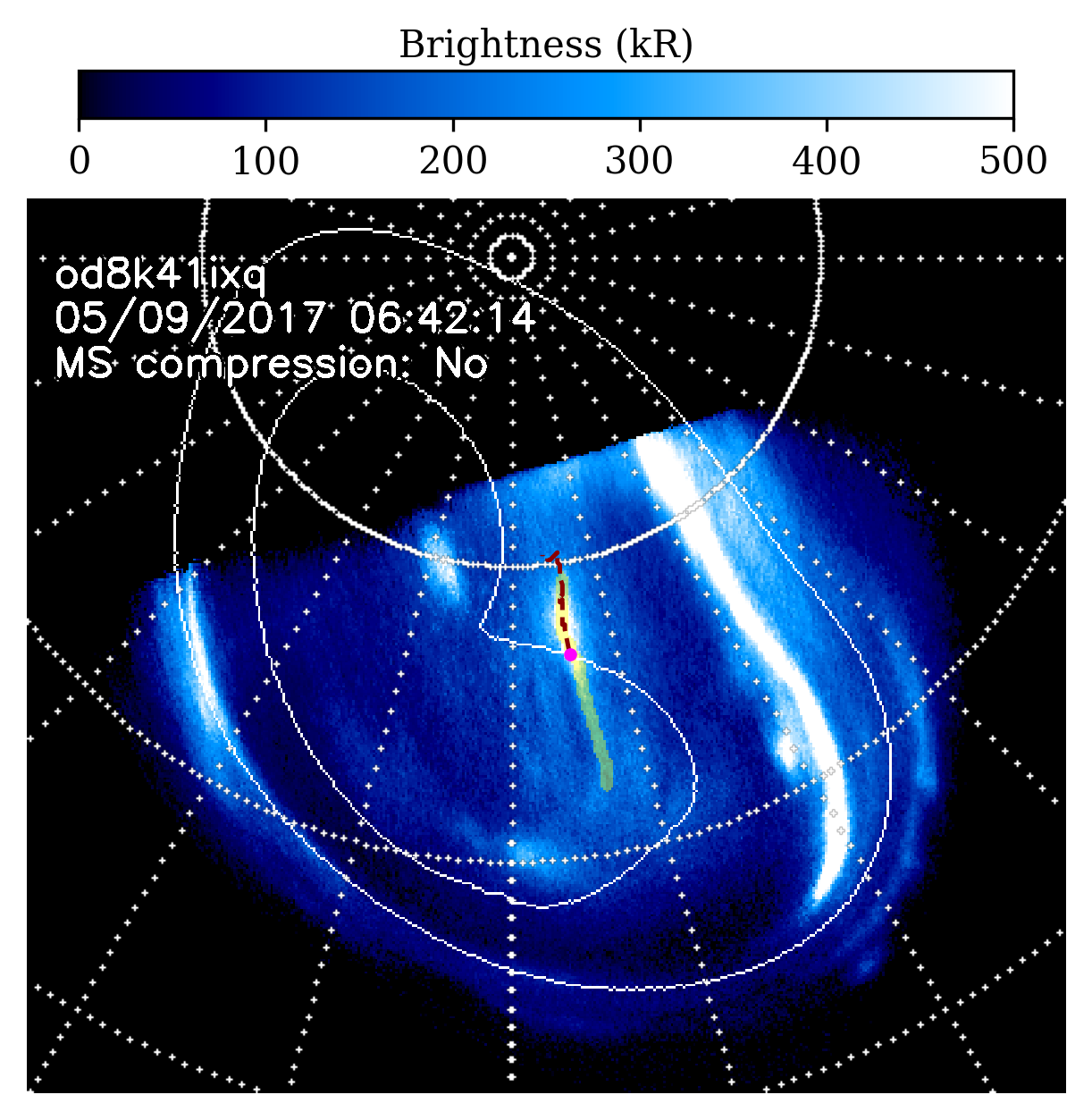}
    }
    \subfloat[]{
        \includegraphics[width=0.18\linewidth]{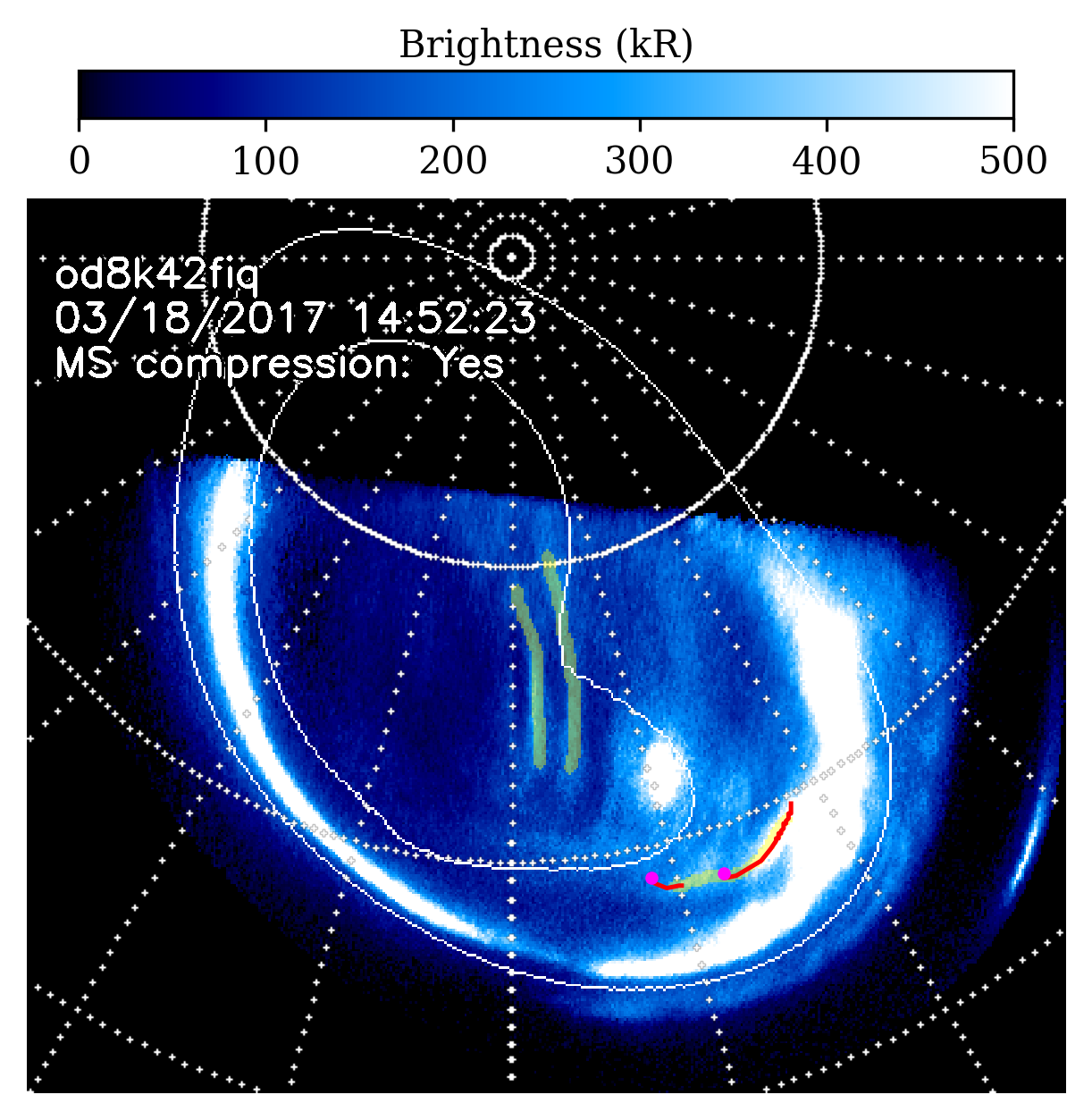}
    } 
    \subfloat[]{
        \includegraphics[width=0.18\linewidth]{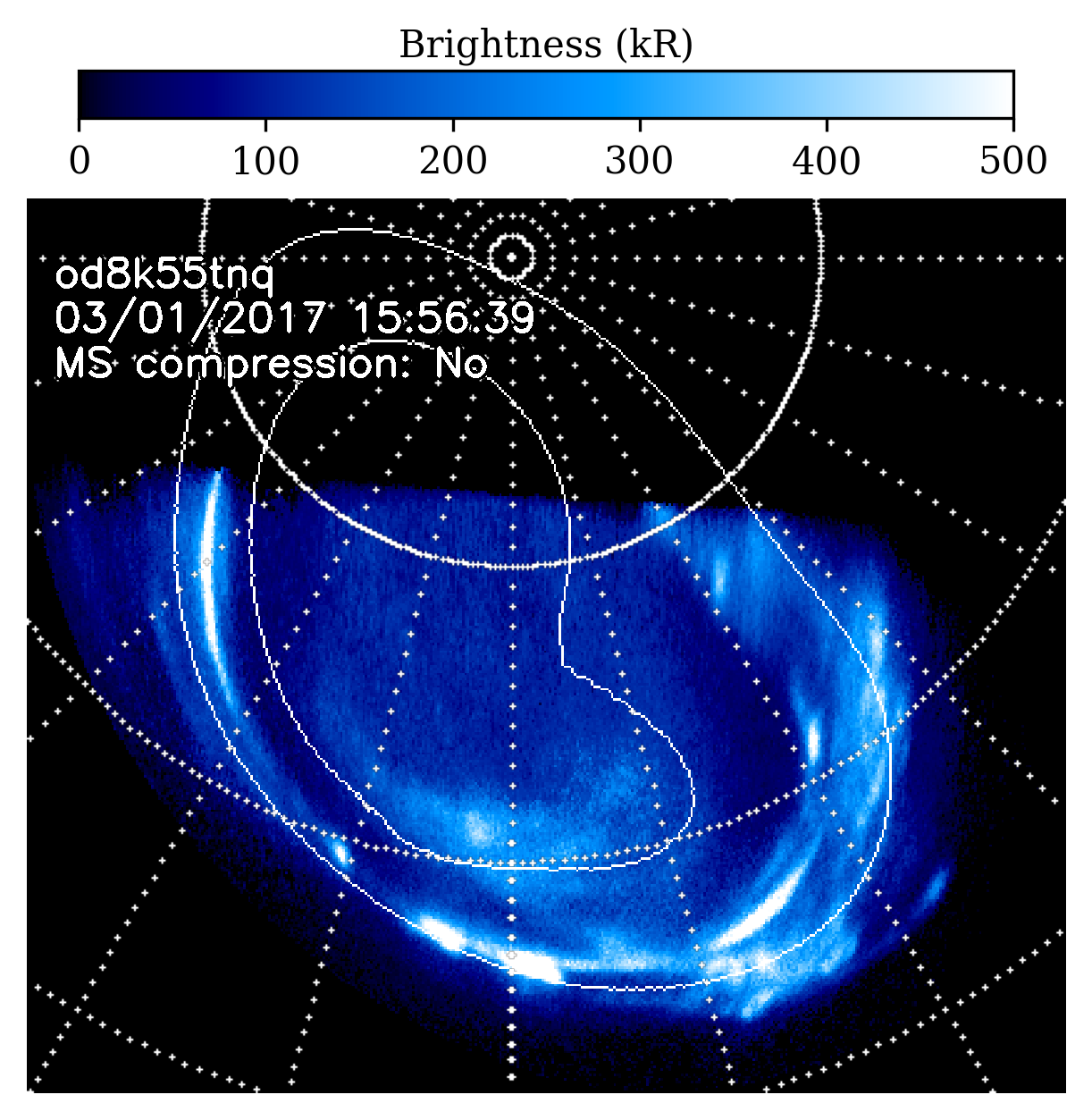}
    }\\
    \subfloat[]{
        \includegraphics[width=0.18\linewidth]{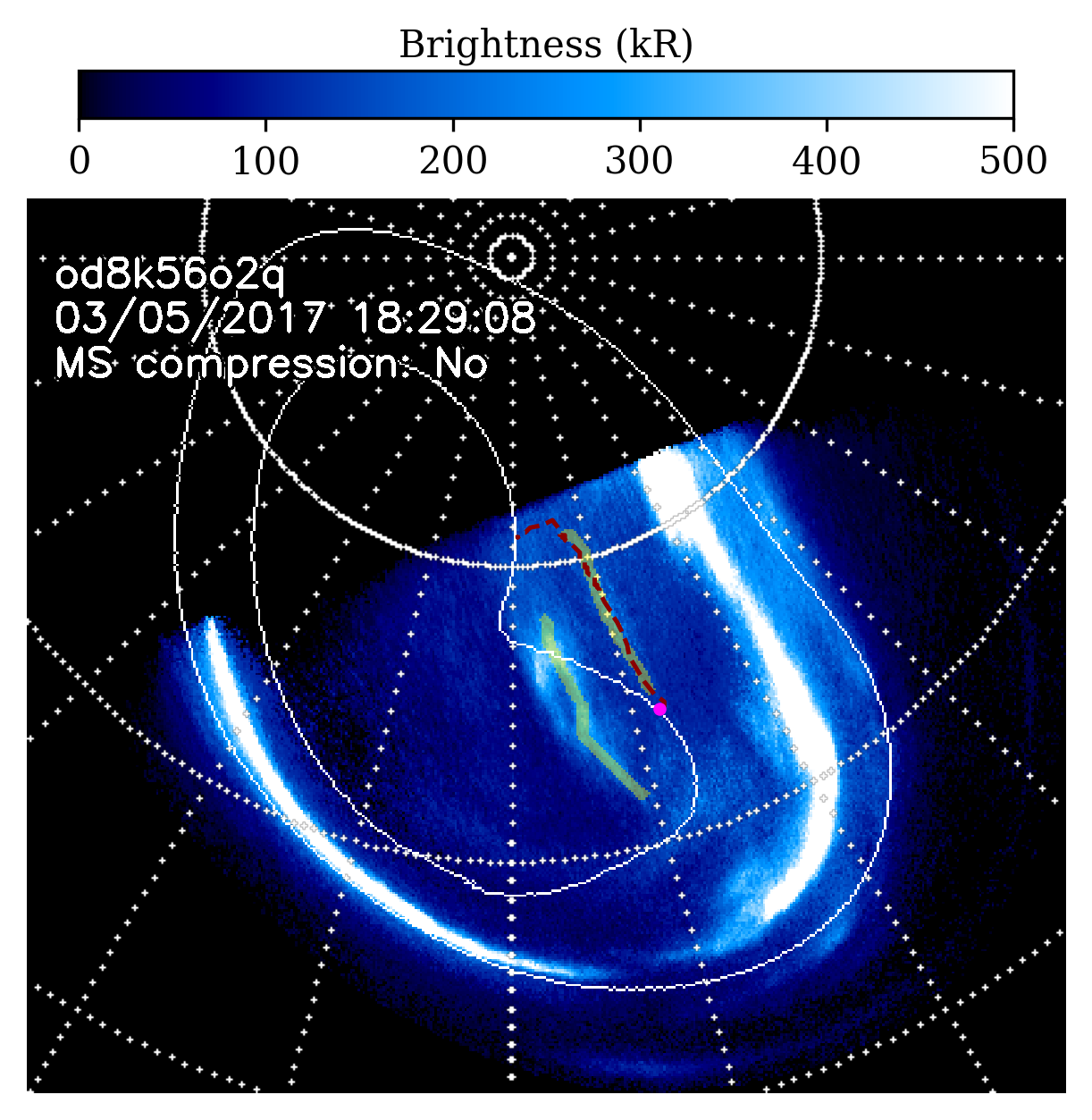}
    } 
    \subfloat[]{
        \includegraphics[width=0.18\linewidth]{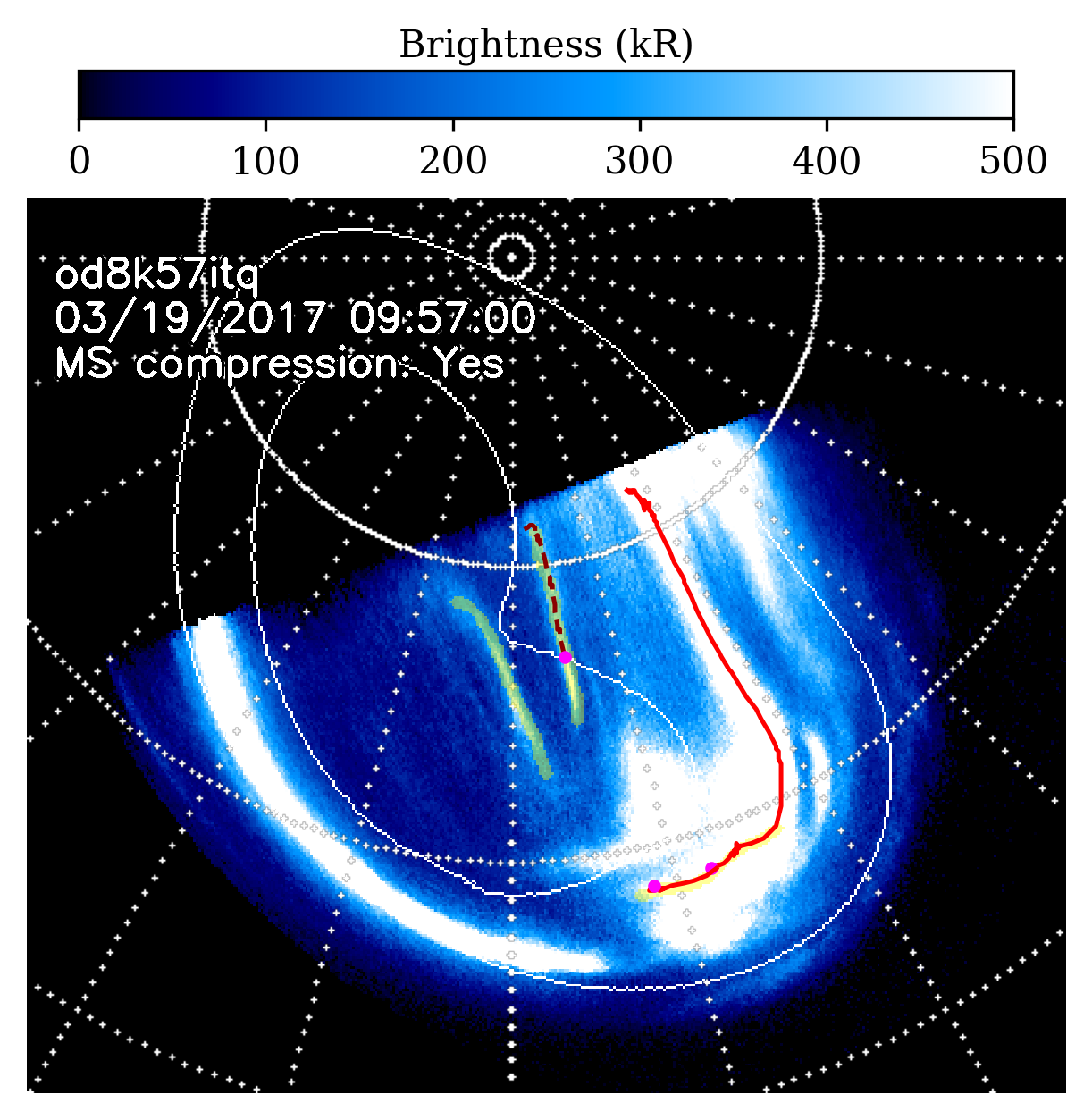}
    }
    \subfloat[]{
        \includegraphics[width=0.18\linewidth]{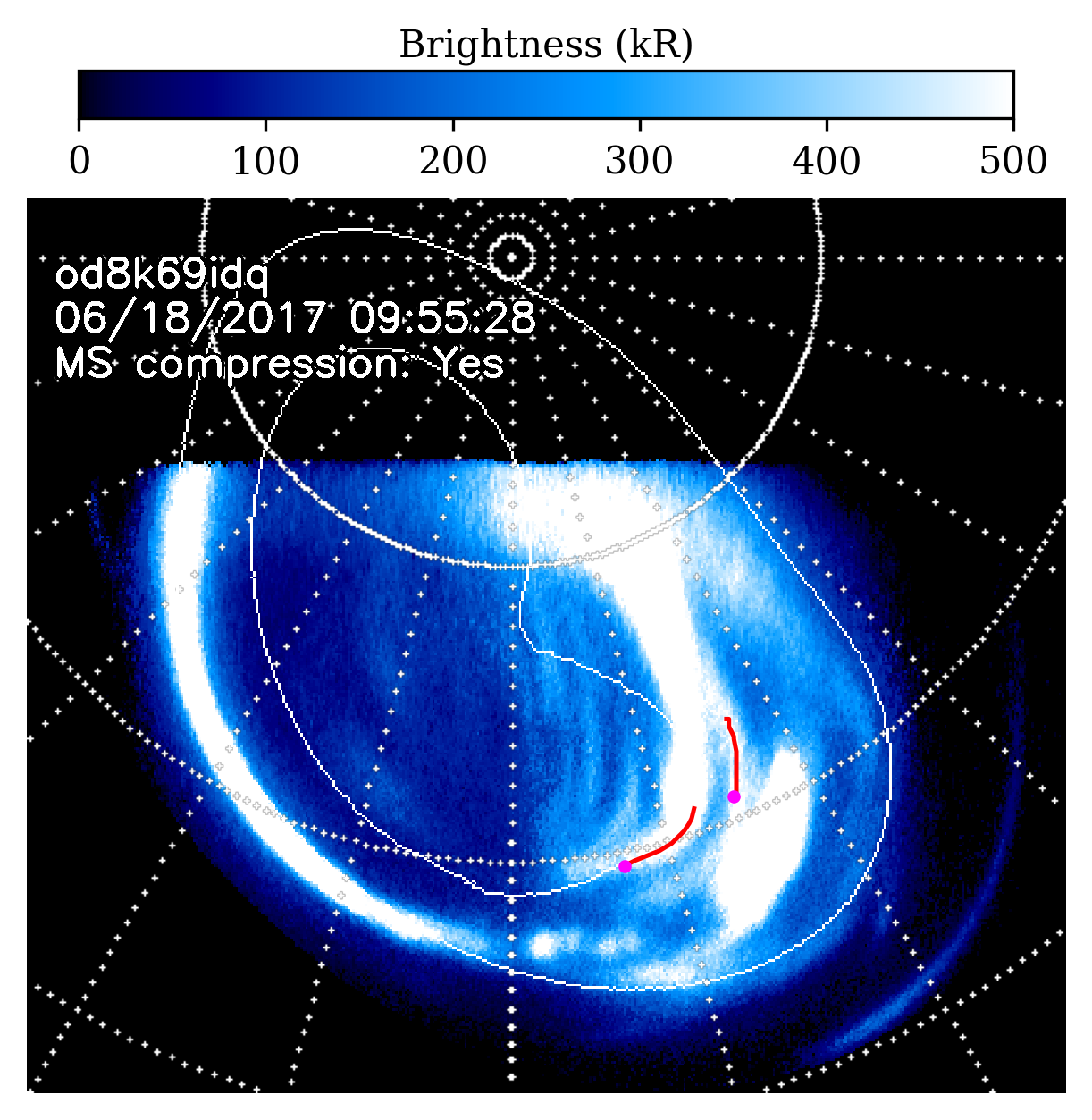}
    }
    \subfloat[]{
        \includegraphics[width=0.18\linewidth]{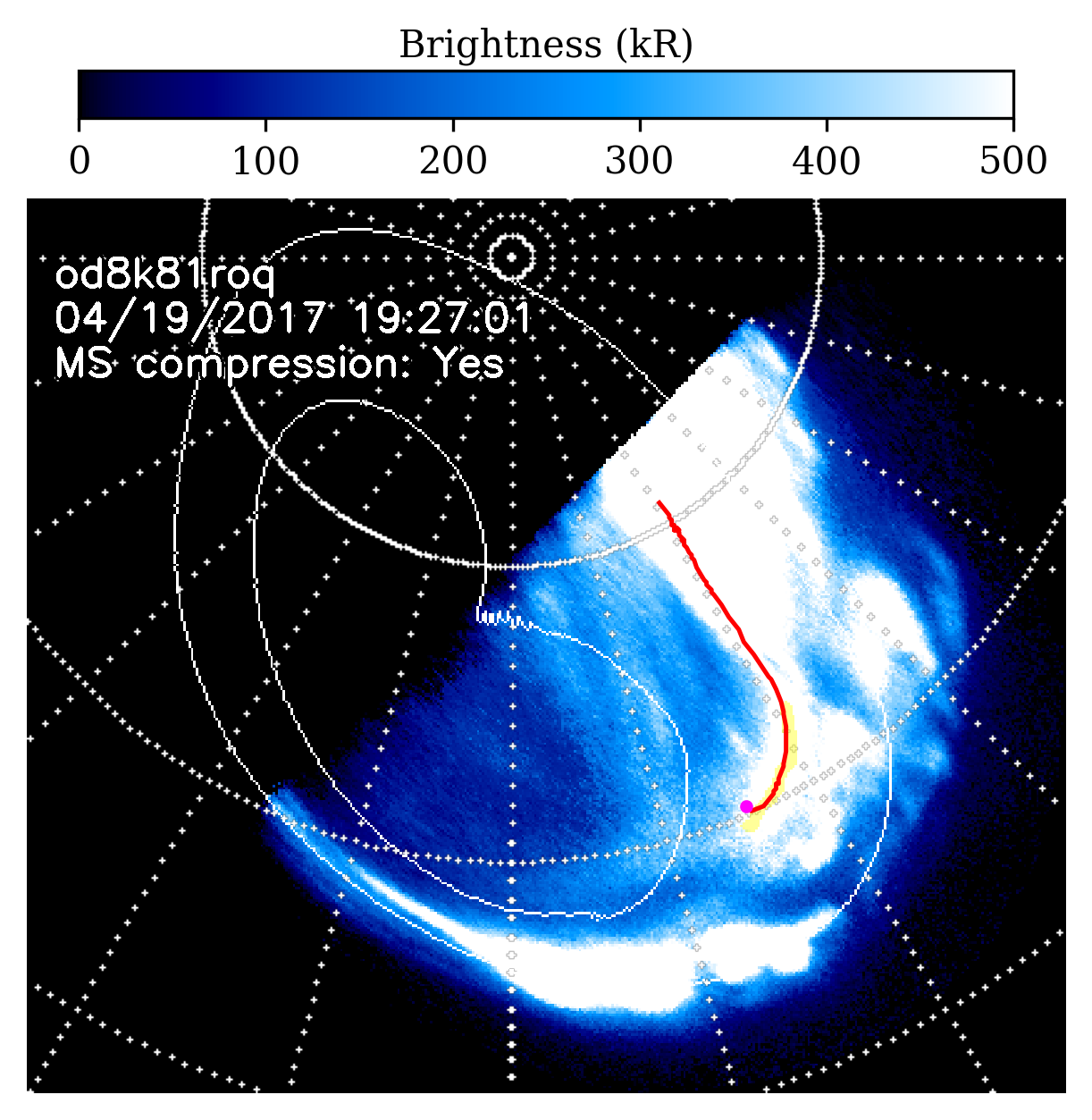}
    }
    \subfloat[]{
        \includegraphics[width=0.18\linewidth]{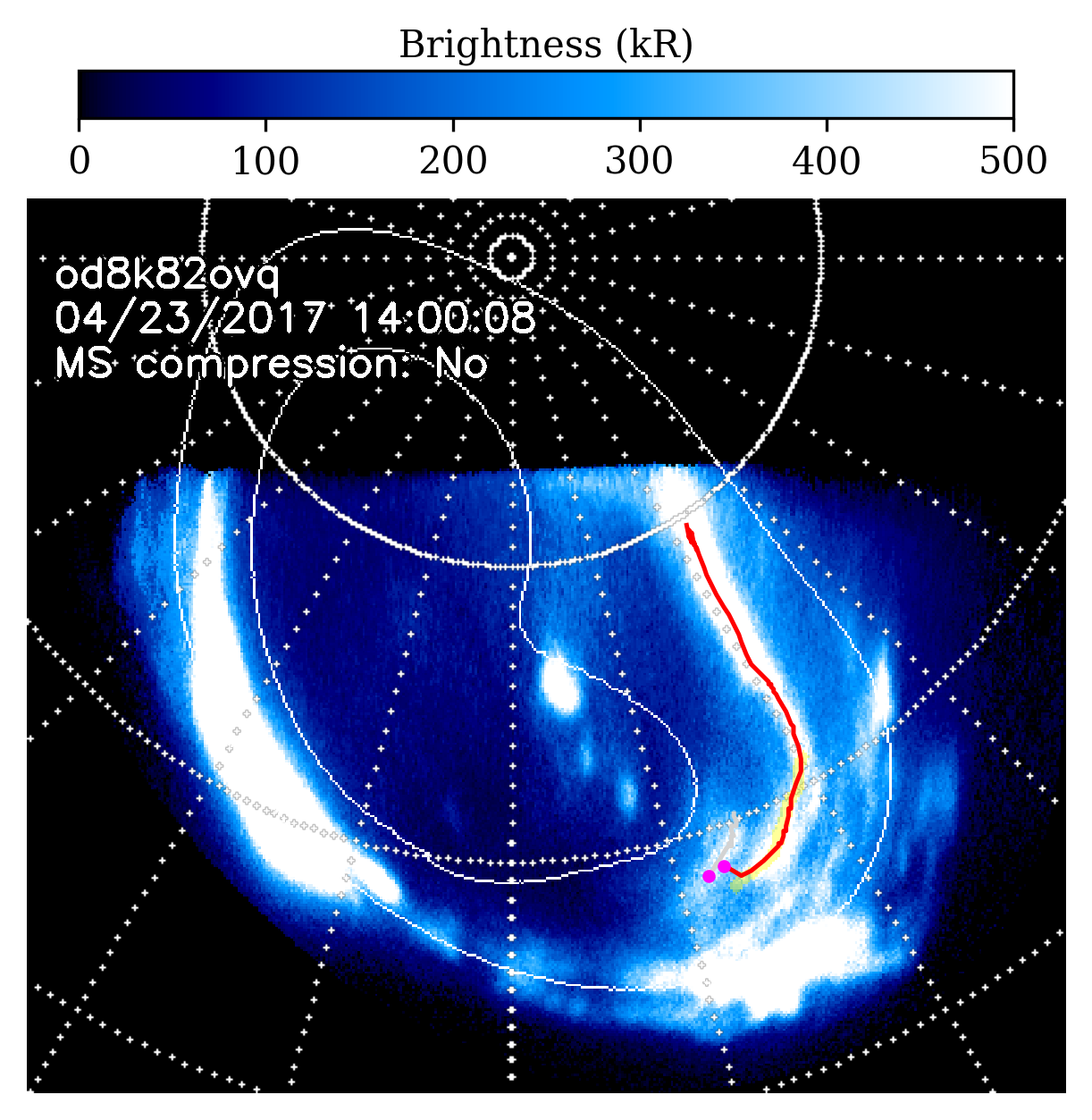}
    } 
    
    \caption{
        Results of the bridge-detection algorithm after filtering for cases where the compression state of the magnetosphere is known \citep{yao+:2022,louis+:2023}. Red lines denote bridge-candidate arcs that correspond to manual designations, and grey the arcs that do not but are nevertheless accepted by the random-forest filter. Dark-red dashed lines are those arcs that are identified as PAFs rather than bridge candidates. The seed point of each arc is given in magenta. Manually designated arcs are given in yellow. White contours give the region of validity of the \citet{vogt+:2011} JRM33 flux-equivalence mapping along closed field lines. HST exposure identifiers, central timestamps, and the compression state of the magnetosphere are given in the top left. 
    }
    \label{fig:compressed_aurora_images}
\end{figure}

\twocolumn
\begin{figure}[]
    \centering
    \captionsetup{width=\linewidth}
    \includegraphics[width=0.75\linewidth]{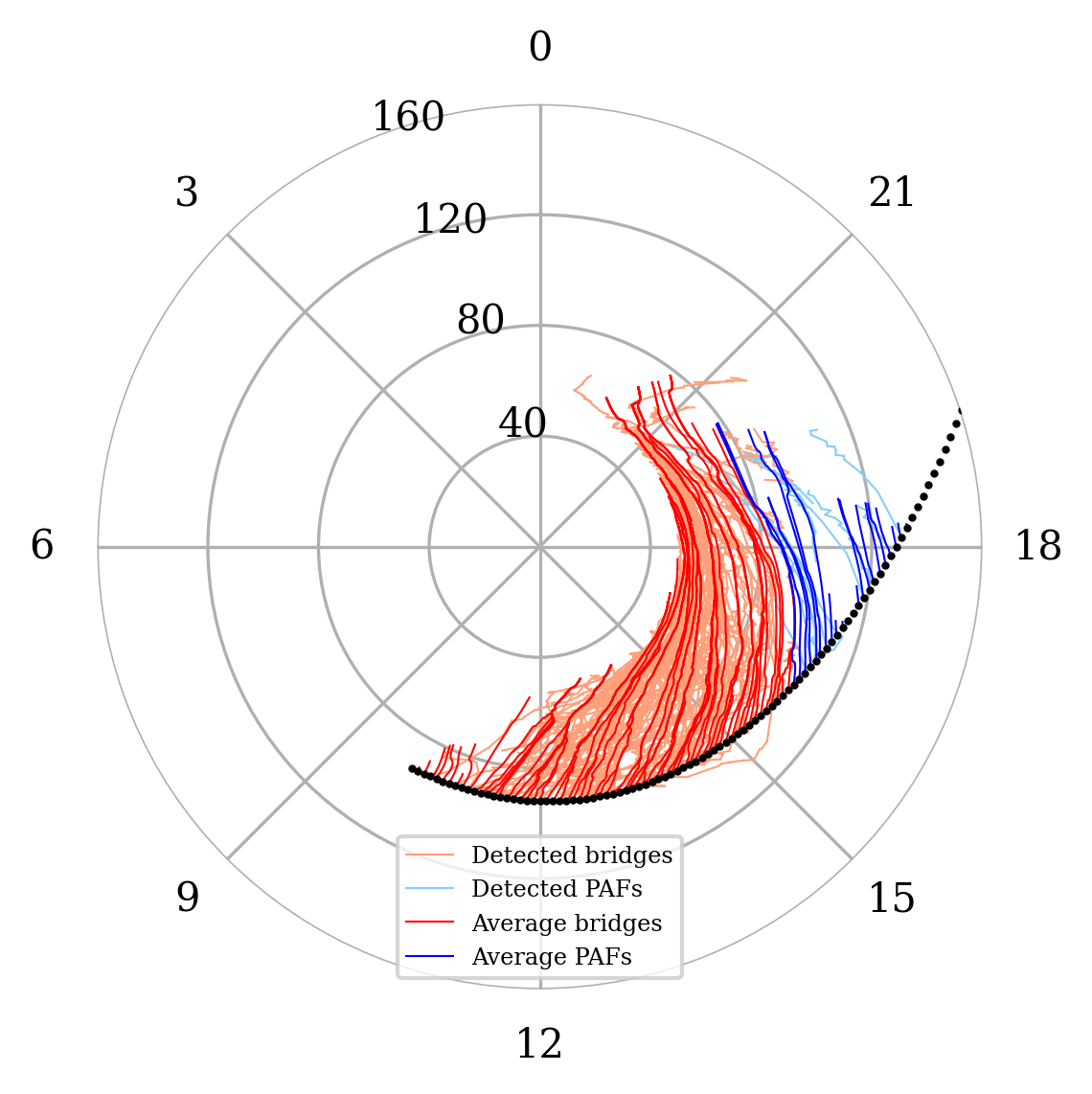}
    \caption{
        As Figure \ref{fig:HST_mapping}, but using field-line tracing of JRM33 \citep{connerney+:2022} and Con2020 \citep{connerney+:2020} to map the bridges from the aurora to the magnetosphere.
    }
    \label{fig:HST_mapping_flt}
\end{figure}
\begin{figure}[]
    \centering
    \captionsetup{width=\linewidth}
    \includegraphics[width=0.8\linewidth]{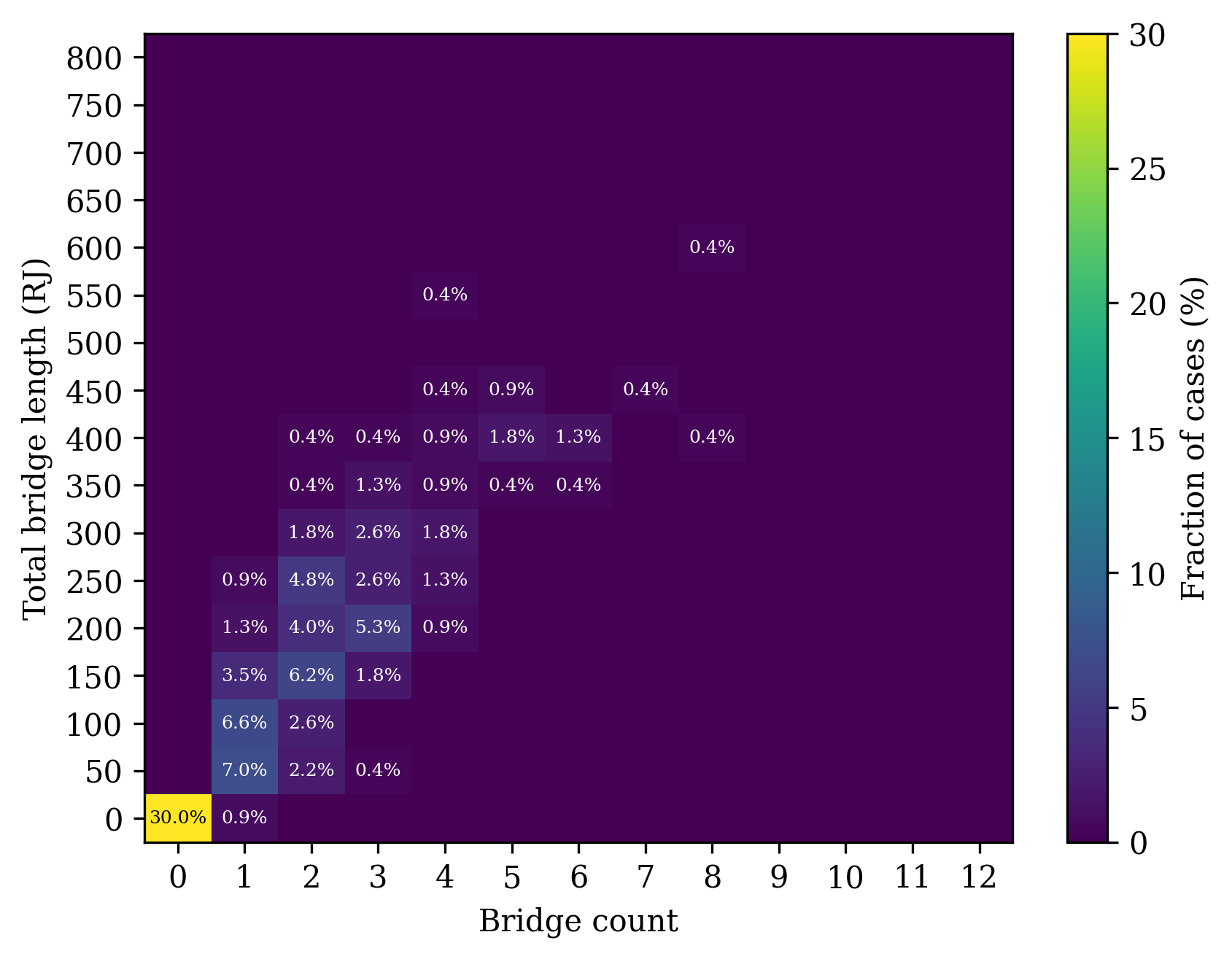}
    \caption{
        Detected bridge count vs total detected bridge magnetosphere-mapped length for the HST-STIS northern-hemisphere cases considered in section \ref{sec:HST_analysis}, expressed as a percentage of the total cases. There is an approximately linear relationship between bridge count and total bridge length.
    }
    \label{fig:bridge_length_vs_count}
\end{figure}
\begin{figure}[]
    \centering
    \captionsetup{width=\linewidth}
    \includegraphics[width=0.8\linewidth]{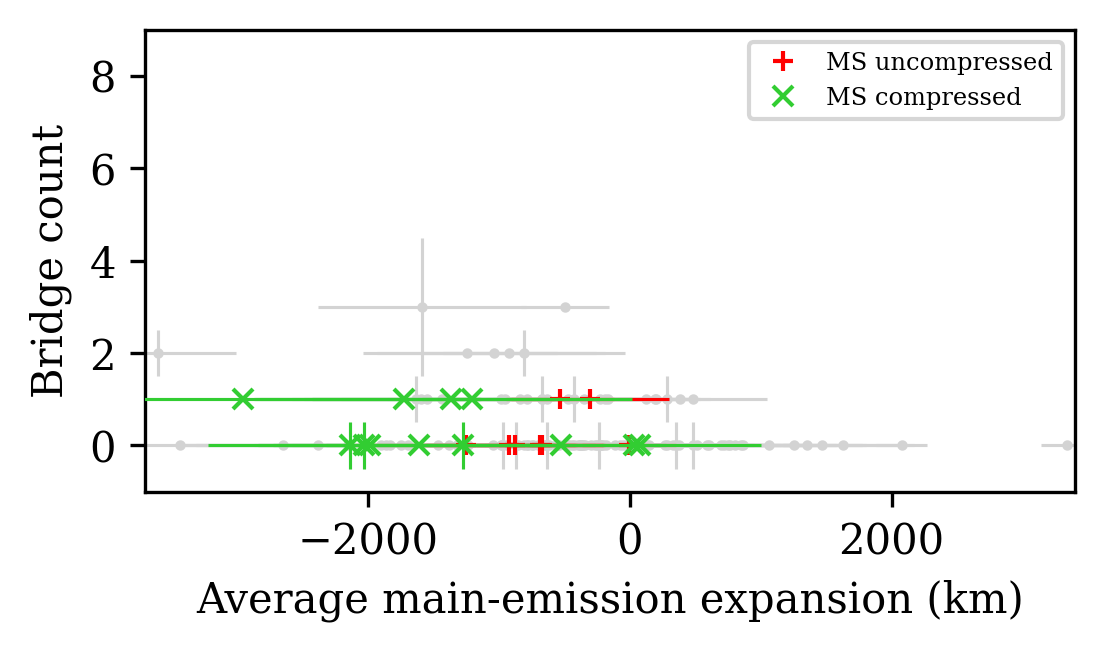}
    \caption{
         Detected average expansion of the ME from \citet{head+:2024} vs the detected PAF count for each northern HST-STIS series considered in this work. Negative expansions imply a contracted ME. Green crosses denote those cases with a compressed magnetosphere, and red pluses those cases with an uncompressed magnetosphere \citep{yao+:2022}; grey points denote cases where the compression state of the magnetosphere is unknown. 
    }
    \label{fig:paf_vs_ms_size}
\end{figure}

\begin{figure}[]
    \centering
    \captionsetup{width=\linewidth}
    \includegraphics[width=0.8\linewidth]{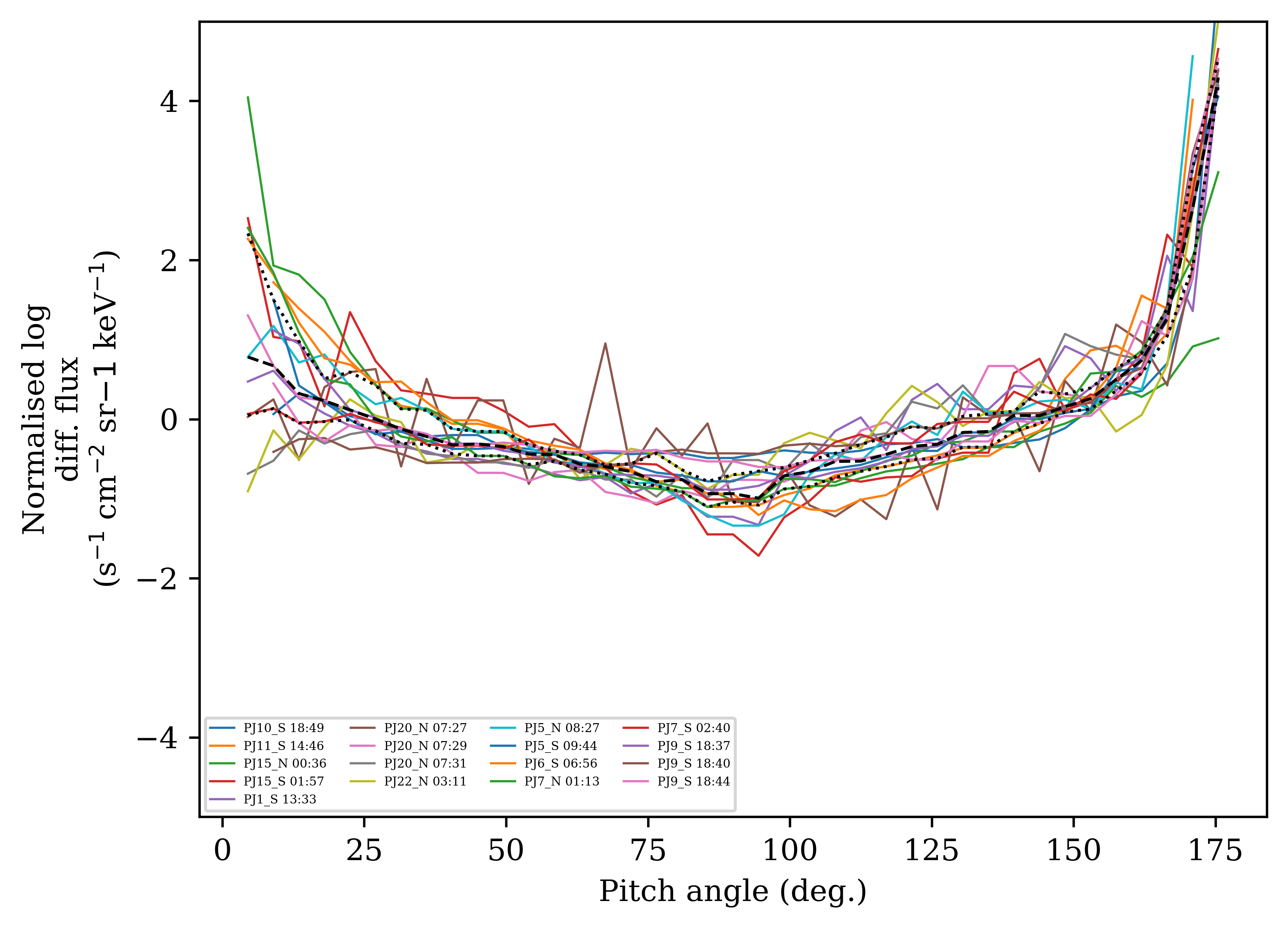}
    \caption{
         The normalised log JEDI electron differential flux pitch-angle profiles used to produce the average profile for bridge crossings in Figure \ref{fig:bridge_averages} (red). The median-average profile is given as a black dashed line. The 25th- and 75th-percentile values are given as black dotted lines.
    }
    \label{fig:pitch_angle_average_bridges}
\end{figure}

\begin{figure}[ht]
    \centering
    \captionsetup{width=\linewidth}
    \includegraphics[width=0.8\linewidth]{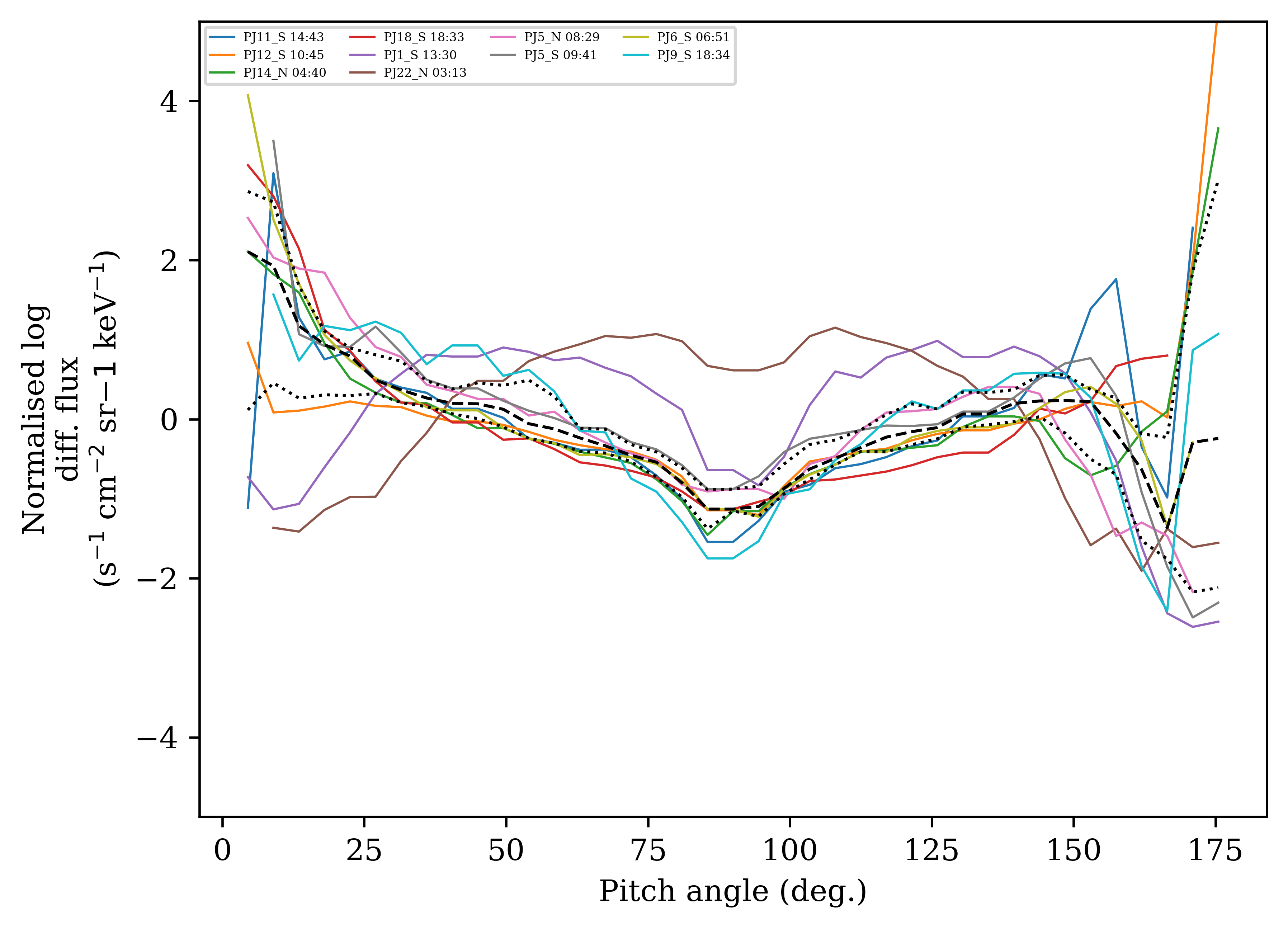}
    \caption{
         The normalised log JEDI electron differential flux pitch-angle profiles used to produce the average profile for main-emission crossings with nearby bridges in Figure \ref{fig:bridge_averages} (green). The median-average profile is given as a black dashed line. The 25th- and 75th-percentile values are given as black dotted lines.
    }
    \label{fig:pitch_angle_average_ME_bridge}
\end{figure}

\begin{figure}[ht]
    \centering
    \captionsetup{width=\linewidth}
    \includegraphics[width=0.8\linewidth]{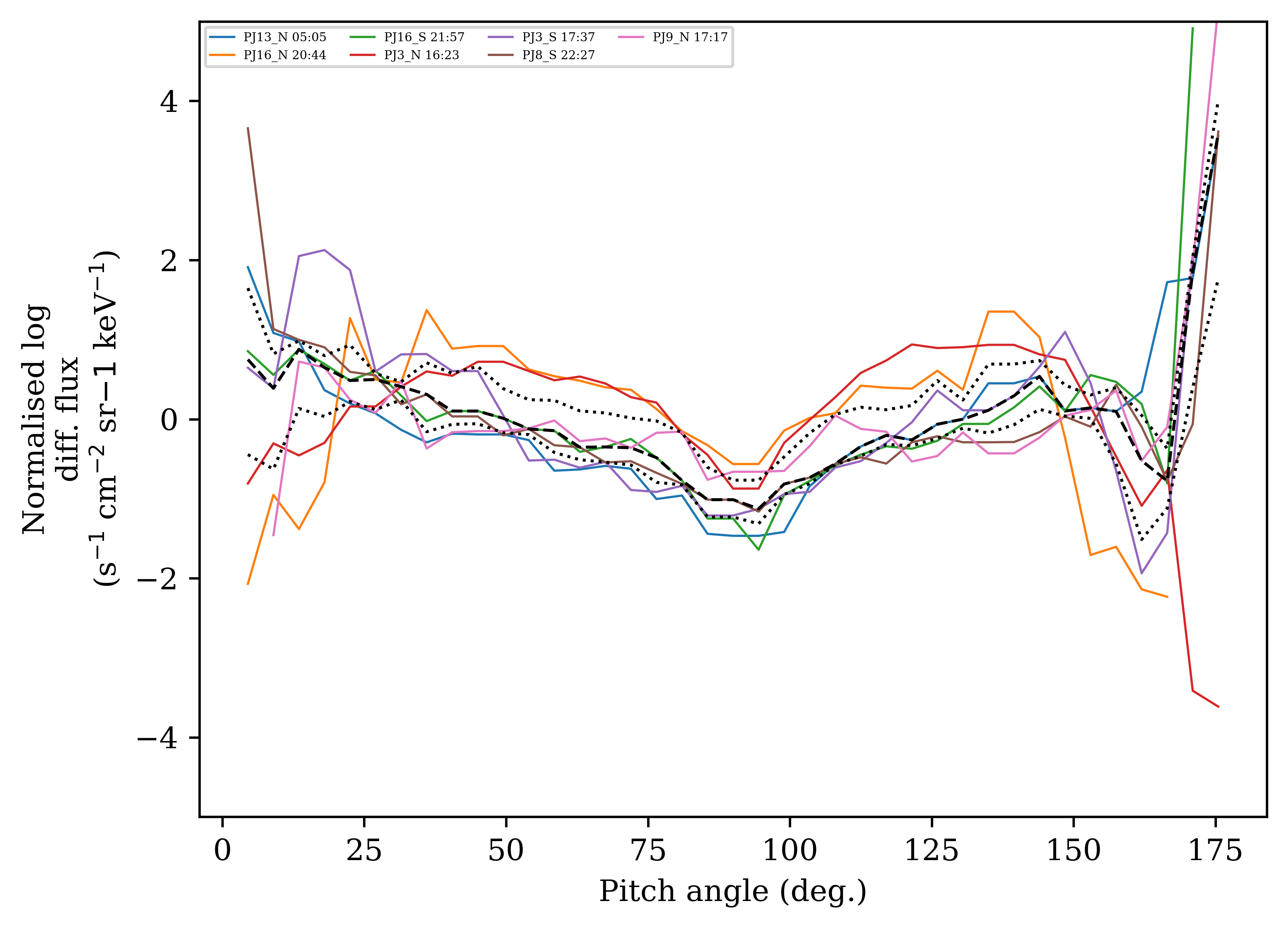}
    \caption{
         The normalised log JEDI electron differential flux pitch-angle profiles used to produce the average profile for main-emission crossings without nearby bridges in Figure \ref{fig:bridge_averages} (orange). The median-average profile is given as a black dashed line. The 25th- and 75th-percentile values are given as black dotted lines.
    }
    \label{fig:pitch_angle_average_ME_nobridge}
\end{figure}

\FloatBarrier

\onecolumn
\section{Supplementary tables}

\begin{ThreePartTable}
    \label{tab:uvs_cases}
    \centering
    \setlength\tabcolsep{4pt}


    \begin{TableNotes}
      \item\footnotesize ``Perijove'' gives the perijove number and hemisphere. ``Date'' gives the date and time of the polar crossing, in the format YYYY-MM-DD hh:mm:ss. ``$\phi_{SS}$'' denotes the left-handed System-III subsolar longitude of Jupiter at this time. ``Crossings'' give the timestamps of the semi-automatically detected arc crossings by Juno; \textcolor{red}{red} denotes bridge crossings and \textcolor{green_custom}{green} denotes main-emission crossings; note that cases with many apparent traversals of the main emission are those cases where the Juno footprint passes along the main-emission arc. ``Figure \ref{fig:bridge_averages} crossings'' further specify those low-altitude, dusk-side crossings used in the analysis presented in Figure \ref{fig:bridge_averages}; \textcolor{red}{red} denotes bridge crossings, \textcolor{green_custom}{green} denotes crossings of the main emission with associated bridges, and \textcolor{orange}{orange} those main-emission crossings without obvious bridges, in line with the key of Figure \ref{fig:bridge_averages}.
    \end{TableNotes}
   
    \begin{longtable}[htbp]{ llr p{5cm} p{5cm} }
        \label{tab:uvs_cases}\\
        \caption{Perijoves used in this work.}\\
    
        \toprule
        Perijove & Date & $\phi_{SS}$ (\textdegree) & Crossings & Figure \ref{fig:bridge_averages}, \ref{fig:waves_e_brightness} crossings\\
        \midrule
        \endhead

        \midrule[\heavyrulewidth]
        \multicolumn{4}{r}{\textit{continued...}}\\
        \endfoot

        \midrule[\heavyrulewidth]
        \insertTableNotes  
        \endlastfoot

PJ1-N & 27/08/2016 12:01 & 154   & \textcolor{green_custom}{2016-08-27 12:08:56 - 12:10:26} \newline \textcolor{red}{2016-08-27 12:09:56 - 12:11:01}  & - \\
PJ1-S & 27/08/2016 14:52 & 257   & \textcolor{green_custom}{2016-08-27 13:30:00 - 13:32:00} \newline \textcolor{red}{2016-08-27 13:33:00 - 13:36:00}  & \textcolor{green_custom}{2016-08-27 13:30:00 - 13:32:00} \newline \textcolor{red}{2016-08-27 13:33:00 - 13:36:00}  \\
PJ3-N & 11/12/2016 16:40 & 72    & \textcolor{green_custom}{2016-12-11 16:23:35 - 16:24:35} \newline \textcolor{green_custom}{2016-12-11 16:25:30 - 16:26:20}  & \textcolor{orange}{2016-12-11 16:23:35 - 16:24:35}  \\
PJ3-S & 11/12/2016 19:11 & 163   & \textcolor{green_custom}{2016-12-11 17:37:27 - 17:38:27}  & \textcolor{orange}{2016-12-11 17:37:27 - 17:38:27}  \\
PJ4-N & 02/02/2017 12:42 & 342   & \textcolor{green_custom}{2017-02-02 11:43:10 - 11:50:25} \newline \textcolor{green_custom}{2017-02-02 12:24:10 - 12:25:15}  & - \\
PJ4-S & 02/02/2017 14:37 & 51    & \textcolor{green_custom}{2017-02-02 13:38:16 - 13:39:51} \newline \textcolor{green_custom}{2017-02-02 15:14:11 - 15:34:01} \newline \textcolor{green_custom}{2017-02-02 15:44:36 - 16:04:26} \newline \textcolor{green_custom}{2017-02-02 16:09:51 - 16:10:16} \newline \textcolor{green_custom}{2017-02-02 16:10:56 - 16:11:26} \newline \textcolor{green_custom}{2017-02-02 16:24:51 - 16:32:31} \newline \textcolor{green_custom}{2017-02-02 16:34:56 - 16:35:41} \newline \textcolor{green_custom}{2017-02-02 16:36:21 - 16:38:26} \newline \textcolor{green_custom}{2017-02-02 16:39:06 - 16:40:01} \newline \textcolor{green_custom}{2017-02-02 16:40:36 - 16:44:36} \newline \textcolor{green_custom}{2017-02-02 17:12:01 - 17:31:56}  & - \\
PJ5-N & 27/03/2017 08:45 & 254   & \textcolor{green_custom}{2017-03-27 07:35:58 - 07:42:53} \newline \textcolor{red}{2017-03-27 08:27:18 - 08:28:43} \newline \textcolor{green_custom}{2017-03-27 08:29:18 - 08:30:28}  & \textcolor{red}{2017-03-27 08:27:18 - 08:28:43} \newline \textcolor{green_custom}{2017-03-27 08:29:18 - 08:30:28}  \\
PJ5-S & 27/03/2017 11:38 & 358   & \textcolor{green_custom}{2017-03-27 09:41:49 - 09:43:59} \newline \textcolor{red}{2017-03-27 09:44:49 - 09:48:49} \newline \textcolor{green_custom}{2017-03-27 12:27:39 - 12:47:34}  & \textcolor{green_custom}{2017-03-27 09:41:49 - 09:43:59} \newline \textcolor{red}{2017-03-27 09:44:49 - 09:48:49}  \\
PJ6-N & 19/05/2017 05:34 & 192   & \textcolor{green_custom}{2017-05-19 02:28:24 - 02:48:19} \newline \textcolor{green_custom}{2017-05-19 03:16:14 - 03:34:04} \newline \textcolor{green_custom}{2017-05-19 03:35:54 - 03:55:44} \newline \textcolor{green_custom}{2017-05-19 05:27:39 - 05:30:39}  & - \\
PJ6-S & 19/05/2017 07:40 & 268   & \textcolor{green_custom}{2017-05-19 06:51:01 - 06:55:11} \newline \textcolor{red}{2017-05-19 06:56:56 - 06:59:11} \newline \textcolor{green_custom}{2017-05-19 10:39:51 - 10:59:46}  & \textcolor{green_custom}{2017-05-19 06:51:01 - 06:55:11} \newline \textcolor{red}{2017-05-19 06:56:56 - 06:59:11}  \\
PJ7-N & 11/07/2017 01:05 & 84    & \textcolor{red}{2017-07-11 01:13:00 - 01:15:00} \newline \textcolor{green_custom}{2017-07-11 01:15:15 - 01:16:39}  & \textcolor{red}{2017-07-11 01:13:00 - 01:15:00}  \\
PJ7-S & 11/07/2017 03:51 & 184   & \textcolor{green_custom}{2017-07-11 02:33:16 - 02:35:19} \newline \textcolor{red}{2017-07-11 02:40:05 - 02:46:03}  & \textcolor{red}{2017-07-11 02:40:05 - 02:46:03}  \\
PJ8-N & 01/09/2017 21:20 & 2     & \textcolor{green_custom}{2017-09-01 21:16:43 - 21:17:33}  & - \\
PJ8-S & 01/09/2017 23:28 & 80    & \textcolor{green_custom}{2017-09-01 22:27:24 - 22:28:49}  & \textcolor{orange}{2017-09-01 22:27:24 - 22:28:49}  \\
PJ9-N & 24/10/2017 17:23 & 274   & \textcolor{green_custom}{2017-10-24 16:41:07 - 16:44:22} \newline \textcolor{green_custom}{2017-10-24 17:17:37 - 17:18:47}  & \textcolor{orange}{2017-10-24 17:17:37 - 17:18:47}  \\
PJ9-S & 24/10/2017 19:22 & 346   & \textcolor{green_custom}{2017-10-24 18:34:08 - 18:36:08} \newline \textcolor{red}{2017-10-24 18:37:13 - 18:39:43} \newline \textcolor{red}{2017-10-24 18:40:58 - 18:43:38} \newline \textcolor{red}{2017-10-24 18:44:58 - 18:48:13} \newline \textcolor{green_custom}{2017-10-24 21:09:03 - 21:27:53} \newline \textcolor{green_custom}{2017-10-24 21:28:23 - 21:28:53} \newline \textcolor{green_custom}{2017-10-24 21:34:08 - 21:35:33} \newline \textcolor{green_custom}{2017-10-24 21:36:08 - 21:38:28} \newline \textcolor{green_custom}{2017-10-24 21:40:03 - 21:54:58} \newline \textcolor{green_custom}{2017-10-24 22:00:03 - 22:09:43} \newline \textcolor{green_custom}{2017-10-24 22:10:18 - 22:10:43} \newline \textcolor{green_custom}{2017-10-24 22:12:13 - 22:12:58} \newline \textcolor{green_custom}{2017-10-24 22:19:08 - 22:44:28}  & \textcolor{green_custom}{2017-10-24 18:34:08 - 18:36:08} \newline \textcolor{red}{2017-10-24 18:37:13 - 18:39:43} \newline \textcolor{red}{2017-10-24 18:40:58 - 18:43:38} \newline \textcolor{red}{2017-10-24 18:44:58 - 18:48:13}  \\
PJ10-N & 16/12/2017 17:55 & 347   & \textcolor{green_custom}{2017-12-16 16:53:58 - 16:57:48}  & - \\
PJ10-S & 16/12/2017 19:36 & 48    & \textcolor{red}{2017-12-16 18:49:00 - 18:51:00}  & \textcolor{red}{2017-12-16 18:49:00 - 18:51:00}  \\
PJ11-N & 07/02/2018 14:00 & 259   & \textcolor{green_custom}{2018-02-07 08:50:15 - 08:56:35} \newline \textcolor{green_custom}{2018-02-07 09:23:15 - 09:28:15} \newline \textcolor{green_custom}{2018-02-07 12:53:45 - 12:59:10} \newline \textcolor{green_custom}{2018-02-07 13:31:50 - 13:32:55}  & - \\
PJ11-S & 07/02/2018 15:43 & 322   & \textcolor{green_custom}{2018-02-07 14:43:53 - 14:46:13} \newline \textcolor{red}{2018-02-07 14:46:13 - 14:49:13} \newline \textcolor{green_custom}{2018-02-07 17:48:48 - 17:50:18} \newline \textcolor{green_custom}{2018-02-07 18:09:03 - 18:28:53}  & \textcolor{green_custom}{2018-02-07 14:43:53 - 14:46:13} \newline \textcolor{red}{2018-02-07 14:46:13 - 14:49:13}  \\
PJ12-N & 01/04/2018 08:57 & 131   & \textcolor{green_custom}{2018-04-01 06:29:37 - 06:31:17} \newline \textcolor{green_custom}{2018-04-01 06:32:47 - 06:34:27} \newline \textcolor{green_custom}{2018-04-01 06:35:02 - 06:36:17} \newline \textcolor{green_custom}{2018-04-01 07:02:32 - 07:13:47} \newline \textcolor{green_custom}{2018-04-01 07:15:22 - 07:35:17} \newline \textcolor{green_custom}{2018-04-01 08:34:37 - 08:52:37} \newline \textcolor{green_custom}{2018-04-01 09:05:17 - 09:09:02}  & - \\
PJ12-S & 01/04/2018 11:25 & 220   & \textcolor{green_custom}{2018-04-01 10:45:19 - 10:48:54} \newline \textcolor{red}{2018-04-01 10:48:58 - 10:53:00}  & \textcolor{green_custom}{2018-04-01 10:45:19 - 10:48:54}  \\
PJ13-N & 24/05/2018 05:22 & 55    & \textcolor{green_custom}{2018-05-24 05:05:51 - 05:06:41}  & \textcolor{orange}{2018-05-24 05:05:51 - 05:06:41}  \\
PJ13-S & 24/05/2018 07:19 & 126   & \textcolor{red}{2018-05-24 06:27:39 - 06:29:54} \newline \textcolor{red}{2018-05-24 06:31:29 - 06:33:29}  & - \\
PJ14-N & 16/07/2018 05:03 & 98    & \textcolor{green_custom}{2018-07-16 03:03:14 - 03:23:09} \newline \textcolor{green_custom}{2018-07-16 03:30:14 - 03:36:09} \newline \textcolor{red}{2018-07-16 04:38:24 - 04:39:49} \newline \textcolor{green_custom}{2018-07-16 04:40:24 - 04:42:59}  & - \\
PJ14-S & 16/07/2018 06:57 & 167   & \textcolor{green_custom}{2018-07-16 06:09:33 - 06:14:28}  & - \\
PJ15-N & 07/09/2018 01:12 & 13    & \textcolor{red}{2018-09-07 00:36:22 - 07:37:40}  & \textcolor{red}{2018-09-07 00:36:22 - 07:37:40}  \\
PJ15-S & 07/09/2018 02:50 & 72    & \textcolor{red}{2018-09-07 01:57:13 - 02:00:08} \newline \textcolor{red}{2018-09-07 02:00:43 - 02:02:58}  & \textcolor{red}{2018-09-07 01:57:13 - 02:00:08}  \\
PJ16-N & 29/10/2018 21:09 & 280   & \textcolor{green_custom}{2018-10-29 20:16:22 - 20:18:47} \newline \textcolor{green_custom}{2018-10-29 20:44:22 - 20:45:17}  & \textcolor{orange}{2018-10-29 20:44:22 - 20:45:17}  \\
PJ16-S & 29/10/2018 22:45 & 338   & \textcolor{green_custom}{2018-10-29 21:57:49 - 21:58:50}  & \textcolor{orange}{2018-10-29 21:57:49 - 21:58:50}  \\
PJ17-N & 21/12/2018 17:02 & 185   & -     & - \\
PJ17-S & 21/12/2018 18:54 & 253   & \textcolor{green_custom}{2018-12-21 18:05:14 - 18:07:59}  & - \\
PJ18-N & 12/02/2019 17:43 & 264   & \textcolor{green_custom}{2019-02-12 16:51:06 - 16:52:56}  & - \\
PJ18-S & 12/02/2019 19:14 & 319   & \textcolor{green_custom}{2019-02-12 18:33:57 - 18:36:27} \newline \textcolor{red}{2019-02-12 18:38:27 - 18:42:07} \newline \textcolor{red}{2019-02-12 20:41:47 - 20:57:02} \newline \textcolor{green_custom}{2019-02-12 21:40:47 - 21:42:32} \newline \textcolor{green_custom}{2019-02-12 21:43:02 - 21:43:27} \newline \textcolor{green_custom}{2019-02-12 21:45:12 - 21:45:42} \newline \textcolor{green_custom}{2019-02-12 21:46:17 - 21:46:42} \newline \textcolor{green_custom}{2019-02-12 21:47:52 - 21:48:47} \newline \textcolor{green_custom}{2019-02-12 22:03:47 - 22:23:37} \newline \textcolor{red}{2019-02-12 22:22:52 - 22:23:32} \newline \textcolor{red}{2019-02-12 22:25:07 - 22:28:17} \newline \textcolor{red}{2019-02-12 22:28:47 - 22:29:52}  & \textcolor{green_custom}{2019-02-12 18:33:57 - 18:36:27}  \\
PJ19-N & 06/04/2019 11:55 & 108   & \textcolor{green_custom}{2019-04-06 09:16:41 - 09:27:46} \newline \textcolor{green_custom}{2019-04-06 09:28:46 - 09:29:11} \newline \textcolor{green_custom}{2019-04-06 09:44:16 - 09:45:36} \newline \textcolor{green_custom}{2019-04-06 09:46:06 - 09:58:06} \newline \textcolor{green_custom}{2019-04-06 09:58:46 - 10:07:41} \newline \textcolor{green_custom}{2019-04-06 10:43:31 - 10:47:56}  & - \\
PJ19-S & 06/04/2019 13:54 & 180   & \textcolor{green_custom}{2019-04-06 13:23:05 - 13:28:20}  & - \\
PJ20-N & 29/05/2019 08:07 & 24    & \textcolor{red}{2019-05-29 07:27:51 - 07:28:52} \newline \textcolor{red}{2019-05-29 07:29:00 - 07:30:00} \newline \textcolor{red}{2019-05-29 07:31:55 - 07:32:56}  & \textcolor{red}{2019-05-29 07:27:51 - 07:28:52} \newline \textcolor{red}{2019-05-29 07:29:00 - 07:30:00} \newline \textcolor{red}{2019-05-29 07:31:55 - 07:32:56}  \\
PJ20-S & 29/05/2019 09:48 & 86    & \textcolor{green_custom}{2019-05-29 08:57:26 - 08:59:31} \newline \textcolor{red}{2019-05-29 12:57:51 - 13:10:41}  & - \\
PJ21-N & 21/07/2019 04:06 & 293   & \textcolor{green_custom}{2019-07-21 03:07:54 - 03:09:59} \newline \textcolor{green_custom}{2019-07-21 03:37:29 - 03:38:19}  & - \\
PJ21-S & 21/07/2019 05:59 & 1     & \textcolor{green_custom}{2019-07-21 04:59:29 - 05:01:29} \newline \textcolor{red}{2019-07-21 08:48:04 - 09:07:09}  & - \\
PJ22-N & 12/09/2019 03:45 & 334   & \textcolor{green_custom}{2019-09-12 02:28:45 - 02:32:00} \newline \textcolor{red}{2019-09-12 03:11:25 - 03:12:20} \newline \textcolor{green_custom}{2019-09-12 03:13:55 - 03:14:40}  & \textcolor{red}{2019-09-12 03:11:25 - 03:12:20} \newline \textcolor{green_custom}{2019-09-12 03:13:55 - 03:14:40}  \\
PJ22-S & 12/09/2019 05:20 & 31    & \textcolor{green_custom}{2019-09-12 04:28:55 - 04:30:40}  & - \\
PJ23-N & 03/11/2019 22:34 & 200   & -     & - \\
PJ23-S & 04/11/2019 00:14 & 261   & \textcolor{green_custom}{2019-11-03 23:25:35 - 23:29:20} \newline \textcolor{red}{2019-11-04 01:14:55 - 01:25:00} \newline \textcolor{red}{2019-11-04 02:54:05 - 03:13:55}  & - \\
PJ24-N & 26/12/2019 17:23 & 67    & \textcolor{green_custom}{2019-12-26 14:46:42 - 14:48:57}  & - \\
PJ24-S & 26/12/2019 20:08 & 166   & \textcolor{green_custom}{2019-12-26 18:52:07 - 19:01:07}  & - \\
PJ25-S & 17/02/2020 19:52 & 210   & \textcolor{red}{2020-02-17 22:22:44 - 22:37:29}  & - \\
PJ26-N & 10/04/2020 13:42 & 41    & -     & - \\
PJ26-S & 10/04/2020 15:27 & 104   & \textcolor{green_custom}{2020-04-10 15:13:02 - 15:25:42} \newline \textcolor{green_custom}{2020-04-10 15:26:17 - 15:26:52}  & - \\
PJ27-N & 02/06/2020 10:22 & 334   & \textcolor{green_custom}{2020-06-02 09:13:35 - 09:21:05}  & - \\
PJ27-S & 02/06/2020 12:21 & 46    & \textcolor{red}{2020-06-02 14:47:40 - 15:07:30}  & - \\
PJ28-N & 25/07/2020 06:25 & 245   & -     & - \\
PJ28-S & 25/07/2020 07:54 & 299   & \textcolor{green_custom}{2020-07-25 07:15:20 - 07:17:40} \newline \textcolor{red}{2020-07-25 09:30:35 - 09:31:55} \newline \textcolor{red}{2020-07-25 09:42:40 - 10:02:35} \newline \textcolor{red}{2020-07-25 10:13:30 - 10:14:10} \newline \textcolor{red}{2020-07-25 10:36:00 - 10:36:35} \newline \textcolor{red}{2020-07-25 10:37:10 - 10:37:40} \newline \textcolor{red}{2020-07-25 10:47:35 - 11:07:30}  & - \\
PJ29-N & 16/09/2020 02:25 & 153   & -     & - \\
PJ29-S & 16/09/2020 04:23 & 225   & \textcolor{green_custom}{2020-09-16 03:26:32 - 03:30:07} \newline \textcolor{red}{2020-09-16 06:18:52 - 06:32:12} \newline \textcolor{red}{2020-09-16 07:07:22 - 07:11:32}  & - \\
PJ30-N & 08/11/2020 01:57 & 190   & \textcolor{green_custom}{2020-11-08 01:20:16 - 01:21:41}  & - \\
PJ30-S & 08/11/2020 03:28 & 246   & \textcolor{green_custom}{2020-11-08 02:52:37 - 02:55:47}  & - \\

    \end{longtable}   
\end{ThreePartTable}
\end{appendix}

\end{document}